\documentclass[pra,twocolumn,superscriptaddress,floatfix,nofootinbib,amsmath,amssymb]{revtex4-1}

\usepackage[english]{babel}
\usepackage[T1]{fontenc}
\usepackage{float}
\usepackage{graphicx}
\usepackage{bm}
\usepackage{braket}
\usepackage{xcolor}
\usepackage{amsfonts}
\usepackage{comment}
\usepackage{dirtytalk}
\usepackage{hyperref}
\usepackage{cleveref}
\usepackage{url}
\usepackage{bbold}
\usepackage{realboxes}
\def\code#1{\texttt{#1}}
\usepackage{enumitem}

\usepackage{algpseudocode}

\newcommand{\fancylink}[2]{\[\footnotesize\Colorbox{bkgd}{\color{orange-red}\href{#1}{\code{#2}}}\]}

\newcommand{\spliteq}[1]{\begin{equation}
\begin{split}
#1
\end{split}
\end{equation}
}

\newcommand{\tr}{\text{tr}}

\usepackage{listings}
\usepackage{xcolor}

\definecolor{bkgd}{RGB}{240,242,246}
\definecolor{ceruleanblue}{rgb}{0.16, 0.32, 0.75}
\definecolor{orange-red}{rgb}{1.0, 0.27, 0.0}
\definecolor{anotherblue}{RGB}{37,92,243}
\definecolor{blackblue}{RGB}{46,60,85}
\definecolor{goldyellow}{RGB}{199,146,12}
\lstdefinestyle{altstyle2}{
    backgroundcolor=\color{bkgd},
    basicstyle=\ttfamily\footnotesize\color{blackblue},
    breakatwhitespace=false,
    breaklines=true,
    captionpos=b,
    commentstyle=\color{goldyellow},
    keepspaces=true,
    keywordstyle=\color{orange-red},
    language=Python,
    numbersep=5pt,
    numberstyle=\tiny\color{ceruleanblue},
    showspaces=false,
    showstringspaces=false,
    showtabs=false,
    stringstyle=\color{anotherblue},
    tabsize=2
}
\begin{document}

\title{TensorFlow Quantum:\\
A Software Framework for Quantum Machine Learning
}

\affiliation{Google Quantum AI, Mountain View, CA
}

\affiliation{Sandbox@Alphabet, Mountain View, CA}

\affiliation{Google, Mountain View, CA
}

\affiliation{Institute for Quantum Computing, University of Waterloo, Waterloo, Ontario, N2L 3G1, Canada
}

\author{Michael Broughton}
\email{mbbrough@google.com}

\affiliation{Google Quantum AI, Mountain View, CA
}

\affiliation{School of Computer Science, University of Waterloo, Waterloo, Ontario, N2L 3G1, Canada
}

\author{Guillaume Verdon}
\email{gverdon@google.com}
\affiliation{Google Quantum AI, Mountain View, CA
}

\affiliation{Sandbox@Alphabet, Mountain View, CA}

\affiliation{Institute for Quantum Computing, University of Waterloo, Waterloo, Ontario, N2L 3G1, Canada
}
\affiliation{Department of Applied Mathematics, University of Waterloo, Waterloo, Ontario, N2L 3G1, Canada
}

\author{Trevor McCourt}

\affiliation{Google Quantum AI, Mountain View, CA
}
\affiliation{Department of Mechanical \& Mechatronics Engineering, University of Waterloo, Waterloo, Ontario, N2L 3G1, Canada
}

\author{Antonio J. Martinez}

\affiliation{Google Quantum AI, Mountain View, CA
}
\affiliation{Sandbox@Alphabet, Mountain View, CA}
 \affiliation{Institute for Quantum Computing, University of Waterloo, Waterloo, Ontario, N2L 3G1, Canada
}

 \affiliation{Department of Physics \& Astronomy, University of Waterloo, Waterloo, Ontario, N2L 3G1, Canada
}

\author{\\Jae Hyeon Yoo}

\affiliation{Sandbox@Alphabet, Mountain View, CA}
\affiliation{Google, Mountain View, CA
}

\author{Sergei V. Isakov}

\affiliation{Google Quantum AI, Mountain View, CA
}

\author{Philip Massey}

\affiliation{Google, Mountain View, CA
}

\author{Ramin Halavati}
\affiliation{Google, Mountain View, CA
}

\author{Murphy Yuezhen Niu}

\affiliation{Google Quantum AI, Mountain View, CA
}

\author{Alexander Zlokapa}
\affiliation{Division of Physics, Mathematics and Astronomy, Caltech, Pasadena, CA 91125}

\affiliation{Google Quantum AI, Mountain View, CA
}

\author{Evan Peters}

\affiliation{Institute for Quantum Computing, University of Waterloo, Waterloo, Ontario, N2L 3G1, Canada
}
\affiliation{Department of Applied Mathematics, University of Waterloo, Waterloo, Ontario, N2L 3G1, Canada
}
\affiliation{Fermi National Accelerator Laboratory, P.O. Box 500, Batavia, IL, 605010
}

\author{Owen Lockwood}
\affiliation{Department of Computer Science, Rensselaer Polytechnic Institute, Troy, NY 12180, USA}

\author{Andrea Skolik}
\affiliation{Data:Lab, Volkswagen Group, Ungererstr. 69, 80805 München, Germany}
\affiliation{Leiden University, Niels Bohrweg 1, 2333 CA Leiden, Netherlands}
\affiliation{Quantum Artificial Intelligence Laboratory, NASA Ames Research Center (QuAIL)}
\affiliation{USRA Research Institute for Advanced Computer Science (RIACS)}

\author{Sofiene Jerbi}
\affiliation{Institute for Theoretical Physics, University of Innsbruck, Technikerstr. 21a, A-6020 Innsbruck, Austria}

\author{Vedran Dunjko}
\affiliation{Leiden University, Niels Bohrweg 1, 2333 CA Leiden, Netherlands}

\author{Martin Leib}
\affiliation{Data:Lab, Volkswagen Group, Ungererstr. 69, 80805 München, Germany}

\author{Michael Streif}
\affiliation{Data:Lab, Volkswagen Group, Ungererstr. 69, 80805 München, Germany}

\affiliation{Quantum Artificial Intelligence Laboratory, NASA Ames Research Center (QuAIL)}
\affiliation{USRA Research Institute for Advanced Computer Science (RIACS)}
\affiliation{University Erlangen-Nürnberg (FAU), Institute of Theoretical Physics, Staudtstr. 7, 91058 Erlangen, Germany
}

\author{David Von Dollen}
\affiliation{Volkswagen Group Advanced Technologies, San Francisco, CA 94108
}

\author{Hongxiang Chen}

\affiliation{Rahko Ltd., Finsbury Park, London,  N4 3JP, United Kingdom}

\affiliation{Department of Computer Science, University College London, WC1E 6BT, United Kingdom}

\author{Shuxiang Cao}

\affiliation{Rahko Ltd., Finsbury Park, London,  N4 3JP, United Kingdom}

\affiliation{Clarendon Laboratory, Department of Physics, University of Oxford. Oxford, OX1 3PU, United Kingdom}

\author{Roeland Wiersema}
\affiliation{Vector Institute, MaRS  Centre,  Toronto,  Ontario,  M5G  1M1,  Canada}
\affiliation{Department of Physics and Astronomy, University of Waterloo, Ontario, N2L 3G1, Canada}

\author{Hsin-Yuan Huang}

\affiliation{Google Quantum AI, Mountain View, CA
}
\affiliation{Institute for Quantum Information and Matter, Caltech, Pasadena, CA, USA}
\affiliation{Department of Computing and Mathematical Sciences, Caltech, Pasadena, CA, USA}

\author{Jarrod R. McClean}

\affiliation{Google Quantum AI, Mountain View, CA
}
\author{Ryan Babbush}

\affiliation{Google Quantum AI, Mountain View, CA
}
\author{Sergio Boixo}

\affiliation{Google Quantum AI, Mountain View, CA
}

\author{Dave Bacon}

\affiliation{Google Quantum AI, Mountain View, CA
}

\author{Alan K. Ho}

\affiliation{Google Quantum AI, Mountain View, CA
}
\author{Hartmut Neven}

\affiliation{Google Quantum AI, Mountain View, CA
}
\author{Masoud Mohseni}
\email{mohseni@google.com}

\affiliation{Google Quantum AI, Mountain View, CA
}

\date{\today}


\begin{abstract}
We introduce TensorFlow Quantum (TFQ), an open source library for the rapid prototyping of hybrid quantum-classical models for classical or quantum data.  This framework offers high-level abstractions for the design and training of both discriminative and generative quantum models under TensorFlow and supports high-performance quantum circuit simulators.  We provide an overview of the software architecture and building blocks through several examples and review the theory of hybrid quantum-classical neural networks. We illustrate TFQ functionalities via several basic applications including supervised learning for quantum classification, quantum control, simulating noisy quantum circuits, and quantum approximate optimization. Moreover, we demonstrate how one can apply TFQ to tackle advanced quantum learning tasks including meta-learning, layerwise learning, Hamiltonian learning, sampling thermal states, variational quantum eigensolvers, classification of quantum phase transitions, generative adversarial networks, and reinforcement learning. We hope this framework provides the necessary tools for the quantum computing and machine learning research communities to explore models of both natural and artificial quantum systems, and ultimately discover new quantum algorithms which could potentially yield a quantum advantage.
\end{abstract}
\maketitle
\tableofcontents

\section{Introduction}
\subsection{Quantum Machine Learning}
Machine learning (ML) is the construction of algorithms and statistical models which can extract information hidden within a dataset. By learning a model from a dataset, one then has the ability to make predictions on unseen data from the same underlying probability distribution. 
For several decades, research in machine learning was focused on models that can provide theoretical guarantees for their performance \cite{murphy2012machine,suykens1999least,wold1987principal,jain2010data}. However, in recent years, methods based on heuristics have become dominant, partly due to an abundance of data and computational resources \cite{lecun2015deep}.

Deep learning is one such heuristic method which has seen great success \cite{ImageNetConv_NIPS2012, goodfellow2016deep}.  Deep learning methods are based on learning a representation of the dataset in the form of networks of parameterized layers. These parameters are then tuned by minimizing a function of the model outputs, called the loss function. This function quantifies the fit of the model to the dataset.

In parallel to the recent advances in deep learning, there has been a significant growth of interest in quantum computing in both academia and industry \cite{preskill2018quantum}.  Quantum computing is the use of engineered quantum systems to perform computations.  Quantum systems are described by a generalization of probability theory allowing novel behavior such as superposition and entanglement, which are generally difficult to simulate with a classical computer \cite{Feynman1982}.  The main motivation to build a quantum computer is to access efficient simulation of these uniquely quantum mechanical behaviors.  Quantum computers could one day accelerate computations for chemical and materials development \cite{cao2019quantum}, decryption \cite{shor1994algorithms}, optimization \cite{farhi2014quantum}, and many other tasks.  Google's recent achievement of quantum supremacy \cite{arute2019quantum} marked the first glimpse of this promised power.

How may one apply quantum computing to practical tasks?  One area of research that has attracted considerable interest is the design of machine learning algorithms that inherently rely on quantum properties to accelerate their performance.  One key observation that has led to the application of quantum computers to machine learning is their ability to perform fast linear algebra on a state space that grows exponentially with the number of qubits. These quantum accelerated linear-algebra based techniques for machine learning can be considered the first generation of quantum machine learning (QML) algorithms tackling a wide range of applications in both supervised and unsupervised learning, including principal component analysis \cite{Mohseni14_qpca}, support vector machines \cite{Mohseni14_support}, k-means clustering \cite{Mohseni13_clustering}, and recommendation systems \cite{QRS2016}. These algorithms often admit exponentially faster solutions compared to their classical counterparts on certain types of quantum data. This has led to a significant surge of interest in the subject \cite{biamonte2017quantum}. However, to apply these algorithms to classical data, the data must first be embedded into quantum states \cite{giovannetti2008quantum}, a process whose scalability is under debate \cite{arunachalam2015robustness}.  Additionally, many common approaches for applying these algorithms to classical data rely on specific structure in the data that can also be exploited by classical algorithms, sometimes precluding the possibility of a quantum speedup \cite{Tang2018}.  Tests based on the structure of a classical dataset have recently been developed that can sometimes determine if a quantum speedup is possible on that data \cite{huang2021_power}.  Continuing debates around speedups and assumptions make it prudent to look beyond classical data for applications of quantum computation to machine learning.

With the availability of Noisy Intermediate-Scale Quantum (NISQ) processors \cite{Preskill18_0}, the second generation of QML has emerged \cite{biamonte2017quantum,farhi2018classification,farhi2014quantum,peruzzo2014variational,killoran2018continuous,wecker2015progress,biamonte2017quantum,zhou2018quantum,mcclean2016theory,hadfield2017quantum,grant2018hierarchical,khatri2019quantum,schuld2019quantum,mcardle2018variational,benedetti2019adversarial,nash2019quantum,jiang2018majorana,steinbrecher2018quantum,fingerhuth2018quantum,larose2018variational,cincio2018learning,situ2019variational,chen2018universal,verdon2017quantum,preskill2018quantum}. In contrast to the first generation, this new trend in QML is based on heuristic methods which can be studied empirically due to the increased computational capability of quantum hardware. This is reminiscent of how machine learning evolved towards deep learning with the advent of new computational capabilities \cite{Mohseni17}. These new algorithms use parameterized quantum transformations called parameterized quantum circuits (PQCs) or Quantum Neural Networks (QNNs) \cite{farhi2018classification,chen2018universal}.  In analogy with classical deep learning, the parameters of a QNN are then optimized with respect to a cost function via either black-box optimization heuristics \cite{Verdon2019metalearning} or gradient-based methods \cite{sweke2019stochastic}, in order to learn a representation of the training data.  In this paradigm, \textit{quantum machine learning is the development of models, training strategies, and inference schemes built on parameterized quantum circuits}.

\subsection{Hybrid Quantum-Classical Models}\label{sec:intro-hqcm}

Near-term quantum processors are still fairly small and noisy, thus quantum models cannot disentangle and generalize quantum data using quantum processors alone. NISQ processors will need to work in concert with classical co-processors to become effective.  We anticipate that investigations into various possible \textit{hybrid quantum-classical machine learning} algorithms will be a productive area of research and that quantum computers will be most useful as hardware accelerators, working in symbiosis with traditional computers.  In order to understand the power and limitations of classical deep learning methods, and how they could be possibly improved by incorporating parameterized quantum circuits, it is worth defining key indicators of learning performance:

\indent \textit{Representation capacity}: the model architecture has the capacity to accurately replicate, or extract useful information from, the underlying correlations in the training data for some value of the model's parameters.

\indent \textit{Training efficiency}: minimizing the cost function  via stochastic optimization heuristics should converge to an approximate minimum of the loss function in a reasonable number of iterations.

\indent \textit{Inference tractability:} the ability to run inference on a given model in a scalable fashion is needed in order to make predictions in the training or test phase.

\indent \textit{Generalization power}: the cost function for a given model should yield a landscape where typically initialized and trained networks find approximate solutions which generalize well to unseen data. 

In principle, any or all combinations of these attributes could be susceptible to possible improvements by quantum computation.  There are many ways to combine classical and quantum computations.  One well-known method is to use classical computers as outer-loop optimizers for QNNs. When training a QNN with a classical optimizer in a quantum-classical loop, the overall algorithm is sometimes referred to as a \textit{Variational Quantum-Classical Algorithm}. Some recently proposed architectures of QNN-based variational quantum-classical algorithms include
Variational Quantum Eigensolvers (VQEs) \cite {McClean_2016,mcclean2016theory}, Quantum Approximate Optimization Algorithms (QAOAs) \cite{farhi2014quantum,zhou2018quantum,verdon2019quantum,wang2019xy}, Quantum Neural Networks (QNNs) for classification \cite{farhi18_QNN,mcclean2018barren}, Quantum Convolutional Neural Networks (QCNN) \cite{Cong_2019}, and Quantum Generative Models \cite{lloyd2018quantum}. Generally, the goal is to optimize over a parameterized class of computations to either generate a certain low energy wavefunction (VQE/QAOA), learn to extract non-local information (QNN classifiers), or learn how to generate a quantum distribution from data (generative models).

It is important to note that in the standard model architecture for these applications, the representation typically resides entirely on the quantum processor, with classical heuristics participating only as optimizers for the tunable parameters of the quantum model. One of major obstacles in training of such quantum models is known as \textit{barren plateaus} ~\cite{mcclean2018barren}, which generally arises when a network architecture is lacking any structure and it is randomly initialized. This unusual flat energy landscape of quantum models could seriously hinder the performance of both gradient-based and gradient-free optimization techniques ~\cite{Coles2020}. We discuss various strategies for overcoming this issue in section \ref{sec:random_circuits}.

While the use of classical processors as outer-loop optimizers for quantum neural networks is promising, the reality is that near-term quantum devices are still fairly noisy, thus limiting the depth of quantum circuit achievable with acceptable fidelity.  This motivates allowing as much of the model as possible to reside on classical hardware.  Several applications of quantum computation have ventured beyond the scope of typical variational quantum algorithms to explore this combination. Instead of training a purely quantum model via a classical optimizer, one then considers scenarios where the model itself is a hybrid between quantum computational building blocks and classical computational building blocks \cite{verdon2018universal,romero2019variational,bergholm2018pennylane, verdon2019quantumVQT} and is trained typically via gradient-based methods. Such scenarios leverage a new form of automatic differentiation that allows the backwards propagation of gradients in between parameterized quantum and classical computations. The theory of such hybrid backpropagation will be covered in section \ref{sec:gradients}.

In summary, \textit{a hybrid quantum-classical model is a learning heuristic in which both the classical and quantum processors contribute to the indicators of learning performance defined above.}

\subsection{Quantum Data}\label{sec:quantum_data}
Although it is not yet proven that heuristic QML can provide a speedup on practical classical ML applications, there is growing  evidence that hybrid quantum-classical machine learning applications on ``\textit{quantum data}'' can provide a quantum advantage over classical-only machine learning for reasons described below.  These results motivate the development of software frameworks that can combine coherent access to quantum data with the power of machine learning.

Abstractly, any data emerging from an underlying quantum mechanical process can be considered quantum data. This can be the classical data resulting from quantum mechanical experiments \cite{huang2021_power}, or data which is directly generated by a quantum device and then fed into an algorithm as input \cite{huang2021information}. A quantum or hybrid quantum-classical model will be at least partially represented by a quantum device, and therefore have the inherent capacity to capture the characteristics of a quantum mechanical process. Concretely, we list practical examples of classes of quantum data, which can be routinely generated or simulated on existing quantum devices or processors:

    \textit{Quantum simulations}: These can include output states of quantum chemistry simulations used to extract information about chemical structures and chemical reactions \cite{google2020hartreefock}. Potential applications include material science, computational chemistry, computational biology, and drug discovery. Another example is data from quantum many-body systems and quantum critical systems in condensed matter physics, which could be used to model and design exotic states of matter which exhibit many-body quantum effects.

    \textit{Quantum communication networks}: Machine learning in this class of systems will be related to distilling small-scale quantum data; e.g., to discriminate among non-orthogonal quantum states \cite{Mohseni2004,chen2018universal}, with application to design and construction of quantum error correcting codes for  quantum repeaters, quantum receivers, and purification units, devices which are critical for the construction of a quantum internet \cite{vandam2020quantuminternet}.
  
    \textit{Quantum metrology}: Quantum-enhanced high precision measurements such as quantum sensing and quantum imaging are inherently done on probes that are small-scale quantum devices and could be designed or improved by variational quantum models \cite{meyer2021parameterestimation}.

    \textit{Quantum control}: Variationally learning hybrid quantum-classical algorithms can lead to new optimal open or closed-loop control \cite{niu2019universal}, calibration, and error mitigation, correction, and verification strategies \cite{Carolan2020} for near-term quantum devices and quantum processors.

Of course, this is not a comprehensive list of quantum data. We hope that, with proper software tooling, researchers will be able to find applications of QML in all of the above areas and other categories of applications beyond what we can currently envision.

\subsection{TensorFlow Quantum}
Today, exploring new hybrid quantum-classical models is a difficult and error-prone task. The engineering effort required to manually construct such models, develop quantum datasets, and set up training and validation stages  decreases a researcher's ability to iterate and discover.  TensorFlow has accelerated the research and understanding of deep learning in part by automating common model building tasks.  Development of software tooling for hybrid quantum-classical models should similarly accelerate research and understanding for quantum machine learning.

To develop such tooling, the requirement of accommodating a heterogeneous computational environment involving both classical and quantum processors is key.  This computational heterogeneity suggested the need to expand TensorFlow, which is designed to distribute computations across CPUs, GPUs, and TPUs \cite{tf_whitepaper}, to also encompass quantum processing units (QPUs).  This project has evolved into TensorFlow Quantum.  TFQ is an integration of Cirq with TensorFlow that allows researchers and students to simulate QPUs while designing, training, and testing hybrid quantum-classical models, and eventually run the quantum portions of these models on actual quantum processors as they come online. A core contribution of TFQ is seamless backpropagation through combinations of classical and quantum layers in hybrid quantum-classical models. This allows QML researchers to directly harness the rich set of tools already available in TF and Keras.

The remainder of this document describes TFQ and a selection of applications demonstrating some of the challenges TFQ can help tackle.  In section \ref{sec:software_arch}, we introduce the software architecture of TFQ.  We highlight its main features including batched circuit execution, automated expectation estimation, estimation of quantum gradients, hybrid quantum-classical automatic differentiation, and rapid model construction, all from within TensorFlow. We also present a simple ``Hello, World" example for binary quantum data classification on a single qubit.  By the end of section \ref{sec:software_arch}, we expect most readers to have sufficient knowledge to begin development with TFQ.  For readers who are interested in a more theoretical understanding of QNNs, we provide in section \ref{sec:theory} an overview of QNN models and hybrid quantum-classical backpropagation.  For researchers interested in applying TFQ to their own projects, we provide various applications in sections \ref{sec:basic_app} and \ref{sec:advanced_applications}.  In section \ref{sec:basic_app}, we describe hybrid quantum-classical CNNs for binary classification of quantum phases, hybrid quantum-classical ML for quantum control, and MaxCut QAOA.  In the advanced applications section \ref{sec:advanced_applications}, we describe meta-learning for quantum approximate optimization, discuss issues with vanishing gradients and how we can overcome them by adaptive layer-wise learning schemes, Hamiltonian learning with quantum graph networks, quantum mixed state generation via classical energy-based models, subspace-Search variational quantum eigensolver for excited states to illustrate an integration with OpenFermion, quantum classification of quantum phase transitions, entangling quantum generative adversarial networks, and quantum reinforcement learning.

We hope that TFQ enables the machine learning and quantum computing communities to work together more closely on important challenges and opportunities in the near-term and beyond.

\section{Software Architecture \& Building Blocks}\label{sec:software_arch}

As stated in the introduction, the goal of TFQ is to bridge the quantum computing and machine learning communities. Google already has well-established products for these communities: Cirq, an open source library for invoking quantum circuits \cite{Cirq}, and TensorFlow, an end-to-end open source machine learning platform \cite{tf_whitepaper}.   However, the emerging community of quantum machine learning researchers requires the capabilities of both products.  The prospect of combining Cirq and TensorFlow then naturally arises.

First, we review the capabilities of Cirq and TensorFlow.  We confront the challenges that arise when one attempts to combine both products.  These challenges inform the design goals relevant when building software specific to quantum machine learning.  We provide an overview of the architecture of TFQ and describe a particular abstract pipeline for building a hybrid model for classification of quantum data.  Then we illustrate this pipeline via the exposition of a minimal hybrid model which makes use of the core features of TFQ.  We conclude with a description of our performant C++ simulator for quantum circuits and provide benchmarks of performance on two complementary classes of dense and sparse quantum circuits.

\subsection{Cirq}\label{sec:cirq}
Cirq is an open-source framework for invoking quantum circuits on near term devices \cite{Cirq}.  It contains the basic structures, such as qubits, gates, circuits, and measurement operators, that are required for specifying quantum computations.  User-specified quantum computations can then be executed in simulation or on real hardware.  Cirq also contains substantial machinery that helps users design efficient algorithms for NISQ machines, such as compilers and schedulers.  Below we show example Cirq code for calculating the expectation value of $\hat{Z}_1 \hat{Z}_2$ for a Bell state:
\begin{lstlisting}
(q1, q2) = cirq.GridQubit.rect(1,2)
c = cirq.Circuit(cirq.H(q1), cirq.CNOT(q1, q2))
ZZ = cirq.Z(q1) * cirq.Z(q2)
bell = cirq.Simulator().simulate(c).final_state
expectation = ZZ.expectation_from_wavefunction(
    bell, dict(zip([q1,q2],[0,1])))
\end{lstlisting}
Cirq uses SymPy \cite{sympycite} symbols to represent free parameters in gates and circuits.  You replace free parameters in a circuit with specific numbers by passing a Cirq \Colorbox{bkgd}{\lstinline{ParamResolver}} object with your circuit to the simulator.   Below we construct a parameterized circuit and simulate the output state for $\theta =1$:
\begin{lstlisting}
theta = sympy.Symbol('theta')
c = cirq.Circuit(cirq.Rx(theta).on(q1))
resolver = cirq.ParamResolver({theta:1})
results = cirq.Simulator().simulate(c, resolver)
\end{lstlisting}

\subsection{TensorFlow}\label{sec:tensorflow}
TensorFlow is a language for describing computations as stateful dataflow graphs \cite{tf_whitepaper}.  Describing machine learning models as dataflow graphs is advantageous for performance during training.  First, it is easy to obtain gradients of dataflow graphs using backpropagation \cite{rumelhart1986learning}, allowing efficient parameter updates.  Second, independent nodes of the computational graph may be distributed across independent machines, including GPUs and TPUs, and run in parallel.  These computational advantages established TensorFlow as a powerful tool for machine learning and deep learning.

TensorFlow constructs this dataflow graph using \textit{tensors} for the directed edges and \textit{operations} (ops) for the nodes.  For our purposes, a rank $n$ tensor is simply an $n$-dimensional array.  In TensorFlow, tensors are additionally associated with a data type, such as integer or string.  Tensors are a convenient way of thinking about data; in machine learning, the first index is often reserved for iteration over the members of a dataset. Additional indices can indicate the application of several filters, e.g., in convolutional neural networks with several feature maps.  

\begin{figure}[H]
    \centering
    \includegraphics[width=0.8\columnwidth]{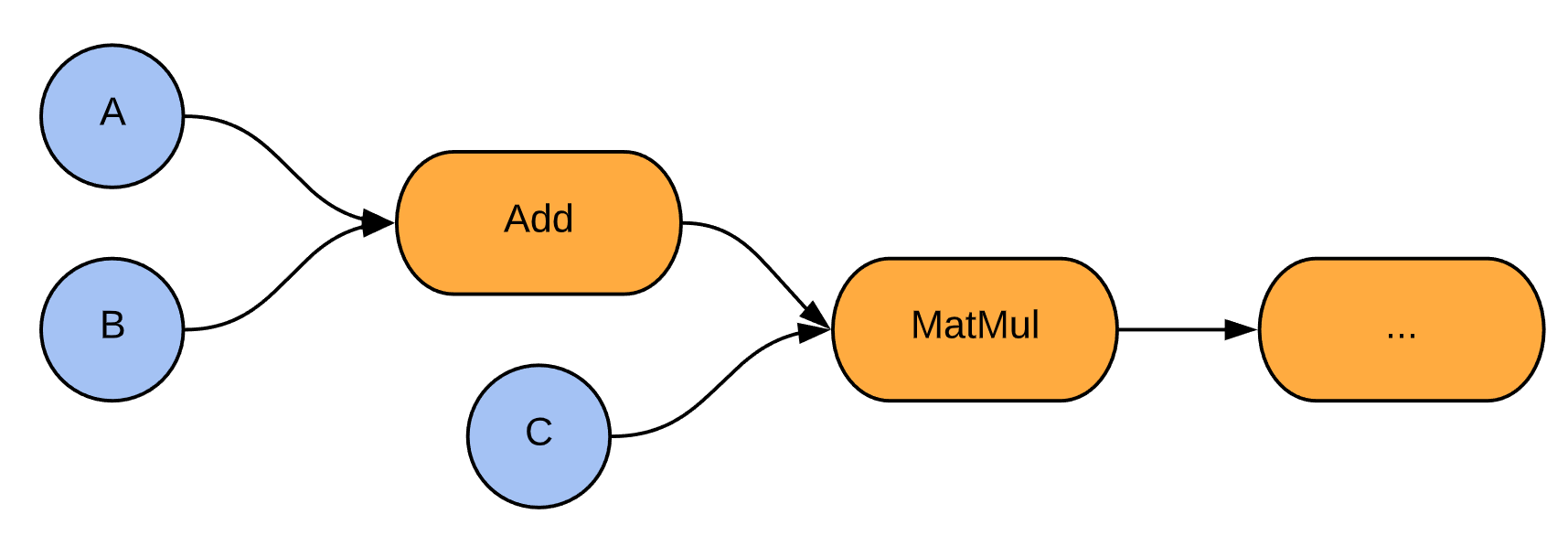}
\caption{A simple example of the TensorFlow computational model.  Two tensor inputs $A$ and $B$ are added and then multiplied against a third tensor input $C$, before flowing on to further nodes in the graph.  Blue nodes are tensor injections (\textit{ops}), arrows are \textit{tensors} flowing through the computational graph, and orange nodes are tensor transformations (\textit{ops}).  Tensor injections are ops in the sense that they are functions which take in zero tensors and output one tensor.}
    \label{fig:tf_simple}
\end{figure}

In general, an op is a function mapping input tensors to output tensors.  Ops may act on zero or more input tensors, always producing at least one tensor as output.  For example, the addition op ingests two tensors and outputs one tensor representing the elementwise sum of the inputs, while a constant op ingests no tensors, taking the role of a root node in the dataflow graph.  The combination of ops and tensors gives the backend of TensorFlow the structure of a directed acyclic graph.  A visualization of the backend structure corresponding to a simple computation in TensorFlow is given in Fig.~\ref{fig:tf_simple}.

It is worth noting that this tensorial data format is not to be confused with Tensor Networks \cite{biamonte2017tensor}, which are a mathematical tool used in condensed matter physics and quantum information science to efficiently represent many-body quantum states and operations. Recently, libraries for building such Tensor Networks in TensorFlow have become available \cite{roberts2019tensornetwork}, we refer the reader to the corresponding blog post for better understanding of the difference between TensorFlow tensors and the tensor objects in Tensor Networks \cite{TNTFblog}.

The recently announced TensorFlow 2 \cite{tf2_effective} takes the dataflow graph structure as a foundation and adds high-level abstractions.  One new feature is the Python function decorator \Colorbox{bkgd}{\lstinline{@tf.function}}, which automatically converts the decorated function into a graph computation.  Also relevant is the native support for Keras \cite{chollet2015keras}, which provides the \Colorbox{bkgd}{\lstinline{Layer}} and \Colorbox{bkgd}{\lstinline{Model}} constructs.  These abstractions allow the concise definition of machine learning models which ingest and process data, all backed by dataflow graph computation.  The increasing levels of abstraction and heterogenous hardware backing which together constitute the TensorFlow stack can be visualized with the orange and gray boxes in our stack diagram in Fig.~\ref{fig:tfq_software_stack}.  The combination of these high-level abstractions and efficient dataflow graph backend makes TensorFlow 2 an ideal platform for data-driven machine learning research.

\subsection{Technical Hurdles in Combining Cirq with TensorFlow}

\begin{figure}[H]
    \centering
    \includegraphics[width=0.8\columnwidth]{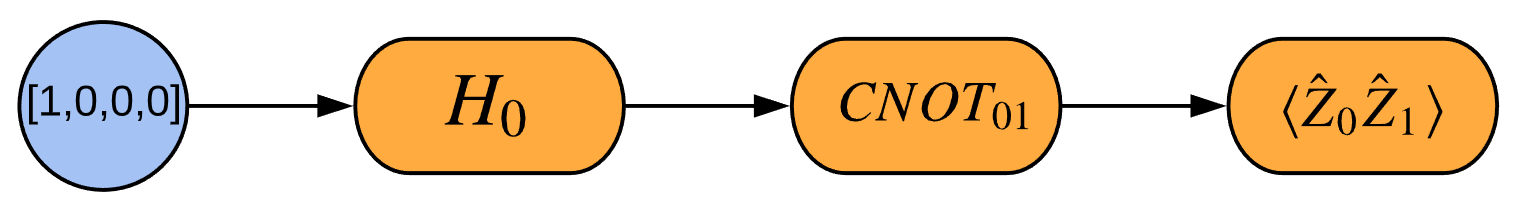}
\caption{The states-as-edges approach to embedding quantum computation in TensorFlow.  Blue nodes are input tensors, arrows are tensors flowing through the graph, and orange nodes are TF Ops transforming the simulated quantum state. Note that the above is not the architecture used in TFQ but rather an alternative which was considered, see Fig.~\ref{fig:expectation_graph} for the equivalent diagram for the true TFQ architecture.}
    \label{fig:states-as-edges}
\end{figure}

There are many ways one could imagine combining the capabilities of Cirq and TensorFlow.  One possible approach is to let graph edges represent quantum states and let ops represent transformations of the state, such as applying circuits and taking measurements.  This approach can be called the ``states-as-edges" architecture.  We show in Fig.~\ref{fig:states-as-edges} how to reformulate the Bell state preparation and measurement discussed in section \ref{sec:cirq} within this proposed architecture.

This architecture may at first glance seem like an attractive option as it is a direct formulation of quantum computation as a dataflow graph.  However, this approach is suboptimal for several reasons.  First, in this architecture, the structure of the circuit being run is static in the computational graph, thus running a different circuit would require the graph to be rebuilt. This is far from ideal for variational quantum algorithms which learn over many iterations with a slightly modified quantum circuit at each iteration. A second problem is the lack of a clear way to embed such a quantum dataflow graph on a real quantum processor: the states would have to remain held in quantum memory on the quantum device itself, and the high latency between classical and quantum processors makes sending transformations one-by-one prohibitive.  Lastly, one needs a way to specify gates and measurements within TF.  One may be tempted to define these directly; however, Cirq already has the necessary tools and objects defined which are most relevant for the near-term quantum computing era.  Duplicating Cirq functionality in TF would lead to several issues, requiring users to re-learn how to interface with quantum computers in TFQ versus Cirq, and adding to the maintenance overhead by needing to keep two separate quantum circuit construction frameworks up-to-date as new compilation techniques arise.  These considerations motivate our core design principles.

\subsection{TFQ architecture}\label{sec:tfq_architecture}
\subsubsection{Design Principles and Overview}
To avoid the aforementioned technical hurdles and in order to satisfy the diverse needs of the research community, we have arrived at the following four design principles:

\begin{enumerate}
\item \textbf{Differentiability}. \label{prin:diff}
As described in the introduction, gradient-based methods leveraging autodifferentiation have become the leading heuristic for optimization of machine learning models.  A software framework for QML must support differentiation of quantum circuits so that hybrid quantum-classical models can participate in backpropagation.
\item \textbf{Circuit Batching}.\label{prin:batch}
Learning on quantum data requires re-running parameterized model circuits on each quantum data point.  A QML software framework must be optimized for running large numbers of such circuits.  Ideally, the semantics should match established TensorFlow norms for batching over data.
\item \textbf{Execution Backend Agnostic}. \label{prin:agnostic}
Experimental quantum computing often involves reconciling perfectly simulated algorithms with the outputs of real, noisy devices. Thus, QML software must allow users to easily switch between running models in simulation and running models on real hardware, such that simulated results and experimental results can be directly compared.

\item \textbf{Minimalism}.\label{prin:min}
Cirq provides an extensive set of tools for preparing quantum circuits. TensorFlow provides a very complete machine learning toolkit through its hundreds of ops and Keras high-level API, with a massive community of active users. Existing functionality in Cirq and TensorFlow should be used as much as possible.  TFQ should serve as a bridge between the two that does not require users to re-learn how to interface with quantum computers or re-learn how to solve problems using machine learning.
\end{enumerate}

First, we provide a bottom-up overview of TFQ to provide intuition on how the framework functions at a fundamental level.  In TFQ, circuits and other quantum computing constructs are tensors, and converting these tensors into classical information via simulation or execution on a quantum device is done by ops.  These tensors are created by converting Cirq objects to TensorFlow string tensors, using the \Colorbox{bkgd}{\lstinline{tfq.convert_to_tensor}} function. This takes in a \Colorbox{bkgd}{\lstinline{cirq.Circuit}} or \Colorbox{bkgd}{\lstinline{cirq.PauliSum}} object and creates a string tensor representation. The \Colorbox{bkgd}{\lstinline{cirq.Circuit}} objects may be parameterized by SymPy symbols.

\begin{figure}
    \centering
    \includegraphics[width=0.8\columnwidth]{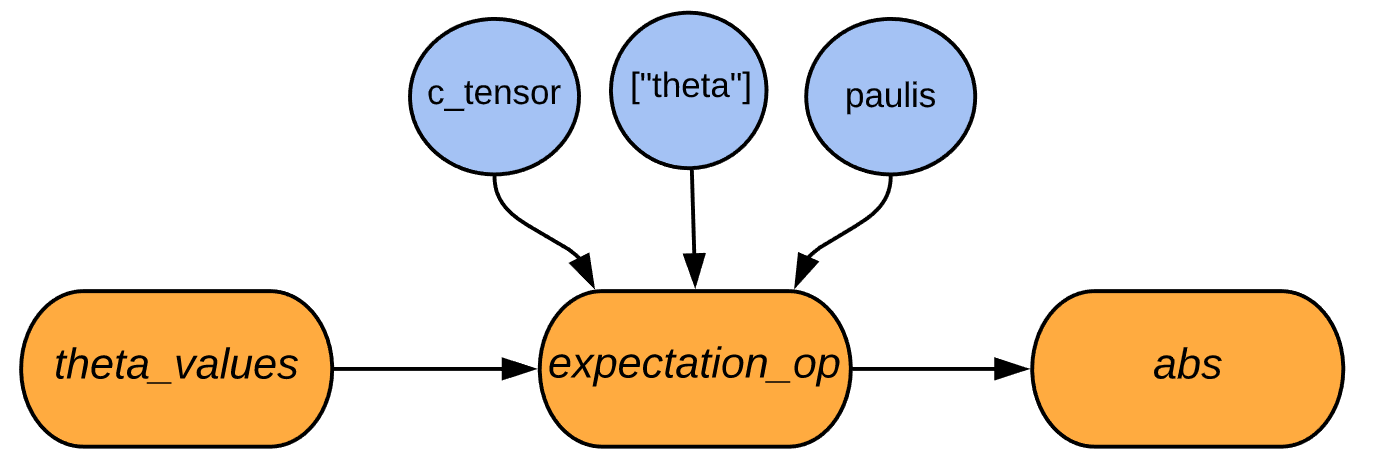}
\caption{The TensorFlow graph generated to calculate the expectation value of a parameterized circuit.  The symbol values can come from other TensorFlow ops, such as from the outputs of a classical neural network.  The output can be passed on to other ops in the graph; here, for illustration, the output is passed to the absolute value op.}
    \label{fig:expectation_graph}
\end{figure}

These tensors are then converted to classical information via state simulation, expectation value calculation, or sampling.  TFQ provides ops for each of these computations.  The following code snippet shows how a simple parameterized circuit may be created using Cirq, and its $\hat{Z}$ expectation evaluated at different parameter values using the tfq expectation value calculation op. We feed the output into the \Colorbox{bkgd}{\lstinline{tf.math.abs}} op to show that tfq ops integrate naively with tf ops.

\begin{lstlisting}
qubit = cirq.GridQubit(0, 0)
theta = sympy.Symbol('theta')
c = cirq.Circuit(cirq.X(qubit) ** theta)
c_tensor = tfq.convert_to_tensor([c] * 3)
theta_values = tf.constant([[0],[1],[2]])
m = cirq.Z(qubit)
paulis = tfq.convert_to_tensor([m] * 3)
expectation_op = tfq.get_expectation_op()
output = expectation_op(
    c_tensor, ['theta'], theta_values, paulis)
abs_output = tf.math.abs(output)
\end{lstlisting}
We supply the expectation op with a tensor of parameterized circuits, a list of symbols contained in the circuits, a tensor of values to use for those symbols, and tensor operators to measure with respect to. Given this, it outputs  a tensor of expectation values. The graph this code generates is given by Fig.~\ref{fig:expectation_graph}.

The expectation op is capable of running circuits on a simulated backend, which can be a Cirq simulator or our native TFQ simulator qsim (described in detail in section \ref{sec:qsim}), or on a real device.  This is configured on instantiation.

The expectation op is fully differentiable. Given that there are many ways to calculate the gradient of a quantum circuit with respect to its input parameters, TFQ allows expectation ops to be configured with one of many built-in differentiation methods using the \Colorbox{bkgd}{\lstinline{tfq.Differentiator}} interface, such as finite differencing, parameter shift rules, and various stochastic methods. The \Colorbox{bkgd}{\lstinline{tfq.Differentiator}} interface also allows users to define their own gradient calculation methods for their specific problem if they desire.

\begin{figure}
    \centering
    \includegraphics[width=0.9\columnwidth]{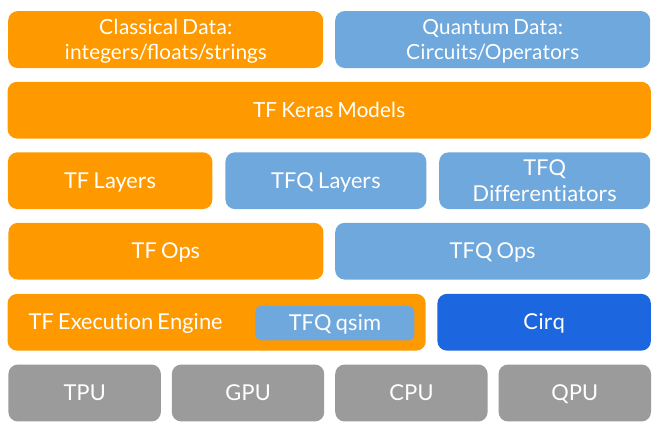}
\caption{The software stack of TFQ, showing its interactions with TensorFlow, Cirq, and computational hardware.  At the top of the stack is the data to be processed.  Classical data is natively processed by TensorFlow; TFQ adds the ability to process quantum data, consisting of both quantum circuits and quantum operators.  The next level down the stack is the Keras API in TensorFlow.  Since a core principle of TFQ is native integration with core TensorFlow, in particular with Keras models and optimizers, this level spans the full width of the stack.  Underneath the Keras model abstractions are our quantum layers and differentiators, which enable hybrid quantum-classical automatic differentiation when connected with classical TensorFlow layers.  Underneath the layers and differentiators, we have TensorFlow ops, which instantiate the dataflow graph.  Our custom ops control quantum circuit execution.  The circuits can be run in simulation mode, by invoking qsim or Cirq, or eventually will be executed on QPU hardware.}
\label{fig:tfq_software_stack}
\end{figure}

The tensor representation of circuits and Paulis along with the execution ops are all that are required to solve any problem in QML. However, as a convenience, TFQ provides an additional op for in-graph circuit construction. This was found to be convenient when solving problems where most of the circuit being run is static and only a small part of it is being changed during training or inference. This functionality is provided by the \Colorbox{bkgd}{\lstinline{tfq.tfq_append_circuit}} op.  It is expected that all but the most dedicated users will never touch these low-level ops, and instead will interface with TFQ using our \Colorbox{bkgd}{\lstinline{tf.keras.layers}} that provide a simplified interface.

The tools provided by TFQ can interact with both core TensorFlow and, via Cirq, real quantum hardware.  The functionality of all three software products and the interfaces between them can be visualized with the help of a ``software-stack" diagram, shown in Fig.~\ref{fig:tfq_software_stack}. Importantly, these interfaces allow users to write a single TensorFlow Quantum program which could then easily be run locally on a workstation, in a highly parallel and distributed setting at the PetaFLOP/s or higher throughput scale \cite{huang2021_power}, or on real QPU device \cite{eqgan}.

In the next section, we describe an example of an abstract quantum machine learning pipeline for hybrid discriminator model that TFQ was designed to support.  Then we illustrate the TFQ pipeline via a Hello Many-Worlds example, which involves building the simplest possible hybrid quantum-classical model for a binary classification task on a single qubit.  More detailed information on the building blocks of TFQ features will be given in section \ref{sec:building_blocks}.

\subsubsection{The Abstract TFQ Pipeline for a specific hybrid discriminator model}\label{sec:abstract_pipeline}
Here, we provide a high-level abstract overview of the computational steps involved in the end-to-end pipeline for inference and training of a hybrid quantum-classical discriminative model for quantum data in TFQ.

\begin{figure}
    \centering
    \includegraphics[width=0.95\columnwidth]{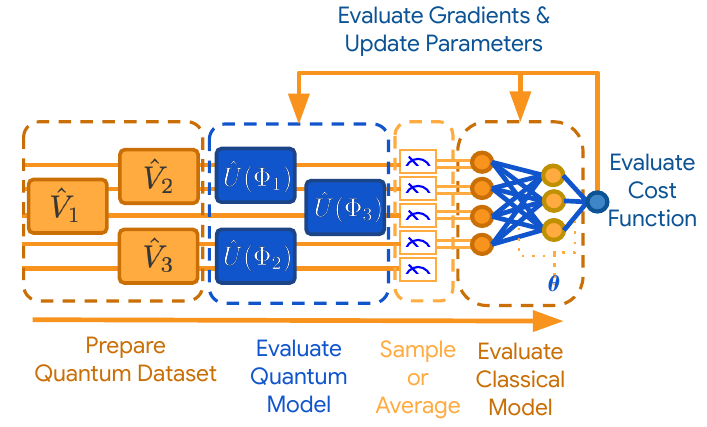}
\caption{Abstract pipeline for inference and training of a hybrid discriminative model in TFQ. Here, $\bm{\Phi}$ represents the quantum model parameters and $\bm{\theta}$ represents the classical model parameters.}
    \label{fig:abstract_pipeline}
\end{figure}

\textbf{(1) Prepare Quantum Dataset:}  In general, this might come from a given black-box source. However, as current quantum computers cannot import quantum data from external sources, the user has to specify quantum circuits which generate the data.  Quantum datasets are prepared using unparameterized \Colorbox{bkgd}{\lstinline{cirq.Circuit}} objects and are injected into the computational graph using \Colorbox{bkgd}{\lstinline{tfq.convert_to_tensor}}.

\textbf{(2) Evaluate Quantum Model:}  Parameterized quantum models can be selected from several categories based on knowledge of the quantum data's structure.  The goal of the model is to perform a quantum computation in order to extract information hidden in a quantum subspace or subsystem. In the case of discriminative learning, this information is the hidden label parameters. To extract a quantum non-local subsystem, the quantum model disentangles the input data, leaving the hidden information encoded in classical correlations, thus making it accessible to local measurements and classical post-processing.  Quantum models are constructed using \Colorbox{bkgd}{\lstinline{cirq.Circuit}} objects containing SymPy symbols, and can be attached to quantum data sources using the \Colorbox{bkgd}{\lstinline{tfq.AddCircuit}} layer.

\textbf{(3) Sample or Average:}
Measurement of quantum states extracts classical information, in the form of samples from a classical random variable. The distribution of values from this random variable generally depends on both the quantum state itself and the measured observable.  As many variational algorithms depend on mean values of measurements, TFQ provides methods for averaging over several runs involving steps (1) and (2).  Sampling or averaging are performed by feeding quantum data and quantum models to the \Colorbox{bkgd}{\lstinline{tfq.Sample}} or \Colorbox{bkgd}{\lstinline{tfq.Expectation}} layers.

\textbf{(4) Evaluate Classical Model:} Once classical information has been extracted, it is in a format amenable to further classical post-processing. As the extracted information may still be encoded in classical correlations between measured expectations, classical deep neural networks can be applied to distill such correlations.  Since TFQ is fully compatible with core TensorFlow, quantum models can be attached directly to classical \Colorbox{bkgd}{\lstinline{tf.keras.layers.Layer}} objects such as \Colorbox{bkgd}{\lstinline{tf.keras.layers.Dense}}.

\textbf{(5) Evaluate Cost Function:}  Given the results of classical post-processing, a cost function is calculated.  This may be based on the accuracy of classification if the quantum data was labeled, or other criteria if the task is unsupervised.  Wrapping the model built in stages (1) through (4) inside a \Colorbox{bkgd}{\lstinline{tf.keras.Model}} gives the user access to all the losses in the \Colorbox{bkgd}{\lstinline{tf.keras.losses}} module.

\textbf{(6) Evaluate Gradients \& Update Parameters:}
After evaluating the cost function, the free parameters in the pipeline is updated in a direction expected to decrease the cost.  This is most commonly performed via gradient descent.  To support gradient descent, TFQ exposes derivatives of quantum operations to the TensorFlow backpropagation machinery via the \Colorbox{bkgd}{\lstinline{tfq.differentiators.Differentiator}} interface.  This allows both the quantum and classical models' parameters to be optimized against quantum data via hybrid quantum-classical backpropagation. See section \ref{sec:theory} for details on the theory.

In the next section, we illustrate this abstract pipeline by applying it to a specific example.  While simple, the example is the minimum instance of a hybrid quantum-classical model operating on quantum data.

\subsubsection{Hello Many-Worlds: Binary Classifier for Quantum Data}
\begin{figure}
    \centering
    \includegraphics[width=0.5\columnwidth]{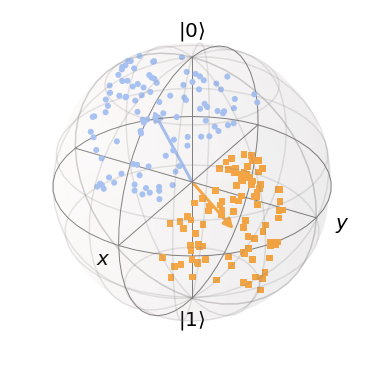}
\caption{Quantum data represented on the Bloch sphere.  States in category $a$ are blue, while states in category $b$ are orange.  The vectors are the states around which the samples were taken.  The parameters used to generate this data are: $\theta_a = 1$, $\theta_b = 4$, and $N = 200$.  The QuTiP package was used to visualize the Bloch sphere \cite{johansson2012qutip}.}
    \label{fig:quantum_dataset}
\end{figure}

Binary classification is a basic task in machine learning that can be applied to quantum data as well.  As a minimal example of a hybrid quantum-classical model, we present here a binary classifier for regions on a single qubit.  In this task, two random vectors in the X-Z plane of the Bloch sphere are chosen.  Around these two vectors, we randomly sample two sets of quantum data points; the task is to learn to distinguish the two sets.  An example quantum dataset of this type is shown in Fig.~\ref{fig:quantum_dataset}.  The following can all be run in-browser by navigating to the Colab example notebook at
\fancylink{https://github.com/tensorflow/quantum/blob/research/binary_classifier/binary_classifier.ipynb}{research/binary\_classifier/binary\_classifier.ipynb}
Additionally, the code in this example can be copy-pasted into a python script after installing TFQ.

To solve this problem, we use the pipeline shown in Fig.~\ref{fig:abstract_pipeline}, specialized to one-qubit binary classification.  This specialization is shown in Fig.~\ref{fig:binary_classifier}.

The first step is to generate the quantum data.  We can use Cirq for this task.  The common imports required for working with TFQ are shown below:
\begin{lstlisting}
import cirq, random, sympy
import numpy as np
import tensorflow as tf
import tensorflow_quantum as tfq
\end{lstlisting}
The function below generates the quantum dataset; labels use a one-hot encoding:
\begin{lstlisting}
def generate_dataset(
    qubit, theta_a, theta_b, num_samples):
  q_data = []
  labels = []
  blob_size = abs(theta_a - theta_b) / 5
  for _ in range(num_samples):
    coin = random.random()
    spread_x, spread_y = np.random.uniform(
      -blob_size, blob_size, 2)
    if coin < 0.5:
      label = [1, 0]
      angle = theta_a + spread_y
    else:
      label = [0, 1]
      angle = theta_b + spread_y
    labels.append(label)
    q_data.append(cirq.Circuit(
      cirq.Ry(-angle)(qubit),
      cirq.Rx(-spread_x)(qubit)))
  return (tfq.convert_to_tensor(q_data),
          np.array(labels))
\end{lstlisting}
We can generate a dataset and the associated labels after picking some parameter values:
\begin{lstlisting}
qubit = cirq.GridQubit(0, 0)
theta_a = 1
theta_b = 4
num_samples = 200
q_data, labels = generate_dataset(
    qubit, theta_a, theta_b, num_samples)
\end{lstlisting}
\begin{figure}
    \centering
    \includegraphics[width=1\columnwidth]{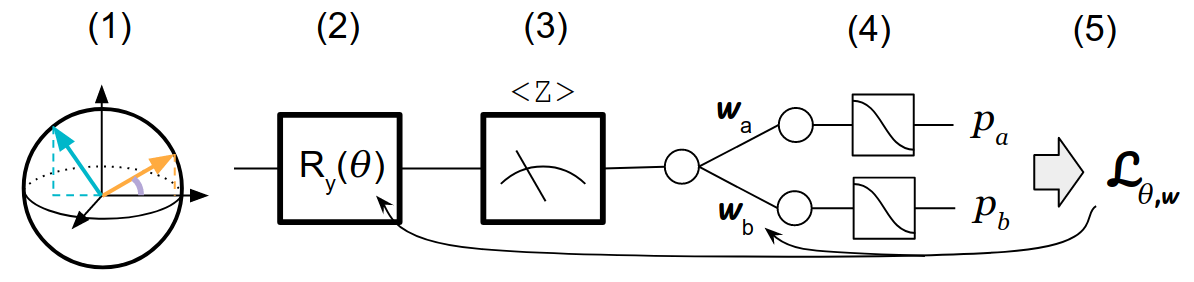}
\caption{
(1) Quantum data to be classified.  (2) Parameterized rotation gate, whose job is to remove superpositions in the quantum data.  (3) Measurement along the Z axis of the Bloch sphere converts the quantum data into classical data.  (4) Classical post-processing is a two-output SoftMax layer, which outputs probabilities for the data to come from category $a$ or category $b$.  (5) Categorical cross entropy is computed between the predictions and the labels.  The Adam optimizer \cite{kingma2014adam} is used to update both the quantum and classical portions of the hybrid model.}
    \label{fig:binary_classifier}
\end{figure}
As our quantum parametric model, we use the simplest case of a universal quantum  discriminator \cite{chen2018universal,Carolan2020}, a single parameterized rotation (linear) and measurement along the $Z$ axis (non-linear):
\begin{lstlisting}
theta = sympy.Symbol('theta')
q_model = cirq.Circuit(cirq.Ry(theta)(qubit))
q_data_input = tf.keras.Input(
    shape=(), dtype=tf.dtypes.string)
expectation = tfq.layers.PQC(
    q_model, cirq.Z(qubit))
expectation_output = expectation(q_data_input)
\end{lstlisting}
The purpose of the rotation gate is to minimize the superposition from the input quantum data such that we can get maximum useful information from the measurement.  This quantum model is then attached to a small classifier NN to complete our hybrid model.  Notice in the code below that quantum layers can appear among classical layers inside a standard Keras model:
\begin{lstlisting}
classifier = tf.keras.layers.Dense(
    2, activation=tf.keras.activations.softmax)
classifier_output = classifier(expectation_output)
model = tf.keras.Model(inputs=q_data_input,
    outputs=classifier_output)
\end{lstlisting}
We can train this hybrid model on the quantum data defined earlier.  Below we use as our loss function the cross entropy between the labels and the predictions of the classical NN; the ADAM optimizer is chosen for parameter updates.
\begin{lstlisting}
optimizer=tf.keras.optimizers.Adam(
    learning_rate=0.1)
loss=tf.keras.losses.CategoricalCrossentropy()
model.compile(optimizer=optimizer, loss=loss)
history = model.fit(
    x=q_data, y=labels, epochs=50)
\end{lstlisting}
Finally, we can use our trained hybrid model to classify new quantum datapoints:
\begin{lstlisting}
test_data, _ = generate_dataset(
    qubit, theta_a, theta_b, 1)
p = model.predict(test_data)[0]
print(f"prob(a)={p[0]:.4f}, prob(b)={p[1]:.4f}")
\end{lstlisting}
This section provided a rapid introduction to just that code needed to complete the task at hand.  The following section reviews the features of TFQ in a more API reference inspired style.

\subsection{TFQ Building Blocks}\label{sec:building_blocks}
Having provided a minimum working example in the previous section, we now seek to provide more details about the components of the TFQ framework.  First, we describe how quantum computations specified in Cirq are converted to tensors for use inside the TensorFlow graph.  Then, we describe how these tensors can be combined in-graph to yield larger models.  Next, we show how circuits are simulated and measured in TFQ.  The core functionality of the framework, differentiation of quantum circuits, is then explored.  Finally, we describe our more abstract layers, which can be used to simplify many QML workflows.

\subsubsection{Quantum Computations as Tensors}\label{sec:q_tensors}
As pointed out in section \ref{sec:cirq}, Cirq already contains the language necessary to express quantum computations, parameterized circuits, and measurements.  Guided by principle \ref{prin:min}, TFQ should allow direct injection of Cirq expressions into the computational graph of TensorFlow.  This is enabled by the \Colorbox{bkgd}{\lstinline{tfq.convert_to_tensor}} function.  We saw the use of this function in the quantum binary classifier, where a list of data generation circuits specified in Cirq was wrapped in this function to promote them to tensors.  Below we show how a quantum data point, a quantum model, and a quantum measurement can be converted into tensors:
\begin{lstlisting}
q0 = cirq.GridQubit(0, 0)
q_data_raw = cirq.Circuit(cirq.H(q0))
q_data = tfq.convert_to_tensor([q_data_raw])

theta = sympy.Symbol('theta')
q_model_raw = cirq.Circuit(
    cirq.Ry(theta).on(q0))
q_model = tfq.convert_to_tensor([q_model_raw])

q_measure_raw = 0.5 * cirq.Z(q0)
q_measure = tfq.convert_to_tensor(
    [q_measure_raw])
\end{lstlisting}
This conversion is backed by our custom serializers.  Once a \Colorbox{bkgd}{\lstinline{Circuit}} or \Colorbox{bkgd}{\lstinline{PauliSum}} is serialized, it becomes a tensor of type \Colorbox{bkgd}{\lstinline{tf.string}}.  This is the reason for the use of \Colorbox{bkgd}{\lstinline{tf.keras.Input(shape=(), dtype=tf.dtypes.string)}} when creating inputs to Keras models, as seen in the quantum binary classifier example.

\subsubsection{Composing Quantum Models}\label{sec:q_models}
After injecting quantum data and quantum models into the computational graph, a custom TensorFlow operation is required to combine them.  In support of guiding principle \ref{prin:batch}, TFQ implements the \Colorbox{bkgd}{\lstinline{tfq.layers.AddCircuit}} layer for combining tensors of circuits.  In the following code, we use this functionality to combine the quantum data point and quantum model defined in subsection \ref{sec:q_tensors}:
\begin{lstlisting}
add_op = tfq.layers.AddCircuit()
data_and_model = add_op(q_data, append=q_model)
\end{lstlisting}
To quantify the performance of a quantum model on a quantum dataset, we need the ability to define loss functions.  This requires converting quantum information into classical information.  This conversion process is accomplished by either \textbf{sampling} the quantum model, which stochastically produces bitstrings according to the probability amplitudes of the model, or by specifying a measurement and taking \textbf{expectation values}.

\subsubsection{Sampling and Expectation Values}\label{sec:quantum_classical_conversion}
Sampling from quantum circuits is an important use case in quantum computing.  The recently achieved milestone of quantum supremacy \cite{arute2019quantum} is one such application, in which the difficulty of sampling from a quantum model was used to gain a computational edge over classical machines.  

TFQ implements \Colorbox{bkgd}{\lstinline{tfq.layers.Sample}}, a Keras layer which enables sampling from batches of circuits in support of design objective \ref{prin:batch}.  The user supplies a tensor of parameterized circuits, a list of symbols contained in the circuits, and a tensor of values to substitute for the symbols in the circuit.  Given  these, the \Colorbox{bkgd}{\lstinline{Sample}} layer produces a \Colorbox{bkgd}{\lstinline{tf.RaggedTensor}} of shape \Colorbox{bkgd}{\lstinline{[batch_size, num_samples, n_qubits]}}, where the n\_qubits dimension is ragged to account for the possibly varying circuit size over the input batch of quantum data.  For example, the following code takes the combined data and model from section \ref{sec:q_models} and produces a tensor of size [1, 4, 1] containing four single-bit samples:
\begin{lstlisting}
sample_layer = tfq.layers.Sample()
samples = sample_layer(
    data_and_model, symbol_names=['theta'], symbol_values=[[0.5]], repetitions=4)
\end{lstlisting}
Though sampling is the fundamental interface between quantum and classical information, differentiability of quantum circuits is much more convenient when using \textbf{expectation values}, as gradient information can then be backpropagated (see section \ref{sec:theory} for more details).

In the simplest case, expectation values are simply averages over samples.  In quantum computing, expectation values are typically taken with respect to a measurement operator $M$.  This involves sampling bitstrings from the quantum circuit as described above, applying $M$ to the list of bitstring samples to produce a list of numbers, then taking the average of the result.  TFQ provides two related layers with this capability: 

In contrast to sampling (which is by default in the \textit{standard computational basis}, the $\hat{Z}$ eigenbasis of all qubits), taking expectation values requires defining a measurement.  As discussed in section \ref{sec:cirq}, these are first defined as \Colorbox{bkgd}{\lstinline{cirq.PauliSum}} objects and converted to tensors.  TFQ implements \Colorbox{bkgd}{\lstinline{tfq.layers.Expectation}}, a Keras layer which enables the extraction of measurement expectation values from quantum models.  The user supplies a tensor of parameterized circuits, a list of symbols contained in the circuits, a tensor of values to substitute for the symbols in the circuit, and a tensor of operators to measure with respect to them.  Given these inputs, the layer outputs a tensor of expectation values.  Below, we show how to take an expectation value of the measurement defined in section \ref{sec:q_tensors}:
\begin{lstlisting}
expectation_layer = tfq.layers.Expectation()
expectations = expectation_layer(
    circuit = data_and_model,
    symbol_names = ['theta'],
    symbol_values = [[0.5]],
    operators = q_measure)
\end{lstlisting}
In Fig.~\ref{fig:expectation_graph}, we illustrate the dataflow graph which backs the expectation layer, when the parameter values are supplied by a classical neural network.  The expectation layer is capable of using either a simulator or a real device for execution, and this choice is simply specified at run time.  While Cirq simulators may be used for the backend, TFQ provides its own native TensorFlow simulator written in performant C++.  A description of our quantum circuit simulation code is given in section \ref{sec:qsim}.

Having converted the output of a quantum model into classical information, the results can be fed into subsequent computations.  In particular, they can be fed into functions that produce a single number, allowing us to define loss functions over quantum models in the same way we do for classical models.

\subsubsection{Differentiating Quantum Circuits}
We have taken the first steps towards implementation of quantum machine learning, having defined quantum models over quantum data and loss functions over those models.  As described in both the introduction and our first guiding principle, differentiability is the critical machinery needed to allow training of these models.  As described in section \ref{sec:tensorflow}, the architecture of TensorFlow is optimized around backpropagation of errors for efficient updates of model parameters; one of the core contributions of TFQ is integration with TensorFlow's backpropagation mechanism.  TFQ implements this functionality with our differentiators module.  The theory of quantum circuit differentiation will be covered in section \ref{sec:gradients}; here, we overview the software that implements the theory.

Since there are many ways to calculate gradients of quantum circuits, TFQ provides the \Colorbox{bkgd}{\lstinline{tfq.differentiators.Differentiator}} interface.  Our \Colorbox{bkgd}{\lstinline{Expectation}} and \Colorbox{bkgd}{\lstinline{SampledExpectation}} layers rely on classes inheriting from this interface to specify how TensorFlow should compute their gradients.  While advanced users can implement their own custom differentiators by inheriting from the interface, TFQ comes with several built-in options, two of which we highlight here.  These two methods are instances of the two main categories of quantum circuit differentiators: the finite difference methods and the parameter shift methods.

The first class of quantum circuit differentiators is the finite difference methods.  This class samples the primary quantum circuit for at least two different parameter settings, then combines them to estimate the derivative.  The \Colorbox{bkgd}{\lstinline{ForwardDifference}} differentiator provides most basic version of this.  For each parameter in the circuit, the circuit is sampled at the current setting of the parameter. Then, each parameter is perturbed separately and the circuit resampled. 

For the 2-local circuits implementable on near-term hardware, methods more sophisticated than finite differences are possible.  These methods involve running an ancillary quantum circuit, from which the gradient of the  primary circuit with respect to some parameter can be directly measured.  One specific method is gate decomposition and parameter shifting \cite{parameter_shift_gradients}, implemented in TFQ as the \Colorbox{bkgd}{\lstinline{ParameterShift}} differentiator.  For in-depth discussion of the theory, see section \ref{sec:gradients_parameter_shift}.

The differentiation rule used by our layers is specified through an optional keyword argument.   Below, we show the expectation layer being called with our parameter shift rule:
\begin{lstlisting}
diff = tfq.differentiators.ParameterShift()
expectation = tfq.layers.Expectation(differentiator=diff)
\end{lstlisting}
For further discussion of the trade-offs when choosing between differentiators, see the gradients tutorial on our GitHub website:
\fancylink{https://github.com/tensorflow/quantum/blob/master/docs/tutorials/gradients.ipynb}{docs/tutorials/gradients.ipynb}

\subsubsection{Simplified Layers}
Some workflows do not require control as sophisticated as our \Colorbox{bkgd}{\lstinline{Expectation}}, \Colorbox{bkgd}{\lstinline{Sample}}, and \Colorbox{bkgd}{\lstinline{SampledExpectation}} layers allow.  For these workflows we provide the \Colorbox{bkgd}{\lstinline{PQC}} and \Colorbox{bkgd}{\lstinline{ControlledPQC}} layers.  Both of these layers allow parameterized circuits to be updated by hybrid backpropagation without the user needing to provide the list of symbols associated with the circuit.  The \Colorbox{bkgd}{\lstinline{PQC}} layer provides automated Keras management of the variables in a parameterized circuit:
\begin{lstlisting}
q = cirq.GridQubit(0, 0)
(a, b) = sympy.symbols("a b")
circuit = cirq.Circuit(
  cirq.Rz(a)(q), cirq.Rx(b)(q))
outputs = tfq.layers.PQC(circuit, cirq.Z(q))
quantum_data = tfq.convert_to_tensor([
  cirq.Circuit(), cirq.Circuit(cirq.X(q))])
res = outputs(quantum_data)
\end{lstlisting}
When the variables in a parameterized circuit will be controlled completely by other user-specified machinery, for example by a classical neural network, then the user can call our \Colorbox{bkgd}{\lstinline{ControlledPQC}} layer:
\begin{lstlisting}
outputs = tfq.layers.ControlledPQC(
  circuit, cirq.Z(bit))
model_params = tf.convert_to_tensor(
  [[0.5, 0.5], [0.25, 0.75]])
res = outputs([quantum_data, model_params])
\end{lstlisting}
Notice that the call is similar to that for PQC, except that we provide parameter values for the symbols in the circuit.  These two layers are used extensively in the applications highlighted in the following sections.

\subsubsection{Basic Quantum Datasets}\label{sec:TFQDatasets}
A major goal of TensorFlow Quantum is to expand the application of machine learning to quantum data.  Towards that goal, here we provide some basic labelled datasets with the \Colorbox{bkgd}{\lstinline{tfq.datasets}} module.

The first dataset is \Colorbox{bkgd}{\lstinline{tfq.datasets.excited_cluster_states}}.  Given a list of qubits, this function builds a dataset of ground state and excited cluster states.  The ground state is labelled with $-1$, while excited states are labelled with $+1$.  With this data, a QML model can be trained to distinguish between the ground and excited states on a ring.  This is the same dataset used in the QCNN tutorial \ref{sec:HQCNN}.

The second dataset offered is \Colorbox{bkgd}{\lstinline{tfq.datasets.tfi_chain}}.  This is a 1D Transverse field Ising-model, which can be written as
\[
\hat{H} = - \sum_{j} (\hat{Z}_j \hat{Z}_{j+1} - g\hat{X}_j).
\]
This dataset contains 81 datapoints, corresponding to the ground states of the 1D TFI chain for $g$ in $[0.2,1.8]$ in increments of $0.2$.  Each datapoint contains a circuit, a label, a Hamiltonian, and some additional metadata.  The circuit prepares the approximate ground state of the Hamiltonian in the datapoint.

This dataset can be used for many purposes.  For one, the labels can be used to train a QML model to distinguish different phases of a chain.  The labels are 0 for the ferromagnetic phase (occurs for $g<1$), 1 for the critical point ($g==1$) and 2 for the paramagnetic phase ($g>1$).  Further, the additional metadata contains the exact ground state energy from expensive exact diagonalization; this can be used as a benchmark for VQE-like models.  If a QML model is successfully trained to prepare the ground state of a Hamiltonian given in the dataset, it should achieve the corresponding ground state energy given in the datapoint.

The third dataset offered is \Colorbox{bkgd}{\lstinline{tfq.datasets.xxz_chain}}.  
\[
\hat{H} = \sum_{j} (\hat{X}_j \hat{X}_{j+1} + \hat{Y}_j \hat{Y}_{j+1} + \Delta\,\hat{Z}_j \hat{Z}_{j+1})
\]
This dataset contains 76 datapoints, corresponding to the ground states of the 1D XXZ chain for $\Delta$ in $[0.3,1.8]$ in increments of $0.2$.

Similar to the TFI dataset, each datapoint contains a circuit, a label, a Hamiltonian, and some additional metadata.  The circuit prepares the approximate ground state of the Hamiltonian in the datapoint.  The labels are 0 for the critical metallic phase ($\Delta<=1$) and 1 for the insulating phase ($\Delta>1$).  As such, this dataset can also be used for classification and optimization benchmarking.

We expect to add more datasets as consensus is reached in the QML community around good benchmark tasks.  As such, we welcome contributions!  Those looking to contribute new quantum datasets should reach out via our GitHub page.

\subsection{High Performance Quantum Circuit Simulation with qsim}\label{sec:qsim}
Concurrently with TFQ, we are open sourcing qsim, a software package for simulating quantum circuits on classical computers.  We have adapted its C++ implementation to work inside TFQ's TensorFlow ops.  The performance of qsim derives from two key ideas that can be seen in the literature on classical simulators for quantum circuits \cite{smelyanskiy2016, haner2017}.  The first idea is the fusion of gates in a quantum circuit with their neighbors to reduce the number of matrix-vector multiplications required when applying the circuit to a wavefunction.  The second idea is to create a set of matrix-vector multiplication functions specifically optimized for the application of two-qubit (or more) gates to state vectors, to take maximal advantage of gate fusion.  We discuss these points in detail below.  To quantify the performance of qsim, we also provide an initial benchmark comparing qsim to Cirq.  We note that qsim is significantly faster.  We further note that the qsim benchmark times include the full TFQ software stack of serializing a circuit to ProtoBuffs in Python, conversion of ProtoBuffs to C++ objects inside the dataflow graph for our custom TensorFlow ops, and the relaying of results back to Python.

\subsubsection{Comment on the Simulation of Quantum Circuits}
To motivate the qsim fusing algorithm, consider how circuits are applied to states in simulation.  Suppose we wish to apply two gates $\hat{G}_1$ and $\hat{G}_2$ to our initial state $\ket{\psi}$, and suppose these gates act on the same two qubits.  Since gate application is associative, we have $(\hat{G}_2 \hat{G}_1) \ket{\psi} = \hat{G}_2 (\hat{G}_1 \ket{\psi})$.  However, as the number of qubits $n$ supporting $\ket{\psi}$ grows, the left side of the equality becomes much faster to compute.  This is because applying a gate to a state requires broadcasting the parameters of the gate to all $2^{n}$ elements of the state vector, so that each gate application incurs a cost scaling as $2^{n}$.  In contrast, multiplying small gate matrices incurs a small cost.  This means a simulation of a circuit will be fastest if we pre-multiply as many gates as possible, while keeping the matrix size small, before applying them to a state.  This pre-multiplication is called \textit{gate fusion} and is accomplished with the qsim fusion algorithm.

\subsubsection{Gate Fusion with qsim}
\begin{figure}
    \centering
    \includegraphics[width=1\columnwidth]{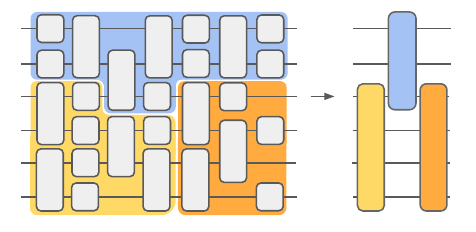}
\caption{Visualization of the qsim fusing algorithm.  In this example, a set of single and two-qubit gates are fused into three or four-qubit gates, increasing the speed of subsequent circuit simulation. qsim supports gate fusions for larger numbers of qubits (up to 6 qubits at a time).}
    \label{fig:qsim}
\end{figure}

The core idea of the fusion algorithm is to construct multi-qubit gates out of smaller gates.  The circuit can be interpreted as a $(d+1)$-dimensional lattice, where $d$ denotes the spatial direction and $1$ denotes the time direction.  Suppose we choose a fusion size $F$ between 2 and 6 inclusive.  The fusion algorithm combines gates that are close in space and time to form larger gates that act on up to $F$ qubits.  There are two steps in the algorithm.  First, we fuse each $q$-qubit gate ($q \geq 2$) with smaller or same size neighbors in time direction that act on the same set of qubits.  Second, we greedily (in increasing time order) combine small gates that are neighbors in space and time to construct the largest gates possible (up to $F$ qubits).  Typically $F=4$ is optimal for many threads and $F=2$ or $F=3$ is optimal for one or two threads.

\subsubsection{Hardware-Acceleration}
With the given quantum circuit fused into the minimal number of up to $F$-qubit gates, we need simulators optimized for applying up to $2^F\times2^F$ matrices to state vectors.  TFQ will adapt to the user's available hardware.  For CPU based simulations, SSE2 instruction set \cite{sse2} and  AVX2 + AVX512 instruction sets \cite{avx2} will be detected and used to increase performance.  If a compatible CUDA GPU is detected TFQ will be able to switch to GPU based simulation as well.  The next section illustrates this power with benchmarks comparing the performance of TFQ to the performance of parallelized Cirq running in simulation mode.  In the future, we hope to expand the range of custom simulation hardware supported to include TPU integration.

\subsubsection{Benchmarks}\label{sec:benchmarks}
Here, we demonstrate the performance of TFQ, backed by qsim, relative to Cirq on two benchmark simulation tasks.  As detailed above, the performance difference is due to circuit pre-processing via gate fusion combined with performant C++ simulators.  The benchmarks were performed on a desktop equipped with an Intel(R) Xeon(R) Gold 6154 CPU (18 cores and 36 threads) and an NVidia V100 GPU.

\begin{figure}
    \centering
    \includegraphics[width=1\columnwidth]{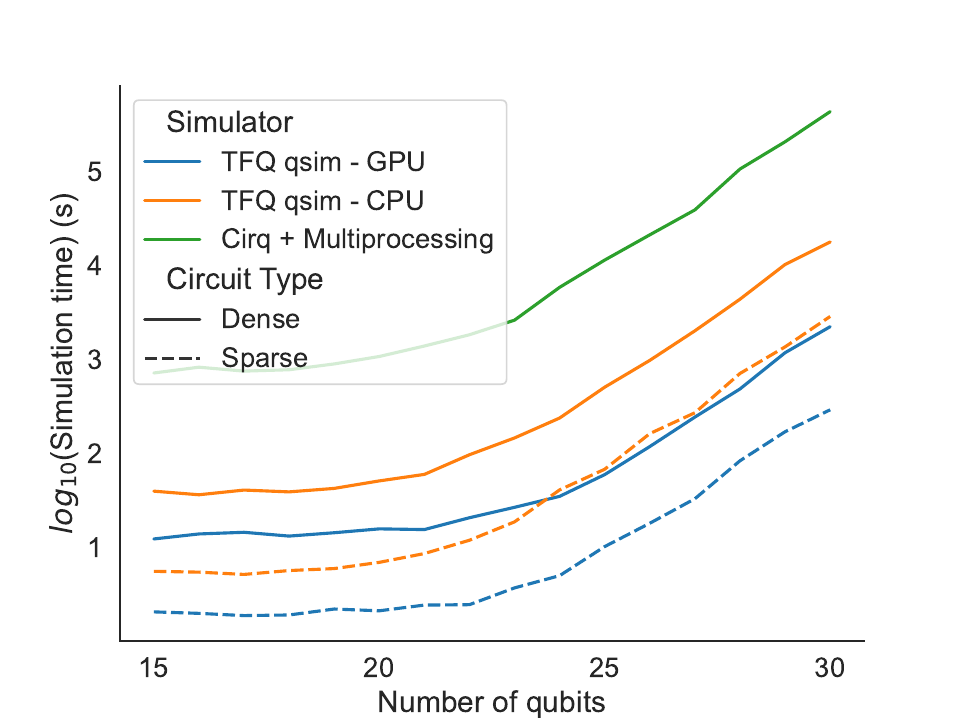}
    \caption{Performance of TFQ and Cirq on simulation tasks.  The plots show the base 10 logarithm of the total time to solution (in seconds) versus the number of qubits simulated.  Simulation of 500 random circuits, taken in batches of 50.  Circuits were of depth 40. In the dense case, we see that TFQ + qsim on CPU is around 10x faster than Cirq. With the GPU this gap grows to approximately 50x. In the case of sparse circuits, qsim gate fusion makes this speed difference even more pronounced reaching well over 100x performance difference on both CPU and GPU.  }
    \label{fig:qsim_benchmark}
\end{figure}

The first benchmark task is simulation of 500 random (early variant of beyond-classical/supremacy-style) circuits batched 50 at a time.  These circuits were generated using the Cirq function \Colorbox{bkgd}{\lstinline{cirq.generate_boixo_2018_supremacy_circuits_v2}}.  These circuits are only tractable for benchmarking due to the small numbers of qubits involved here.  These circuits involve dense (subject to a constraint~\cite{boixo2018characterizing}) interleaved two-qubit gates to generate entanglement as quickly as possible.  In summary, at the largest benchmarked problem size of 30 qubits, qsim achieves an approximately 10-fold improvement in simulation time over Cirq.  The performance curves are shown in Fig.~\ref{fig:qsim_benchmark}.

When the simulated circuits have more sparse structure, the Fusion algorithm allows us to achieve a larger performance boost by reducing the number of gates that ultimately need to be simulated.  The circuits for this task are a factorized version of the supremacy-style circuits which generate entanglement only on small subsets of the qubits.  In summary, for these circuits, we find a roughly 100 times improvement in simulation time in TFQ versus Cirq.  The performance curves are shown in Fig.~\ref{fig:qsim_benchmark}.

Thus we see that in addition to our core functionality of implementing native TensorFlow gradients for quantum circuits, TFQ also provides a significant boost in performance over Cirq when running in simulation mode.  Additionally, as noted before, this performance boost is despite the additional overhead of serialization between the TensorFlow frontend and qsim proper.

\subsubsection{Large-scale simulation for quantum machine learning}

Recently, we have provided a series of learning-theoretic tests for evaluating whether a quantum machine learning model can predict more accurately than classical machine learning models \cite{huang2021_power}. One of the key findings is that various existing quantum machine learning models perform slightly better than standard classical machine learning models in small system sizes. However, these quantum machine learning models perform substantially worse than classical machine learning models in larger system sizes (more than $10$ to $15$ qubits). Note that performing better in small system sizes is not useful since a classical algorithm can efficiently simulate the quantum machine learning model. The above observation shows that understanding the performance of quantum machine learning models in large system sizes is very crucial for assessing whether the quantum machine learning model will eventually provide an advantage over classical models.
Ref. \cite{huang2021_power} also provide improvements to existing quantum machine learning models to improve the prediction performance in large system sizes.

In the numerical experiments conducted in \cite{huang2021_power}, we consider the simulation of these quantum machine learning models up to $30$ qubits. The large-scale simulation allows us to gauge the potential and limitations of different quantum machine learning models better. We utilize the qsim software package in TFQ to perform large-scale quantum simulations using Google Cloud Platform. The simulation reaches a peak throughput of up to $1.1$ quadrillion floating-point operations per second
(petaflop/s). Trends of approximately $300$ teraflop/s for quantum simulation and $800$ teraflop/s for classical analysis were observed up to the maximum experiment size with the overall floating-point operations across all experiments totaling approximately two quintillions (exaflop).

\subsubsection{Noise in qsim}\label{sec:QsimOtherFeatures}
The study of noise in quantum circuits is an important use-case for simulators to support.  To address this, qsim supports simulation of all the common channels provided by Cirq.  Further, TFQ supports the serialization of circuits containing such channels, allowing users to apply the many features of TFQ to the study of noisy circuits.

The two main choices for simulation of noisy circuits are density matrix simulation and trajectory simulation.  There are trade-offs between these choices.  Density matrix simulation keeps a full density matrix in memory, and at each timestep applies all the Kraus operators associated with a channel; at the end of simulation, properties of the state can be computed against this density matrix.  For a circuit on $N$ qubits, a full density matrix simulation requires memory of size $2^{2N}$. In contrast, trajectory simulations keep only a pure state in memory, and at each time step, a single Kraus operator in the channel is selected probabilistically and applied to the state; properties of interest are measured against the resulting pure state.  The process must be run many times, and the results averaged; but, the memory required at any one time is only $2^{N}$.  When a circuit is not too noisy, it is often more efficient to average over trajectories than it is to simulate the full density matrix, so we choose this method for TFQ.  For information on how to use this feature, see \ref{sec:SimulatingNoisyCircuits}.

\section{Theory of Hybrid Quantum-Classical Machine Learning}\label{sec:theory}

In the previous section, we reviewed the building blocks required for use of TFQ.  In this section, we consider the theory behind the software.  
We define quantum neural networks as products of parameterized unitary matrices. 
Samples and expectation values are defined by expressing the loss function as an inner product.  With quantum neural networks and expectation values defined, we can then define gradients.  We finally combine quantum and classical neural networks and formalize hybrid quantum-classical backpropagation, one of core components of TFQ.

\subsection{Quantum Neural Networks}\label{sec:2qnn}
A Quantum Neural Network ansatz can generally be written as a product of layers of unitaries in the form
\begin{equation}\label{eq:full_par_circ}
    \hat{U}(\bm{\theta}) = \prod_{\ell=1}^L\hat{V}^{\ell}\hat{U}^{\ell}(\bm{\theta}^{\ell}),
\end{equation}
where the $\ell^\text{th}$ layer of the QNN consists of the product of $\hat{V}^{\ell}$, a non-parametric unitary,  and $\hat{U}^{\ell}(\bm{\theta}^{\ell})$ a unitary with variational parameters (note the superscripts here represent indices rather than exponents). The multi-parameter unitary of a given layer can itself be generally comprised of multiple unitaries $\{\hat{U}_{j}^{\ell}(\theta^{\ell}_{j})\}_{j=1}^{M_\ell}$ applied in parallel:
\begin{equation}\label{eq:layer_blocks}
    \hat{U}^{\ell}(\bm{\theta}^{\ell})\equiv \bigotimes_{j=1}^{M_\ell} \hat{U}_{j}^{\ell}(\theta^{\ell}_{j}).
\end{equation}
Finally, each of these unitaries $\hat{U}_{j}^{\ell}$ can be expressed as the exponential of some generator $\hat{g}_{j\ell}$, which itself can be any Hermitian operator on $n$ qubits (thus expressible as a linear combination of $n$-qubit Pauli's),
\begin{equation}\label{eq:Pauli_decomp}
  \hat{U}^{\ell}_{j}(\theta^{\ell}_{j})=   e^{-i\theta^{\ell}_{j} \hat{g}_j^{\ell }}, \quad \hat{g}_j^{\ell} = \sum_{k=1}^{K_{j\ell}} \beta^{j\ell}_k \hat{P}_k,
\end{equation}
 where $\hat{P}_k \in \mathcal{P}_n$, here $\mathcal{P}_n$ denotes the Paulis on $n$-qubits \cite{Gottesman1997}, and $\beta^{j\ell}_k\in \mathbb{R}$ for all $k,j,\ell$. For a given $j$ and $\ell$, in the case where all the Pauli terms commute, i.e.  $[\hat{P}_k,\hat{P}_m]=0$ for all $m,k$ such that $\beta^{j\ell}_m,\beta^{j\ell}_k\neq 0$, one can simply decompose the unitary into a product of exponentials of each term,
 \begin{equation}\label{eq:}
    \hat{U}^{\ell}_{j}(\theta^{\ell}_{j}) = \prod_k e^{-i\theta^{\ell}_{j} \beta^{j\ell}_k \hat{P}_k}.
 \end{equation}
Otherwise, in instances where the various terms do not commute, one may apply a Trotter-Suzuki decomposition of this exponential \cite{suzuki1990fractal}, or other quantum simulation methods \cite{Campbell_2019}. 
Note that in the above case where the unitary of a given parameter is decomposable as the product of exponentials of Pauli terms, one can explicitly express the layer as 
\begin{equation}\label{eq:cs_decomp}
    \hat{U}^{\ell}_{j}(\theta^{\ell}_{j}) = \prod_k \left[\cos({\theta^{\ell}_{j} \beta^{j\ell}_k })\hat{I} -i \sin({\theta^{\ell}_{j} \beta^{j\ell}_k })\hat{P}_k\right].
 \end{equation}
The above form will be useful for our discussion of gradients of expectation values below. 

\begin{figure}
    \centering
    \includegraphics[width=0.9\columnwidth]{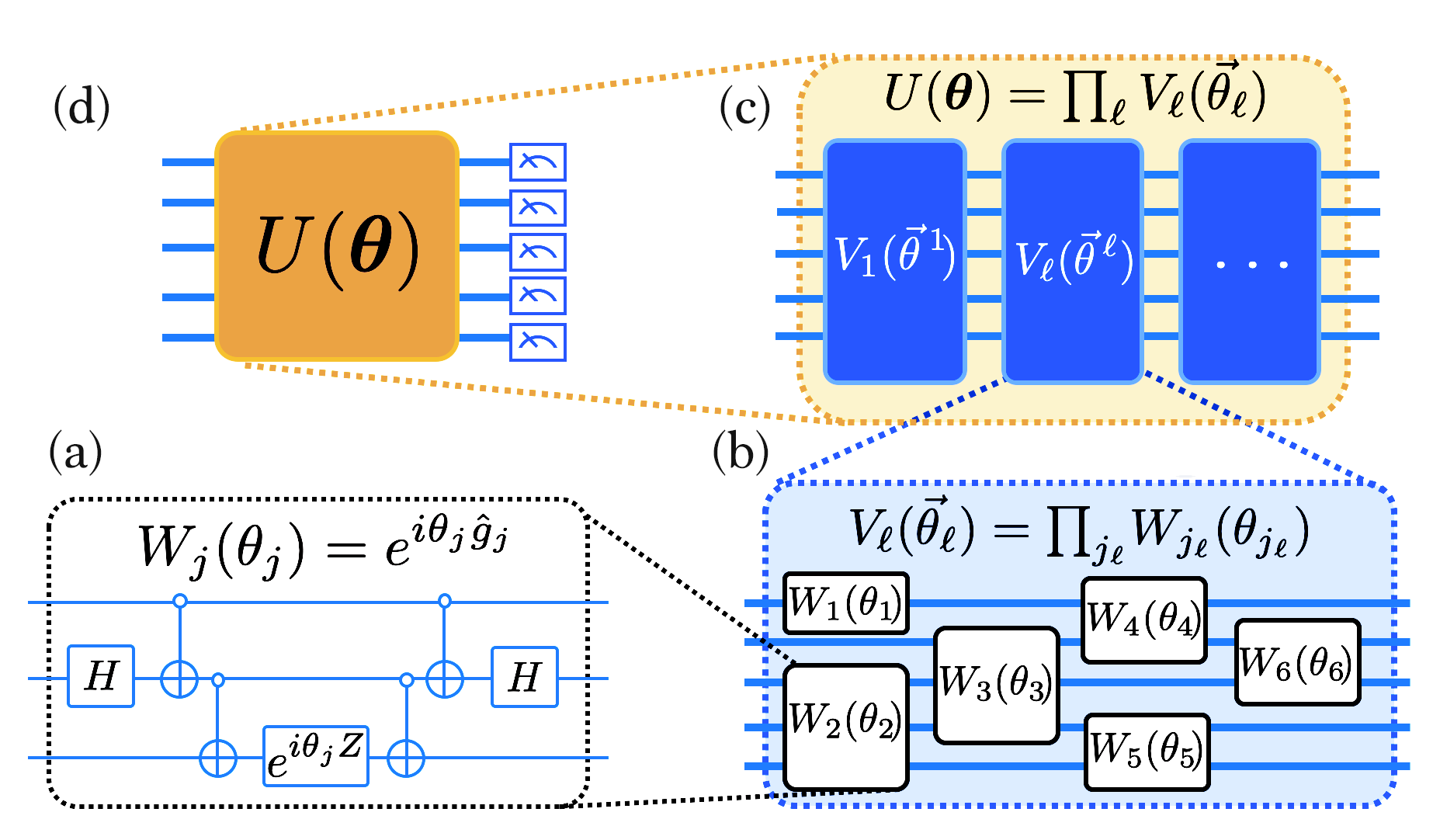}
\caption{
High-level depiction of a multilayer quantum neural network (also known as a parameterized quantum circuit), at various levels of abstraction. (a) At the most detailed level we have both fixed and parameterized quantum gates, any parameterized operation is compiled into a combination of parameterized single-qubit operations. (b) Many singly-parameterized gates $W_j(\theta_j)$ form a multi-parameterized unitary $V_l(\Vec{\theta}_l)$ which performs a specific function. (c) The product of multiple unitaries $V_l$ generates the quantum model $U(\bm{\theta})$ shown in (d). }
    \label{fig:qnn_compiled}
\end{figure}

\subsection{Sampling and Expectations}
To optimize the parameters of an ansatz from equation \eqref{eq:full_par_circ}, we need a cost function to optimize. In the case of standard variational quantum algorithms, this cost function is most often chosen to be the expectation value of a cost Hamiltonian, 
\begin{equation}\label{eq:cost}
    f(\bm{\theta}) = \braket{\hat{H}}_{\bm{\theta}} \equiv  \bra{\Psi_0}\hat{U}^\dagger(\bm{\theta})\hat{H}\hat{U}(\bm{\theta})\ket{\Psi_0}
\end{equation}
where $\ket{\Psi_0}$ is the input state to the parameterized circuit. In general, the cost Hamiltonian can be expressed as a linear combination of operators, e.g. in the form
\begin{equation}\label{eq:cost_lincomb}
    \hat{H} = \sum_{k=1}^N \alpha_k\hat{h}_k \equiv \bm{\alpha}\cdot \bm{\hat{h}},
\end{equation}
where we defined a vector of coefficients $\bm{\alpha}\in \mathbb{R}^N$ and a vector of $N$ operators $\bm{\hat{h}}$.  Often this decomposition is chosen such that each of these sub-Hamiltonians is in the $n$-qubit Pauli group $\hat{h}_k \in\mathcal{P}_n$. The expectation value of this Hamiltonian is then generally evaluated via quantum expectation estimation, i.e. by taking the linear combination of expectation values of each term 
\begin{equation}\label{eq:qee_lincomb}
    f(\bm{\theta}) =\braket{\hat{H}}_{\bm{\theta}} = \sum_{k=1}^N \alpha_k \braket{\hat{h}_k}_{\bm{\theta}} \equiv \bm{\alpha}\cdot \bm{h}_{\bm{\theta}},
\end{equation}
where we introduced the vector of expectations $\bm{h}_{\bm{\theta}} \equiv \braket{\bm{\hat{h}}}_{\bm{\theta}} $.
In the case of non-commuting terms, the various expectation values $\braket{\hat{h}_k}_{\bm{\theta}}$ are estimated over separate runs. 
 
Note that, in practice, each of these quantum expectations is estimated via sampling of the output of the quantum computer \cite{mcclean2016theory}. Even assuming a perfect fidelity of quantum computation, sampling measurement outcomes of eigenvalues of observables from the output of the quantum computer to estimate an expectation will have some non-negligible variance for any finite number of samples. Assuming each of the Hamiltonian terms of equation \eqref{eq:cost_lincomb} admit a Pauli operator decomposition as 
\begin{equation}
    \hat{h}_j = \sum_{k=1}^{J_{j}} \gamma^{j}_k \hat{P}_k,
\end{equation}
where the $\gamma^{j}_k$'s are real-valued coefficients and the $\hat{P}_j$'s are Paulis that are Pauli operators \cite{Gottesman1997}, then to get an estimate of the expectation value $\braket{ \hat{h}_j }$ within an accuracy $\epsilon$, one needs to take a number of measurement samples scaling as $\sim\mathcal{O}(\lVert \hat{h}_k\rVert_*^2/\epsilon^2)$, where $\lVert \hat{h}_k\rVert_* = \sum_{k=1}^{J_{j}} |\gamma^{j}_k|$ is the Pauli coefficient norm of each Hamiltonian term. Thus, to estimate the expectation value of the full Hamiltonian \eqref{eq:cost_lincomb} accurately within a precision $\varepsilon = \epsilon\sum_k|\alpha_k|^2$, we would need on the order of \(\sim\mathcal{O}(\tfrac{1}{\epsilon^2}\sum_k|\alpha_k|^2\lVert \hat{h}_k\rVert_*^2) \) measurement samples in total \cite{rubin2018application,wecker2015progress}, as we would need to measure each expectation independently if we are following the quantum expectation estimation trick of \eqref{eq:qee_lincomb}. 
 
This is in sharp contrast to classical methods for gradients involving backpropagation, where gradients can be estimated to numerical precision; i.e. within a precision $\epsilon$ with $\sim\mathcal{O}(\text{PolyLog}(\epsilon))$ overhead. Although there have been attempts to formulate a backpropagation principle for quantum computations \cite{verdon2018universal}, these methods also rely on the measurement of a quantum observable, thus also requiring $\sim\mathcal{O}(\tfrac{1}{\epsilon^2})$ samples.
 
As we will see in the following section \ref{sec:gradients}, estimating gradients of quantum neural networks on quantum computers involves the estimation of several expectation values of the cost function for various values of the parameters. One trick that was recently pointed out \cite{harrow2019low,sweke2019stochastic} and has been proven to be successful both theoretically and empirically to estimate such gradients is the stochastic selection of various terms in the quantum expectation estimation. This can greatly reduce the number of measurements needed per gradient update, we will cover this in subsection \ref{sec:stoc_ge}.

\subsection{Gradients of Quantum Neural Networks}\label{sec:gradients}

Now that we have established how to evaluate the loss function, let us describe how to obtain gradients of the cost function with respect to the parameters. Why should we care about gradients of quantum neural networks? In classical deep learning, the most common family of optimization heuristics for the minimization of cost functions are gradient-based techniques \cite{lecun1998gradient,bottou2010large,ruder2016overview}, which include stochastic gradient descent and its variants. To leverage gradient-based techniques for the learning of multilayered models, the ability to rapidly differentiate error functionals is key. For this, the backwards propagation of errors \cite{lecun1988theoretical} (colloquially known as \textit{backprop}), is a now canonical method to progressively calculate gradients of parameters in deep networks. In its most general form, this technique is known as \textit{automatic differentiation} \cite{baydin2017automatic}, and has become so central to deep learning that this feature of \textit{differentiability} is at the core of several frameworks for deep learning, including of course TensorFlow (TF) \cite{abadi2016tensorflow}, JAX \cite{frostig2018compiling}, and several others. 

To be able to train hybrid quantum-classical models (section \ref{sec:hqcm-th}), the ability to take gradients of quantum neural networks is key. Now that we understand the greater context, let us describe a few techniques below for the estimation of these gradients.

\subsubsection{Finite difference methods}\label{sec:gradients_finite_difference}
 A simple approach is to use simple finite-difference methods, for example, the central difference method, 
 \begin{equation}\label{eq:central_diff}
     \partial_k f(\bm{\theta}) = \frac{f(\bm{\theta} + \varepsilon\bm{\Delta}_k)-f(\bm{\theta} - \varepsilon\bm{\Delta}_k)}{2\varepsilon} +\mathcal{O}(\varepsilon^2)
 \end{equation}
 which, in the case where there are $M$ continuous parameters, involves $2M$ evaluations of the objective function, each evaluation varying the parameters by $\varepsilon$ in some direction, thereby giving us an estimate of the gradient of the function with a precision $\mathcal{O}(\varepsilon^2)$. Here the $\bm{\Delta}_k$ is a unit-norm perturbation vector in the $k^{\text{th}}$ direction of parameter space, $(\bm{\Delta}_k)_j = \delta_{jk}$. In general, one may use lower-order methods, such as forward difference with $\mathcal{O}(\varepsilon)$ error from $M+1$ objective queries \cite{farhi2018classification}, or higher order methods, such as a five-point stencil method, with $\mathcal{O}(\varepsilon^4)$ error from $4M$ queries \cite{abramowitz20061965}. 
 
 \subsubsection{Parameter shift methods}\label{sec:gradients_parameter_shift}
 As recently pointed out in various works \cite{schuld2018evaluating,harrow2019low}, given knowledge of the form of the ansatz (e.g. as in \eqref{eq:Pauli_decomp}), one can measure the analytic gradients of the expectation value of the circuit for Hamiltonians which have a single-term in their Pauli decomposition \eqref{eq:Pauli_decomp} (or, alternatively, if the Hamiltonian has a spectrum  $\{\pm \lambda\}$ for some positive $\lambda$). For multi-term Hamiltonians, in \cite{schuld2018evaluating} a method to obtain the analytic gradients is proposed which uses a linear combination of unitaries. Here, instead, we will simply use a change of variables and the chain rule to obtain analytic gradients of parametric unitaries of the form \eqref{eq:cs_decomp} without the need for ancilla qubits or additional unitaries. 
 
 For a parameter of interest $\theta^{\ell}_{j}$ appearing in a layer $\ell$ in a parametric circuit in the form \eqref{eq:cs_decomp}, consider the change of variables $\eta_{k}^{j\ell} \equiv \theta^{\ell}_{j}\beta_k^{j\ell}$, then from the chain rule of calculus \cite{newton1999principia}, we have
 \begin{equation}\label{eq:chain_rule}
     \frac{\partial f}{\partial \theta^{\ell}_{j}} = \sum_k \frac{\partial f}{\partial \eta_{k}^{j\ell}} \frac{\partial \eta_{k}^{j\ell}}{\partial \theta^{\ell}_{j}}  = \sum_k \beta_k^{j\ell} \frac{\partial f}{\partial \eta_k}.
 \end{equation}
 Thus, all we need to compute are the derivatives of the cost function with respect to $\eta^{j\ell}_k$. Due to this change of variables, we need to reparameterize our unitary $\hat{U}(\bm{\theta})$ from \eqref{eq:full_par_circ} as
 \begin{equation}
      \hat{U}^{\ell}(\bm{\theta}^{\ell}) \mapsto  \hat{U}^{\ell}(\bm{\eta}^{\ell})  \equiv \bigotimes_{j\in \mathcal{I}_{\ell}} \Big(\prod_k e^{-i\eta^{j\ell}_k \hat{P}_k}\Big),
 \end{equation}
 where $\mathcal{I}_{\ell} \equiv \{1,\ldots, M_{\ell} \}$ is an index set for the QNN layers. One can then expand each exponential in the above in a similar fashion to \eqref{eq:cs_decomp}: \begin{equation}
    e^{-i\eta^{j\ell}_k \hat{P}_k} = \cos(\eta^{j\ell}_k)\hat{I} -i \sin(\eta^{j\ell}_k)\hat{P}_k.
 \end{equation}
As can be shown from this form, the analytic derivative of the expectation value $f(\bm{\eta}) \equiv  \bra{\Psi_0}\hat{U}^\dagger(\bm{\eta})\hat{H}\hat{U}(\bm{\eta})\ket{\Psi_0} $ with respect to a component $\eta_k^{j\ell}$ can be reduced to following parameter shift rule~\cite{mitarai2018quantum,harrow2019low,sweke2019stochastic}: 
 \begin{equation}\label{eq:parshift_grad}
    \tfrac{\partial}{\partial \eta^{j\ell}_k} f(\bm{\eta}) = f(\bm{\eta}+\tfrac{\pi}{4}\bm{\Delta}_{k}^{j\ell})-f(\bm{\eta}-\tfrac{\pi}{4}\bm{\Delta}_{k}^{j\ell})
 \end{equation}
 where $\bm{\Delta}_{k}^{j\ell}$ is a vector representing unit-norm perturbation of the variable $\eta^{j\ell}_k$ in the positive direction. We thus see that this shift in parameters can generally be much larger than that of the numerical differentiation parameter shifts as in equation \eqref{eq:central_diff}. In some cases this is useful as one does not have to resolve as fine-grained a difference in the cost function as an infinitesimal shift, hence requiring less runs to achieve a sufficiently precise estimate of the value of the gradient. 
 
 Note that in order to compute through the chain rule in \eqref{eq:chain_rule} for a parametric unitary as in \eqref{eq:Pauli_decomp}, we need to evaluate the expectation function $2K_\ell$ times to obtain the gradient of the parameter $\theta_j^\ell$. Thus, in some cases where each parameter generates an exponential of many terms, although the gradient estimate is of higher precision, obtaining an analytic gradient can be too costly in terms of required queries to the objective function. To remedy this additional overhead, Harrow et al. \cite{harrow2019low} proposed to stochastically select terms according to a distribution weighted by the coefficients of each term in the generator, and to perform gradient descent from these stochastic estimates of the gradient. Let us review this stochastic gradient estimation technique as it is implemented in TFQ.

\subsubsection{Stochastic Parameter Shift Gradient Estimation}\label{sec:stoc_ge}

Consider the full analytic gradient from \eqref{eq:parshift_grad} and \eqref{eq:chain_rule}, if we have $M_{\ell}$ parameters and $L$ layers, there are $\sum_{\ell=1}^L M_{\ell}$ terms of the following form to estimate:
\begin{equation}\label{eq:expansion1}
    \frac{\partial f}{\partial \theta^{\ell}_{j}}  = \sum_{k=1}^{K_{j\ell}} \beta_k^{j\ell} \frac{\partial f}{\partial \eta_k} = \sum_{k=1}^{K_{j\ell}}   \Big[\sum_{\pm}\pm\beta_k^{j\ell} f(\bm{\eta}\pm\tfrac{\pi}{4}\bm{\Delta}_{k}^{j\ell})\Big] .
\end{equation}

 These terms come from the components of the gradient vector itself which has the dimension equal to that of the total number of free parameters in the QNN, $\text{dim}(\bm{\theta})$. For each of these components, for the $j^\text{th}$ component of the $\ell^\text{th}$ layer, there $2K_{j\ell}$ parameter-shifted expectation values to evaluate, thus in total there are $\sum_{\ell=1}^L 2K_{j\ell}M_{\ell}$ parameterized expectation values of the cost Hamiltonian to evaluate.
 
 For practical implementation of this estimation procedure, we must expand this sum further. Recall that, as the cost Hamiltonian generally will have many terms, for each quantum expectation estimation of the cost function for some value of the parameters, we have
\begin{equation}\label{eq:expansion2} f(\bm{\theta}) =\braket{\hat{H}}_{\bm{\theta}} = \sum_{m=1}^N \alpha_m \braket{\hat{h}_m}_{\bm{\theta}} = \sum_{m=1}^N\sum_{q=1}^{J_{m}}  \alpha_m\gamma^{m}_{q} \braket{\hat{P}_{qm}}_{\bm{\theta}},\end{equation}
 which has $\sum_{j=1}^NJ_{j}$ terms. Thus, if we consider that all the terms in \eqref{eq:expansion1} are of the form of \eqref{eq:expansion2}, we see that we have a total number of \(\sum_{m=1}^N\sum_{\ell=1}^L 2J_{m}K_{j\ell}M_{\ell}\) expectation values to estimate the gradient. Note that one of these sums comes from the total number of appearances of parameters in front of Paulis in the generators of the parameterized quantum circuit, the second sum comes from the various terms in the cost Hamiltonian in the Pauli expansion.
 
 As the cost of accurately estimating all these terms one by one and subsequently linearly combining the values such as to yield an estimate of the total gradient may be prohibitively expensive in terms of numbers of runs, instead, one can \textit{stochastically} estimate this sum, by randomly picking terms according to their weighting \cite{harrow2019low,sweke2019stochastic}. 
 
 One can sample a distribution over the appearances of a parameter in the QNN, 
 $k\sim\text{Pr}(k|j,\ell) = |\beta_k^{j\ell}|/(\sum_{o=1}^{K_{j\ell}} |\beta_o^{j\ell}|)$, one then estimates the two parameter-shifted terms corresponding to this index in \eqref{eq:expansion1} and averages over samples. We consider this case to be \textit{simply stochastic gradient estimation} for the gradient component corresponding to the parameter  $\theta^{\ell}_{j}$. One can go even further in this spirit, for each of these sampled expectation values, by also sampling terms from \eqref{eq:expansion2} according to a similar distribution determined by the magnitude of the Pauli expansion coefficients. Sampling the indices $ \{q,m\} \sim\text{Pr}(q,m) = |\alpha_m\gamma^{m}_{q}|/( \sum_{d=1}^N\sum_{r=1}^{J_{d}}  |\alpha_d\gamma^{d}_{r}|)$ and estimating the expectation $\braket{\hat{P}_{q_m}}_{\bm{\theta}}$ for the appropriate parameter-shifted values sampled from the terms of \eqref{eq:expansion1} according to the procedure outlined above. This is considered \textit{doubly stochastic gradient estimation}. In principle, one could go one step further, and per iteration of gradient descent, randomly sample indices representing subsets of \textit{parameters} for which we will estimate the gradient component, and set the non-sampled indices corresponding gradient components to 0 for the given iteration. The distribution we sample in this case is given by $\theta^{\ell}_{j} \sim \text{Pr}(j,\ell) =  \sum_{k=1}^{K_{j\ell}}|\beta_k^{j\ell}|/(\sum_{u=1}^L
\sum_{i=1}^{M_u}\sum_{o=1}^{K_{iu}} |\beta_o^{iu}|)$. This is, in a sense, akin to the SPSA algorithm \cite{bhatnagarstochastic}, in the sense that it is a gradient-based method with a stochastic mask. The above component sampling, combined with doubly stochastic gradient descent, yields what we consider to be \textit{triply stochastic gradient descent}. This is equivalent to simultaneously sampling $\{j,\ell,k,q,m\} \sim \text{Pr}(j,\ell,k,q,m) = \text{Pr}(k|j,\ell)\text{Pr}(j,\ell)\text{Pr}(q,m)$ using the probabilities outlined in the paragraph above, where $j$ and $\ell$ index the parameter and layer, $k$ is the index from the sum in equation \eqref{eq:expansion1}, and $q$ and $m$ are the indices of the sum in equation \eqref{eq:expansion2}.

In TFQ, all three of the stochastic averaging methods above can be turned on or off independently for stochastic parameter-shift gradients. See the details in the \Colorbox{bkgd}{\lstinline{Differentiator}} module of TFQ on GitHub.

\subsubsection{Adjoint Gradient Backpropagation in Simulations}\label{sec:adjoint}

For experiments with tractably classically simulatable system sizes, the derivatives can be computed entirely in simulation, using an analogue of backpropagation called \textit{Adjoint Differentiation} \cite{pontryagin1987mathematical,luo2020yao}. This is a high-performance way to obtain gradients of deep circuits with many parameters. Although it is not possible to perform this technique in quantum hardware, let us outline here how it is implemented in TFQ for numerical simulator backends such as qsim.

Suppose we are given a parameterized quantum circuit in the form of \eqref{eq:full_par_circ}
\[
   \hat{U}(\bm{\theta}) = \prod_{\ell=1}^L\hat{V}^{\ell}\hat{U}^{\ell}(\bm{\theta}^{\ell}) = \prod_{\ell=1}^L\hat{W}_{\bm{\theta}^\ell}^\ell  = \hat{W}_{\bm{\theta}^n}^n \hat{W}_{\bm{\theta}^{n-1}}^{n-1} \cdots \hat{W}_{\bm{\theta}^1}^1 ,
\]where for compactness of notation, we denoted the parameterized layers in a more succinct form, 
\[
\hat{W}_{\bm{\theta}^\ell}^\ell =\hat{V}^{\ell}\hat{U}^{\ell}(\bm{\theta}^{\ell}).
\]

\begin{figure*}[ht]
    \centering
    \includegraphics[width=0.75\textwidth]{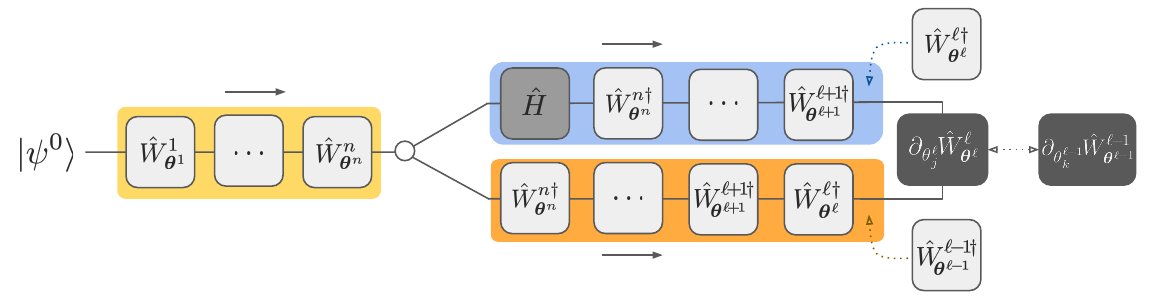}
\caption{Tensor network contraction diagram \cite{biamonte2017tensor,roberts2019tensornetwork} depicting the adjoint backpropagation process for the computation of the gradient of an expectation value at the output of multilayer QNN with respect to a parameter in the $\ell^\text{th}$ layer. Yellow channel depicts the forward pass, which is then duplicated into two copies of the final state. The blue and orange channels depict the backwards pass, performed via layer-wise recursive uncomputation. Both the orange and blue backwards pass outputs are then contracted with the gradient of the $\ell^\text{th}$ layer's unitary. The arrows with dotted lines indicate how to modify the contraction to compute a gradient with respect to a parameter in the $(\ell-1)^\text{th}$ layer.}
    \label{fig:adjoint}
\end{figure*}

Let us label the initial state $\ket{\psi^0}$ final quantum state $\ket{\psi^n_{\bm{\theta}^n}}$, with
\[
\ket{\psi^\ell_{\bm{\theta}^\ell}} = \hat{W}_{\bm{\theta}^\ell}^\ell\ket{\psi^{\ell-1}_{\bm{\theta}^{\ell-1}}} 
\]
being the recursive definition of the Schr\"odinger-evolved state vector at layer $\ell$.
The derivatives of the state with respect to $\bm{\theta}^\ell$, the $j^\text{th}$ parameter of the $\ell^\text{th}$ layer, is given by
\[
\partial_{\theta_j^\ell} \ket{\psi^n_{\bm{\theta}^n}}\! =\! \hat{W}_{\bm{\theta}^n}^n  \cdots \hat{W}_{\bm{\theta}^{\ell+1}}^{\ell+1} \partial_{\theta_j^\ell}  \hat{W}_{\bm{\theta}^{\ell}}^{\ell} \hat{W}_{\bm{\theta}^{\ell-1}}^{\ell-1}\cdots \hat{W}_{\bm{\theta}^{1}}^{1} \ket{\psi^0}.
\]
This gradient of the $\ell^\text{th}$ layer parameters, assuming the QNN is structured as in equations \eqref{eq:full_par_circ}, \eqref{eq:layer_blocks}, and \eqref{eq:Pauli_decomp} is given analytically by
\[\partial_{\theta_j^\ell}  \hat{W}_{\bm{\theta}^{\ell}}^{\ell} = \hat{V}^{\ell} (-i\hat{g}_j^{\ell}) \hat{U}^{\ell}(\bm{\theta}^{\ell})\]
which is an operator that can be computed numerically and inserted in the circuit when dealing with a classical simulation of the quantum circuit.

A key trick for adjoint backpropagation is leveraging the fact that we can reverse unitary layers, so that we do not have to store the history of states; we can access the state at any inner layer by uncomputing the later layers:
\[
\hat{W}_{\bm{\theta}^{\ell-1}}^{\ell-1} \cdots \hat{W}_{\bm{\theta}^{1}}^{1} \ket{\psi_0} = \hat{W}_{\bm{\theta}^{\ell}}^{\dagger \, \ell} \hat{W}_{\bm{\theta}^{\ell+1}}^{\dagger\, \ell+1} \cdots \hat{W}_{\bm{\theta}^{n}}^{\dagger \, n} \ket{\psi^n_{\bm{\theta}^n}}.
\]
We can leverage this trick in order to evaluate gradients of the quantum state with respect to parameters of intermediate layers of the QNN,

\[
\partial_{\theta_j^\ell} \ket{\psi^n_{\bm{\theta}^n}}\! =\! \hat{W}_{\bm{\theta}^n}^n  \cdots \hat{W}_{\bm{\theta}^{\ell+1}}^{\ell+1} \partial_{\theta_j^\ell}  \hat{W}_{\bm{\theta}^{\ell}}^{\ell}  \hat{W}_{\bm{\theta}^{\ell}}^{\dagger \, \ell}  \cdots \hat{W}_{\bm{\theta}^{n}}^{\dagger \, n} \ket{\psi^n_{\bm{\theta}^n}},
\]
this allows us to compute gradients layer-wise starting from the final layer via a backwards pass, following of course a forward pass to compute the final state at the output layer.

In most contexts, we typically want to take gradients of expectation values at the output of several QNN layers. Given a Hermitian observable $\hat{H}$, the expectation value of this observable with respect to our final layer's parameterized state is
\[
\braket{\hat{H}}_{\bm{\theta}^n} = \bra{\psi^n_{\bm{\theta}^n}}\hat{H}\ket{\psi^n_{\bm{\theta}^n}},
\]
and the derivative of this expectation value is given by

\begin{align*}
\partial_{\theta_j^\ell} \braket{\hat{H}}_{\bm{\theta}^n}  &= (\partial_{\theta_j^\ell}\bra{\psi^n_{\bm{\theta}^n}})\hat{H}\ket{\psi^n_{\bm{\theta}^n}} + \bra{\psi^n_{\bm{\theta}^n}}\hat{H}(\partial_{\theta_j^\ell}\ket{\psi^n_{\bm{\theta}^n}}) \\
& = 2 \Re\left[\bra{\psi^n_{\bm{\theta}^n}}\hat{H}(\partial_{\theta_j^\ell}\ket{\psi^n_{\bm{\theta}^n}}) \right],
\end{align*}
where we employed the fact that the observable is Hermitian.  Thus the computational task of gradient evaluation reduces to the evaluation of the modified expectation $\bra{\psi^n_{\bm{\theta}^n}}\hat{H}(\partial_{\theta_j^\ell}\ket{\psi^n_{\bm{\theta}^n}})$ for each parameter.

In order to compute these gradients, we can proceed in a recursive fashion during the backwards pass, starting from the final layer's output. Denoting this recursion using nested parentheses, the gradient evaluation is given by
\begin{align*}\partial_{\theta_j^\ell} \braket{\hat{H}}_{\bm{\theta}^n} &= \left(\cdots\left(\left(\bra{\psi^n_{\bm{\theta}^n}}\hat{H}\right)\hat{W}_{\bm{\theta}^n}^n \right) \cdots \hat{W}_{\bm{\theta}^{\ell+1}}^{\ell+1}\right) \\&\qquad \cdot \left( \partial_{\theta_j^\ell}  \hat{W}_{\bm{\theta}^{\ell}}^{\ell} \right)\cdot\left( \hat{W}_{\bm{\theta}^{\ell}}^{\dagger \, \ell} \cdots \left(\hat{W}_{\bm{\theta}^{n}}^{\dagger \, n} (\ket{\psi^n_{\bm{\theta}^n}})\right)\ldots\right)\\
&= \bra{\tilde{\psi}^{\ell}_{\bm{\theta}^{\ell}}} \left( \partial_{\theta_j^\ell}  \hat{W}_{\bm{\theta}^{\ell}}^{\ell} \right)\ket{\psi^{\ell-1}_{\bm{\theta}^{\ell-1}}}
\end{align*}

This consists of recursively computing the backwards propagated state, 
\[\ket{\psi^{\ell-1}_{\bm{\theta}^{\ell-1}}} = \hat{W}_{\bm{\theta}^\ell}^{\ell\dagger}\ket{\psi^\ell_{\bm{\theta}^\ell}}  \]

and contracting it with both the gradient of the $\ell^{\text{th}}$ layer's unitary \( \partial_{\theta_j^\ell}  \hat{W}_{\bm{\theta}^{\ell}}^{\ell}\) and the backwards propagated contraction of the final state with the observable:
\[\ket{\tilde{\psi}^{\ell-1}_{\bm{\theta}^{\ell-1}}} \equiv \hat{W}_{\bm{\theta}^\ell}^{\ell\dagger}\ket{\tilde{\psi}^\ell_{\bm{\theta}^\ell}},\quad  \ket{\tilde{\psi}^n_{\bm{\theta}^n}} \equiv  \hat{H}\ket{\psi^n_{\bm{\theta}^n}}.\] By only storing \(\ket{\psi^{\ell-1}_{\bm{\theta}^{\ell-1}}}\) and \(\ket{\tilde{\psi}^{\ell}_{\bm{\theta}^{\ell}}}\) in memory for gradient evaluations of the $\ell^{\text{th}}$ layer during the backwards pass, we saved the need for far more forward passes while only using twice the memory. See Figure \ref{fig:adjoint} for a depiction of the adjoint-backpropagation-based gradient computation described above.  This is to be contrasted with, for example, finite-difference gradients, which would require $2M$ forward passes for $M$ parameters. Thus, adjoint differentiation is a useful method for rapid prototyping of QNN's using classical simulators of quantum circuits such as qsim in TFQ.

\subsection{Hybrid Quantum-Classical Computational Graphs}\label{sec:hqcm-th}

Now that we have reviewed various ways of obtaining gradients of expectation values, let us consider how to go beyond basic variational quantum algorithms and consider fully hybrid quantum-classical neural networks. As we will see, our general framework of gradients of cost Hamiltonians will carry over.

\begin{figure}
    \centering
    \includegraphics[width=\columnwidth]{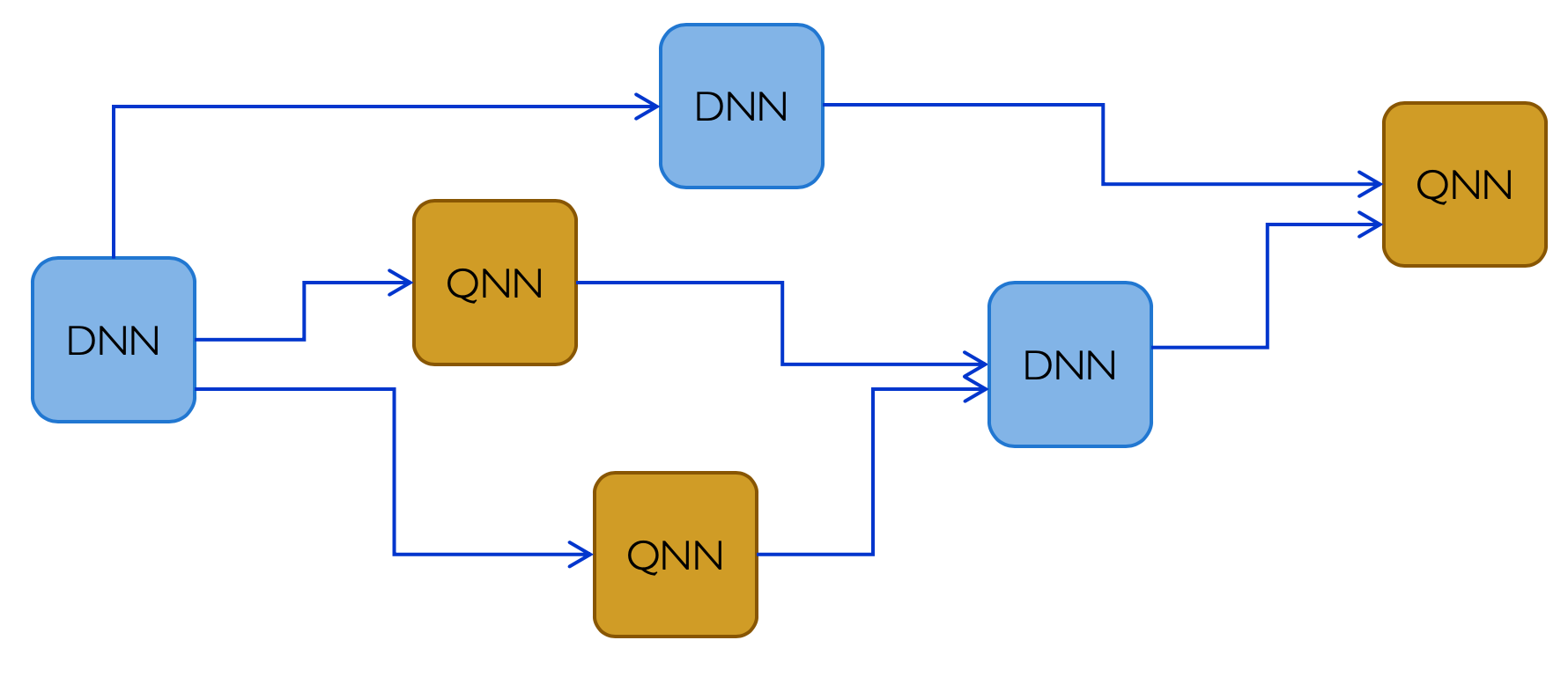}
\caption{High-level depiction of a quantum-classical neural network. Blue blocks represent Deep Neural Network (DNN) function blocks and orange boxes represent Quantum Neural Network (QNN) function blocks. Arrows represent the flow of information during the feedforward (inference) phase. For an example of the interface between quantum and classical neural network blocks, see Fig.~\ref{fig:backprop}. }
    \label{fig:hqc-graph}
\end{figure}

\subsubsection{Hybrid Quantum-Classical Neural Networks}
Now, we are ready to formally introduce the notion of Hybrid Quantum-Classical Neural Networks (HQCNN's). HQCNN's are meta-networks comprised of quantum and classical neural network-based function blocks composed with one another in the topology of a directed graph. We can consider this a rendition of a hybrid quantum-classical computational graph where the inner workings (variables, component functions) of various functions are abstracted into boxes (see Fig.~\ref{fig:hqc-graph} for a depiction of such a graph). The edges then simply represent the flow of classical information through the meta-network of quantum and classical functions. The key will be to construct parameterized (differentiable) functions $f_{\bm{\theta}}:\mathbb{R}^M\rightarrow \mathbb{R}^N$ from expectation values of parameterized quantum circuits, then creating a meta-graph of quantum and classical computational nodes from these blocks. Let us first describe how to create these functions from expectation values of QNN's. 

As we saw in equations \eqref{eq:cost} and \eqref{eq:cost_lincomb}, we get a differentiable cost function $f: \mathbb{R}^M \mapsto \mathbb{R}$ from taking the expectation value of a single Hamiltonian at the end of the parameterized circuit,  $f(\bm{\theta}) =\braket{\hat{H}}_{\bm{\theta}}$. As we saw in equations \eqref{eq:cost_lincomb} and \eqref{eq:qee_lincomb}, to compute this expectation value, as the readout Hamiltonian is often decomposed into a linear combination of operators $\hat{H} = \bm{\alpha}\cdot \bm{\hat{h}}$ (see \eqref{eq:cost_lincomb}), then the function is itself a linear combination of expectation values of multiple terms (see \eqref{eq:qee_lincomb}), 
$\braket{\hat{H}}_{\bm{\theta}} = \bm{\alpha}\cdot \bm{h}_{\bm{\theta}}
$
where $\bm{h}_{\bm{\theta}} \equiv \braket{\bm{\hat{h}}}_{\bm{\theta}} \in \mathbb{R}^N$ is a vector of expectation values. Thus, before the scalar value of the cost function is evaluated, QNN's naturally are evaluated as a vector of expectation values, $\bm{h}_{\bm{\theta}}$.

Hence, if we would like the QNN to become more like a classical neural network block, i.e. mapping vectors to vectors $ \bm{f}:\mathbb{R}^M \rightarrow \mathbb{R}^N$, we can obtain a vector-valued differentiable function from the QNN by considering it as a function of the parameters which outputs a vector of expectation values of different operators, 
\begin{equation}
 \bm{f}:\bm{\theta}\mapsto   \bm{h}_{\bm{\theta}} 
\end{equation}
where
\begin{equation}\label{eq:exp_vec}
      (\bm{h}_{\bm{\theta}})_k = \braket{\hat{h}_k}_{\bm{\theta}} \equiv  \bra{\Psi_0}\hat{U}^\dagger(\bm{\theta})\hat{h}_k\hat{U}(\bm{\theta})\ket{\Psi_0}.
\end{equation}
We represent such a QNN-based function in Fig.~\ref{fig:backprop}. Note that, in general, each of these $\hat{h}_k$'s could be comprised of multiple terms themselves,
\begin{equation}
    \hat{h}_k = \sum_{t=1}^{N_k} \gamma_t \hat{m}_t
\end{equation}
hence, one can perform Quantum Expectation Estimation to estimate the expectation of each term as $\braket{\hat{h}_k}= \sum_t \gamma_t\braket{\hat{m}_t}$. 

Note that, in some cases, instead of the expectation values of the set of operators $\{\hat{h}_k\}_{k=1}^M$, one may instead want to relay the histogram of measurement results obtained from multiple measurements of the eigenvalues of each of these observables. This case can also be phrased as a vector of expectation values, as we will now show. First, note that the histogram of the measurement results of some $\hat{h}_k$ with eigendecomposition $\hat{h}_k = \sum_{j=1}^{r_k}\lambda_{jk} \ket{\lambda_{jk}}\bra{\lambda_{jk}}$ can be considered as a vector of expectation values where the observables are the eigenstate projectors $\ket{\lambda_{jk}}\!\bra{\lambda_{jk}}$. Instead of obtaining a single real number from the expectation value of $\hat{h}_k$, we can obtain a vector $ \bm{h}_k\in \mathbb{R}^{r_k}$, where $r_k = \text{rank}(\hat{h}_k)$ and the components are given by 
\(
    ( \bm{h}_k)_j \equiv \braket{\ket{\lambda_{jk}}\!\bra{\lambda_{jk}}}_{\bm{\theta}} 
\). We are then effectively considering the categorical (empirical) distribution as our vector.

Now, if we consider measuring the eigenvalues of multiple observables $\{\hat{h}_k\}_{k=1}^M$ and collecting the measurement result histograms, we get a 2-dimensional tensor $(\bm{h}_{\bm{\theta}})_{jk} = \braket{\ket{\lambda_{jk}}\!\bra{\lambda_{jk}}}_{\bm{\theta}} $. Without loss of generality, we can flatten this array into a vector of dimension $\mathbb{R}^R$ where $R = \sum_{k=1}^M r_k$. Thus, considering vectors of expectation values is a relatively general way of representing the output of a quantum neural network. In the limit where the set of observables considered forms an informationally-complete set of observables \cite{Gross_2010}, then the array of measurement outcomes would fully characterize the wavefunction, albeit at an overhead exponential in the number of qubits.

We should mention that in some cases, instead of expectation values or histograms, single samples from the output of a measurement can be used for direct feedback-control on the quantum system, e.g. in quantum error correction \cite{Gottesman1997}. At least in the current implementation of TFQuantum, since quantum circuits are built in Cirq and this feature is not supported in the latter, such scenarios are currently out-of-scope. Mathematically, in such a scenario, one could then consider the QNN and measurement as map from quantum circuit parameters $\bm{\theta}$ to the conditional random variable $\Lambda_{\bm{\theta}}$ valued over $\mathbb{R}^{N_k}$ corresponding to the measured eigenvalues $\lambda_{jk}$ with a probability density $\text{Pr}[(\Lambda_{\bm{\theta}})_{k}\equiv \lambda_{jk}]= p(\lambda_{jk}|\bm{\theta})$ which corresponds to the measurement statistics distribution induced by the Born rule, $ p(\lambda_{jk}|\bm{\theta}) = \braket{\ket{\lambda_{jk}}\!\bra{\lambda_{jk}}}_{\bm{\theta}} $. This QNN and single measurement map from the parameters to the conditional random variable $\bm{f}:\bm{\theta}\mapsto \Lambda_{\bm{\theta}}$ can be considered as a classical stochastic map (classical conditional probability distribution over output variables given the parameters). In the case where only expectation values are used, this stochastic map reduces to a deterministic node through which we may backpropagate gradients, as we will see in the next subsection. In the case where this map is used dynamically per-sample, this remains a stochastic map, and though there exists some algorithms for backpropagation through stochastic nodes \cite{schulman2015gradient}, these are not currently supported natively in TFQ.

\subsection{Autodifferentiation through hybrid quantum-classical backpropagation}\label{sec:qnn_grad}

As described above, hybrid quantum-classical neural network blocks take as input a set of real parameters $\bm{\theta}\in \mathbb{R}^M$, apply a circuit $\hat{U}(\bm{\theta})$ and take a set of expectation values of various observables \[(\bm{h}_{\bm{\theta}})_k = \braket{\hat{h}_k}_{\bm{\theta}}.\] The result of this parameter-to-expected value map is a function $ \bm{f}:\mathbb{R}^M \rightarrow \mathbb{R}^N$ which maps parameters to a real-valued vector, \[ \bm{f}:\bm{\theta}\mapsto   \bm{h}_{\bm{\theta}} .\] This function can then be composed with other parameterized function blocks comprised of either quantum or classical neural network blocks, as depicted in Fig.~\ref{fig:hqc-graph}.

\begin{figure}
    \centering
    \includegraphics[width=\columnwidth]{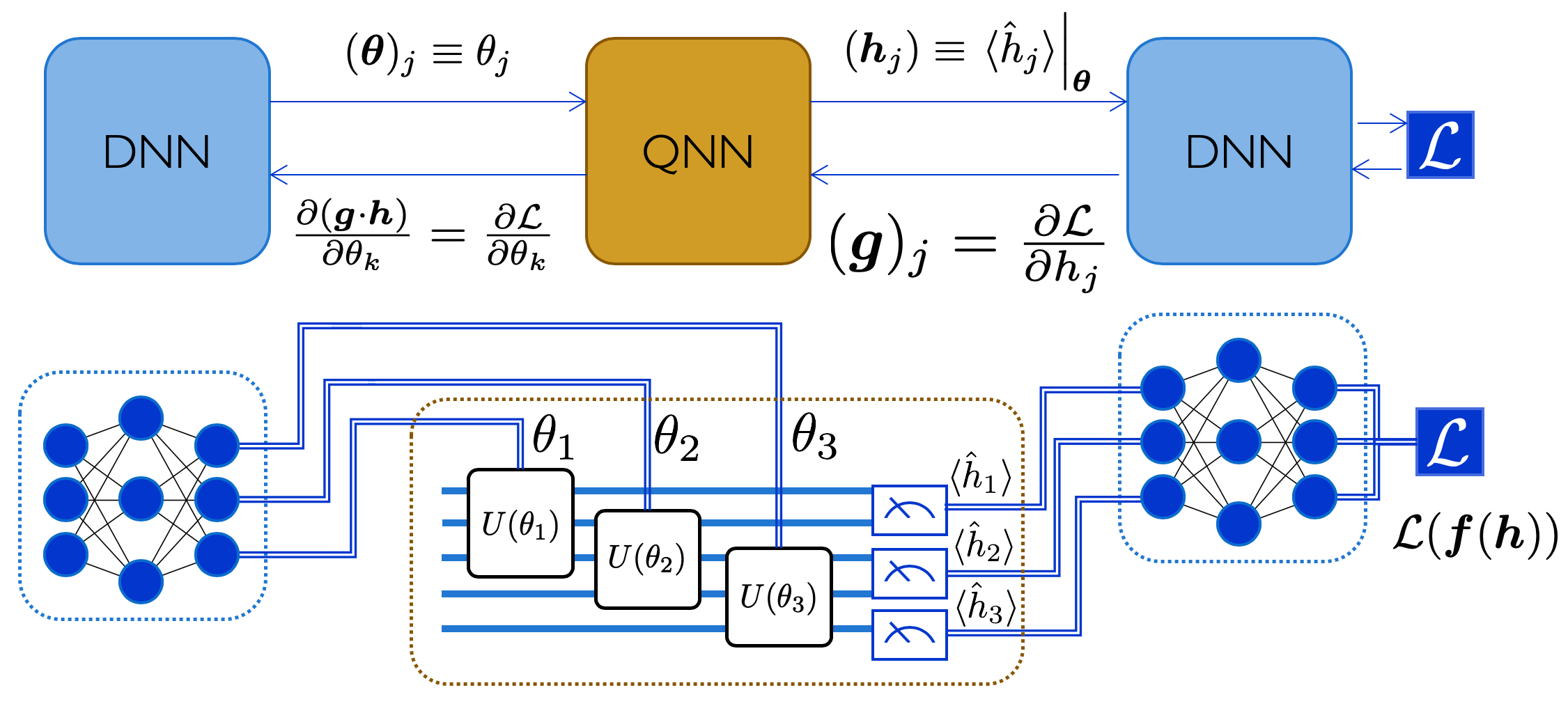}
    \caption{Example of inference and hybrid backpropagation at the interface of a quantum and classical part of a hybrid computational graph. Here we have classical deep neural networks (DNN) both preceding and postceding the quantum neural network (QNN). In this example, the preceding DNN outputs a set of parameters $\bm{\theta}$ which are used as then used by the QNN as parameters for inference. The QNN outputs a vector of expectation values (estimated through several runs) whose components are $(\bm{h}_{\bm{\theta}})_k = \braket{\hat{h}_k}_{\bm{\theta}}$. This vector is then fed as input to another (post-ceding) DNN, and the loss function $\mathcal{L}$ is computed from this output. For backpropagation through this hybrid graph, one first backpropagates the gradient of the loss through the post-ceding DNN to obtain $g_k = \partial\mathcal{L}/\partial{h_k}$. Then, one takes the gradient of the following functional of the output of the QNN:  $f_{\bm{\theta}} = \bm{g}\cdot\bm{h}_{\bm{\theta}} = \sum_k g_k h_{\bm{\theta}\!,k} = \sum_k g_k \braket{\hat{h}_k}_{\bm{\theta}}$ with respect to the QNN parameters $\bm{\theta}$ (which can be achieved with any of the methods for taking gradients of QNN's described in previous subsections of this section).
    This completes the backpropagation of gradients of the loss function through the QNN, the preceding DNN can use the now computed $\partial{\mathcal{L}}/\partial{\bm{\theta}}$ to further backpropagate gradients to preceding nodes of the hybrid computational graph. }
    \label{fig:backprop}
\end{figure}

To be able to backpropagate gradients through general meta-networks of quantum and classical neural network blocks, we simply have to figure out how to backpropagate gradients through a quantum parameterized block function when it is composed with other parameterized block functions. Due to the partial ordering of the quantum-classical computational graph, we can focus on how to backpropagate gradients through a QNN in the scenario where we consider a simplified quantum-classical network where we combine all function blocks that precede and postcede the QNN block into monolithic function blocks. Namely, we can consider a scenario where we have $\bm{f}_\text{pre}:\bm{x}_{\text{in}}\mapsto \bm{\theta}$ ($\bm{f}_\text{pre}:\mathbb{R}^{\text{in}}\rightarrow \mathbb{R}^M$) as the block preceding the QNN, the QNN block as $\bm{f}_{\text{qnn}}:\bm{\theta}\mapsto   \bm{h}_{\bm{\theta}}$, ($\bm{f}_{\text{qnn}}:\mathbb{R}^M \rightarrow \mathbb{R}^N$), the post-QNN block as $\bm{f}_\text{post}:\bm{h}_{\bm{\theta}}\mapsto\bm{y}_{\text{out}}$ ($\bm{f}_\text{post}:\mathbb{R}^M \rightarrow \mathbb{R}^N_{\text{out}}$) and finally the loss function for training the entire network being computed from this output $\mathcal{L}:\mathbb{R}^N_{\text{out}}\rightarrow \mathbb{R}$. The composition of functions from the input data to the output loss is then the sequentially composited function $(\mathcal{L}\circ \bm{f}_\text{post} \circ \bm{f}_{\text{qnn}} \circ \bm{f}_\text{pre})$. This scenario is depicted in Fig.~\ref{fig:backprop}.

Now, let us describe the process of backpropagation through this composition of functions. As is standard in backpropagation of gradients through feedforward networks, we begin with the loss function evaluated at the output units and work our way back through the several layers of functional composition of the network to get the gradients. The first step is to obtain the gradient of the loss function $\partial\mathcal{L}/\partial \bm{y}_{\text{out}}$ and to use classical (regular) backpropagation of gradients to obtain the gradient of the loss with respect to the output of the QNN, i.e. we get $\partial(\mathcal{L}\circ \bm{f}_\text{post})/\partial \bm{h}$ via the usual use of the chain rule for backpropagation, i.e., the contraction of the Jacobian with the gradient of the subsequent layer, $\partial(\mathcal{L}\circ \bm{f}_\text{post})/\partial \bm{h} = \tfrac{\partial \mathcal{L}}{\partial{\bm{y} }}\cdot\tfrac{ \partial{\bm{f}_\text{post} }}{\partial{\bm{h} }}$. 

Now, let us label the evaluated gradient of the loss function with respect to the QNN's expectation values as 
\begin{equation}\label{eq:grad}
    \bm{g} \equiv \left.\tfrac{\partial(\mathcal{L}\circ \bm{f}_\text{post})}{\partial \bm{h}}\right|_{\bm{h} = \bm{h}_{\bm{\theta}}}.
\end{equation}
We can now consider this value as effectively a constant.
Let us then define an \textit{effective backpropagated Hamiltonian} with respect to this gradient as
\[\hat{H}_{\bm{g}} \equiv \sum_k g_k\hat{h}_k,\] where $g_k$ are the components of \eqref{eq:grad}. Notice that expectations of this Hamiltonian are given by \[\braket{\hat{H}_{\bm{g}}}_{\bm{\theta}} = \bm{g} \cdot \bm{h}_{\bm{\theta}},\]
and so, the gradients of the expectation value of this Hamiltonian are given by
\[\tfrac{\partial}{\partial{\theta_j}}\braket{\hat{H}_{\bm{g}}}_{\bm{\theta}} = \tfrac{\partial}{\partial{\theta_j}}(\bm{g} \cdot \bm{h}_{\bm{\theta}}) = \sum_k g_k \tfrac{\partial h_{\bm{\theta}\!,k}}{\partial \theta_j}\]
which is exactly the Jacobian of the QNN function $\bm{f}_{\text{qnn}}$ contracted with the backpropagated gradient of previous layers. Explicitly, 
\[\partial(\mathcal{L}\circ \bm{f}_\text{post}\circ \bm{f}_\text{qnn})/\partial \bm{\theta} = \tfrac{\partial}{\partial{\bm{\theta}}}\braket{\hat{H}_{\bm{g}}}_{\bm{\theta}}.\]

Thus, by taking gradients of the expectation value of the backpropagated effective Hamiltonian, we can get the gradients of the loss function with respect to QNN parameters, thereby successfully backpropagating gradients through the QNN. Further backpropagation through the preceding function block $\bm{f}_\text{pre}$ can be done using standard classical backpropagation by using this evaluated QNN gradient. 

Note that for a given value of $\bm{g}$, the effective backpropagated Hamiltonian is simply a fixed Hamiltonian operator, as such, taking gradients of the expectation of a single multi-term operator can be achieved by any choice in a multitude of the methods for taking gradients of QNN's described earlier in this section.

Backpropagation through parameterized quantum circuits is enabled by our \Colorbox{bkgd}{\lstinline{Differentiator}} interface.

We offer both finite difference, regular parameter shift, and stochastic parameter shift gradients, while the general interface allows users to define custom gradient methods.

\section{Basic Quantum Applications}\label{sec:basic_app}

The following examples show how one may use the various features of TFQ to reproduce and extend existing results in second generation quantum machine learning.  Each application has an associated Colab notebook which can be run in-browser to reproduce any results shown.  Here, we use snippets of code from those example notebooks for illustration; please see the example notebooks for full code details.

\subsection{Hybrid Quantum-Classical Convolutional Neural Network Classifier}\label{sec:HQCNN}
To run this example in the browser through Colab, follow the link:
\fancylink{https://github.com/tensorflow/quantum/blob/master/docs/tutorials/qcnn.ipynb}{docs/tutorials/qcnn.ipynb}

\subsubsection{Background}\label{sec:HQCNN_bg}
Supervised classification is a canonical task in classical machine learning. Similarly, it is also one of the most well-studied applications for QNNs~\cite{chen2018universal,havlivcek2019supervised,grant2018hierarchical,Miles_TN,Mohseni14_support}.  As such, it is a natural starting point for our exploration of applications for quantum machine learning.  Discriminative machine learning with hierarchical models can be understood as a form of compression to isolate the information containing the label \cite{tishby2015deep}.  In the case of quantum data, the hidden classical parameter (real scalar in the case of regression, discrete label in the case of classification) can be embedded in a non-local subsystem or subspace of the quantum system. One then has to perform some disentangling quantum transformation to extract information from this non-local subspace.

To choose an architecture for a neural network model, one can draw inspiration from the symmetries in the training data.  For example, in computer vision, one often needs to detect corners and edges regardless of their position in an image; we thus postulate that a neural network to detect these features should be invariant under translations. In classical deep learning, an example of such translationally-invariant neural networks are \textit{convolutional neural networks}.  These networks tie parameters across space, learning a shared set of filters which are applied equally to all portions of the data.

To the best of the authors' knowledge, there is no strong indication that we should expect a quantum advantage for the classification of classical data using QNNs in the near term. For this reason, we focus on classifying quantum data as defined in section \ref{sec:quantum_data}.  There are many kinds of quantum data with translational symmetry.  One example of such quantum data are  cluster states.  These states are important because they are the initial states for measurement-based quantum computation \cite{walther2005, clusterstate2006}.  In this example we will tackle the problem of detecting errors in the preparation of a simple cluster state.  We can think of this as a supervised classification task: our training data will consist of a variety of correctly and incorrectly prepared cluster states, each paired with their label.  This classification task can be generalized to condensed matter physics and beyond, for example to the classification of phases near quantum critical points, where the degree of entanglement is high.

Since our simple cluster states are translationally invariant, we can extend the spatial parameter-tying of convolutional neural networks to quantum neural networks, using recent work by Cong, et al. \cite{Cong_2019}, which introduced a Quantum Convolutional Neural Network (QCNN) architecture. QCNNs are essentially a quantum circuit version of a MERA (Multiscale Entanglement Renormalization Ansatz) network \cite{Vidal_2008}. MERA has been extensively studied in condensed matter physics. It is a hierarchical representation of highly entangled wavefunctions. The intuition is that as we go higher in the network, the wavefunction's entanglement gets renormalized (coarse grained) and simultaneously a compressed representation of the wavefunction is formed. This is akin to the compression effects encountered in deep neural networks \cite{shwartzziv2017opening}. 

Here we extend the QCNN architecture to include classical neural network postprocessing, yielding a Hybrid Quantum Convolutional Neural Network (HQCNN).  We perform several low-depth quantum operations in order to begin to extract hidden parameter information from a wavefunction, then pass the resulting statistical information to a classical neural network for further processing.  Specifically, we will apply one layer of the hierarchy of the QCNN. This allows us to partially disentangle the input state and obtain statistics about values of multi-local observables. In this strategy, we deviate from the original construction of Cong et al.~\cite{Cong_2019}. Indeed, we are more in the spirit of classical convolutional networks, where there are several families of filters, or \textit{feature maps}, applied in a translationally-invariant fashion. Here, we apply a single QCNN layer followed by several feature maps.  The outputs of these feature maps can then be fed into classical convolutional network layers, or in this particular simplified example directly to fully-connected layers.  

\vspace{\baselineskip}
\noindent\textbf{Target problems}:
\begin{enumerate}[noitemsep]
    \item Learn to extract classical information hidden in correlations of a quantum system
    \item Utilize shallow quantum circuits via hybridization with classical neural networks to extract information
\end{enumerate}
\noindent\textbf{ Required TFQ functionalities}:
      \begin{enumerate}[noitemsep]
        \item Hybrid quantum-classical network models
        \item Batch quantum circuit simulator
        \item Quantum expectation based backpropagation
        \item Fast classical gradient-based optimizer
    \end{enumerate}

\subsubsection{Implementations}
As discussed in section \ref{sec:abstract_pipeline}, the first step in the QML pipeline is the preparation of quantum data.  In this example, our quantum dataset is a collection of correctly and incorrectly prepared cluster states on 8 qubits, and the task is to classify theses states.  The dataset preparation proceeds in two stages; in the first stage, we generate a correctly prepared cluster state:
\begin{lstlisting}
def cluster_state_circuit(bits):
    circuit = cirq.Circuit()
    circuit.append(cirq.H.on_each(bits))
    for this_bit, next_bit in zip(
        bits, bits[1:] + [bits[0]]):
        circuit.append(
            cirq.CZ(this_bit, next_bit))
    return circuit
\end{lstlisting}
Errors in cluster state preparation will be simulated by applying $R_x(\theta)$ gates that rotate a qubit about the X-axis of the Bloch sphere by some amount $0 \leq \theta \leq 2\pi$. These excitations will be labeled 1 if the rotation is larger than some threshold, and -1 otherwise.  Since the correctly prepared cluster state is always the same, we can think of it as the initial state in the pipeline and append the excitation circuits corresponding to various error states:
\begin{lstlisting}
cluster_state_bits = cirq.GridQubit.rect(1, 8)
excitation_input = tf.keras.Input(
    shape=(), dtype=tf.dtypes.string)
cluster_state = tfq.layers.AddCircuit()(
    excitation_input, prepend=cluster_state_circuit(cluster_state_bits))
\end{lstlisting}
Note how \Colorbox{bkgd}{\lstinline{excitation_input}} is a standard Keras data ingester.  The datatype of the input is \Colorbox{bkgd}{\lstinline{string}} to account for our circuit serialization mechanics described in section \ref{sec:q_tensors}.

\begin{figure}
    \centering
    \includegraphics[width=0.5\columnwidth]{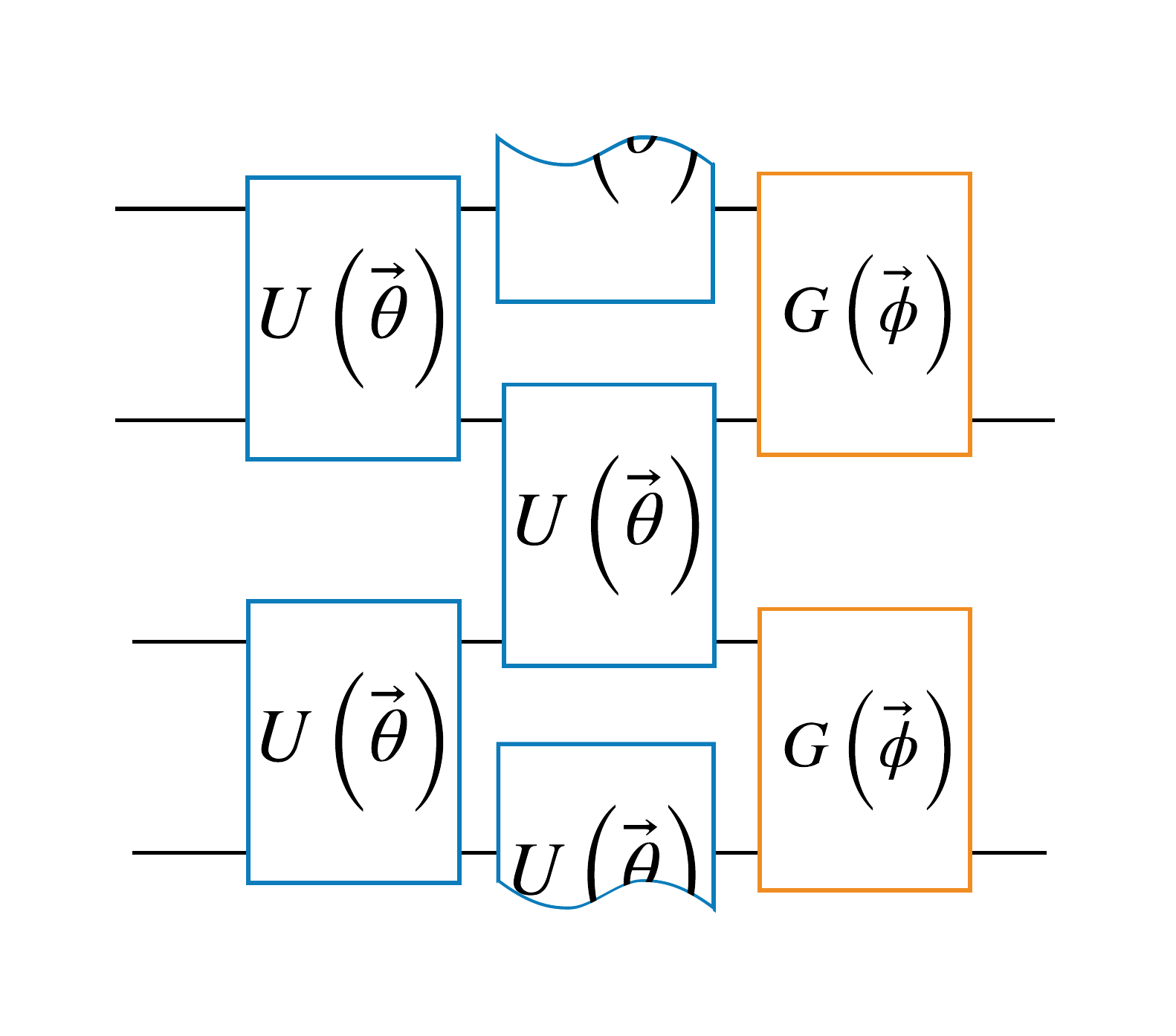}
\caption{The quantum portion of our classifiers, shown on 4 qubits.  The combination of quantum convolution (blue) and quantum pooling (orange) reduce the system size from 4 qubits to 2 qubits.}
    \label{fig:conv_pool}
\end{figure}

Having prepared our dataset, we begin construction of our model.  The quantum portion of all the models we consider in this section will be made of the same operations: \textit{quantum convolution} and \textit{quantum pooling}.  A visualization of these operations on 4 qubits is shown in Fig.~\ref{fig:conv_pool}.  Quantum convolution layers are enacted by applying a 2 qubit unitary $U( \vec{\theta} )$ with a stride of 1.  In analogy with classical convolutional layers, the parameters of the unitaries are tied, so that the same operation is applied to every nearest-neighbor pair of qubits.  Pooling is achieved using a different 2 qubit unitary $G( \vec{\phi})$ designed to disentangle, allowing information to be projected from 2 qubits down to 1.  The code below defines the quantum convolution and quantum pooling operations:

\begin{lstlisting}
def quantum_conv_circuit(bits, syms):
    circuit = cirq.Circuit()
    for a, b in zip(bits[0::2], bits[1::2]):
        circuit += two_q_unitary([a, b], syms)
    for a, b in zip(bits[1::2], bits[2::2] + [bits[0]]):
        circuit += two_q_unitary([a, b], syms)
    return circuit
        
def quantum_pool_circuit(srcs, snks, syms):
    circuit = cirq.Circuit()
    for src, snk in zip(srcs, snks):
        circuit += two_q_pool(src, snk, syms)
    return circuit
\end{lstlisting}
In the code, \Colorbox{bkgd}{\lstinline{two_q_unitary}} constructs a general parameterized two qubit unitary \cite{sousa2006universal}, while \Colorbox{bkgd}{\lstinline{two_q_pool}} represents a CNOT with general one qubit unitaries on the control and target qubits, allowing for variational selection of control and target basis.  

With the quantum portion of our model defined, we move on to the third and fourth stages of the QML pipeline, measurement and classical post-processing.  We consider three classifier variants, each containing a different degree of hybridization with classical networks:
\begin{enumerate}[noitemsep]
    \item Purely quantum CNN
    \item Hybrid CNN in which the outputs of a truncated QCNN are fed into a standard densely connected neural net
    \item Hybrid CNN in which the outputs of multiple truncated QCNNs are fed into a standard densely connected neural net
\end{enumerate}

\begin{figure}
    \centering
    \includegraphics[width=0.9\columnwidth]{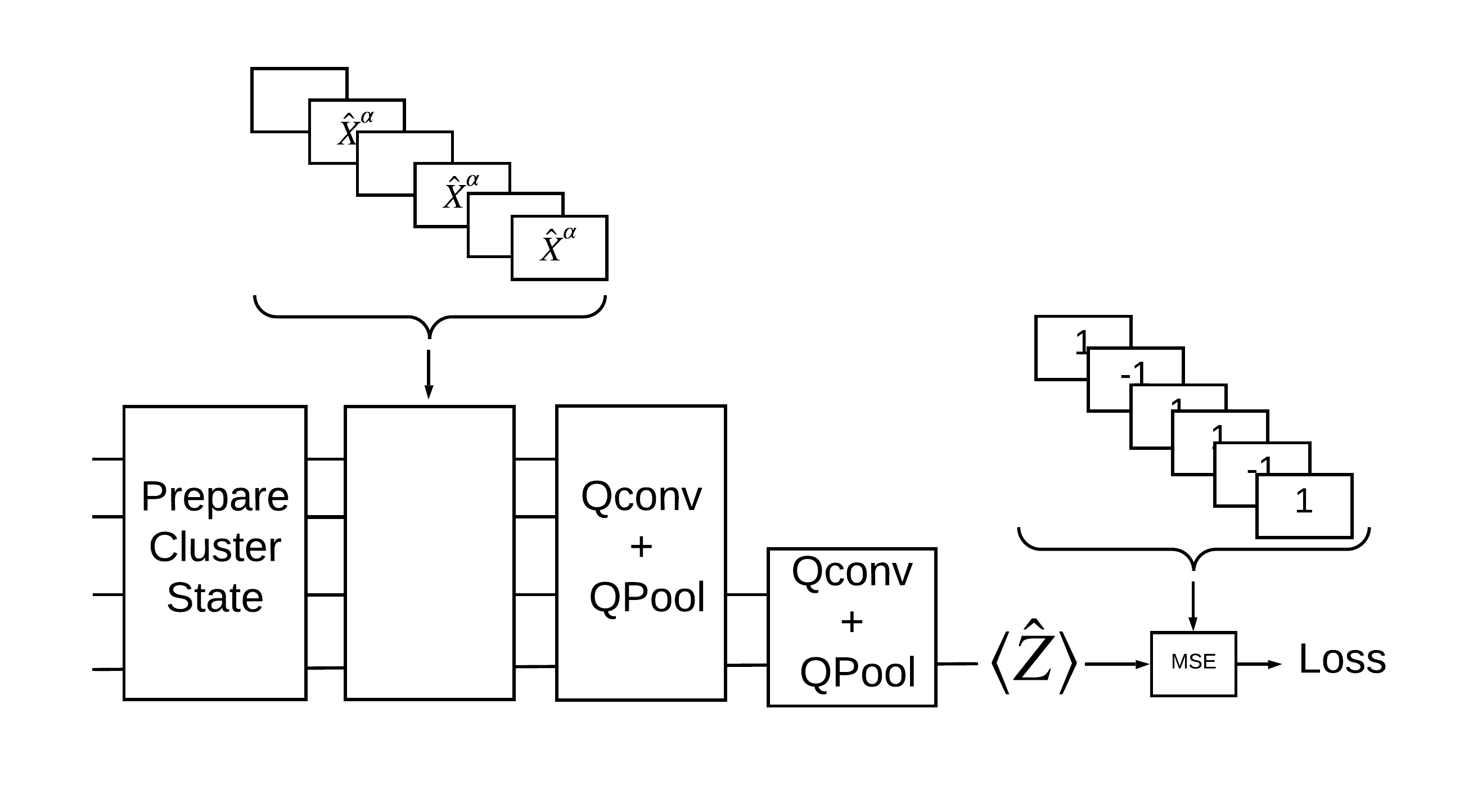}
\caption{Architecture of the purely quantum CNN for detecting excited cluster states.}
    \label{fig:quantum_cnn}
\end{figure}

The first model we construct uses only quantum operations to decorrelate the inputs.  After preparing the cluster state dataset on $N=8$ qubits, we repeatedly apply the quantum convolution and pooling layers until the system size is reduced to 1 qubit.  We average the output of the quantum model by measuring the expectation of Pauli-Z on this final qubit.  Measurement and parameter control are enacted via our \Colorbox{bkgd}{\lstinline{tfq.layers.PQC}} object.  The code for this model is shown below:
\begin{lstlisting}
readout_operators = cirq.Z(
    cluster_state_bits[-1])
quantum_model = tfq.layers.PQC(
    create_model_circuit(cluster_state_bits),
    readout_operators)(cluster_state)
qcnn_model = tf.keras.Model(
    inputs=[excitation_input],
    outputs=[quantum_model])
\end{lstlisting}
In the code, \Colorbox{bkgd}{\lstinline{create_model_circuit}} is a function which applies the successive layers of quantum convolution and quantum pooling.  A simplified version of the resulting model on 4 qubits is shown in Fig.~\ref{fig:quantum_cnn}.

With the model constructed, we turn to training and validation.  These steps can be accomplished using standard Keras tools.  During training, the model output on each quantum datapoint is compared against the label; the cost function used is the mean squared error between the model output and the label, where the mean is taken over each batch from the dataset.  The training and validation code is shown below:
\begin{lstlisting}
qcnn_model.compile(optimizer=tf.keras.Adam,
                   loss=tf.losses.mse)

(train_excitations, train_labels,
    test_excitations, test_labels
) = generate_data(cluster_state_bits)

history = qcnn_model.fit(
    x=train_excitations,
    y=train_labels,
    batch_size=16,
    epochs=25,
    validation_data=(
        test_excitations, test_labels))
\end{lstlisting}
In the code, the \Colorbox{bkgd}{\lstinline{generate_data}} function builds the excitation circuits that are applied to the initial cluster state input, along with the associated labels.  The loss plots for both the training and validation datasets can be generated by running the associated example notebook.

\begin{figure}
    \centering
    \includegraphics[width=\columnwidth]{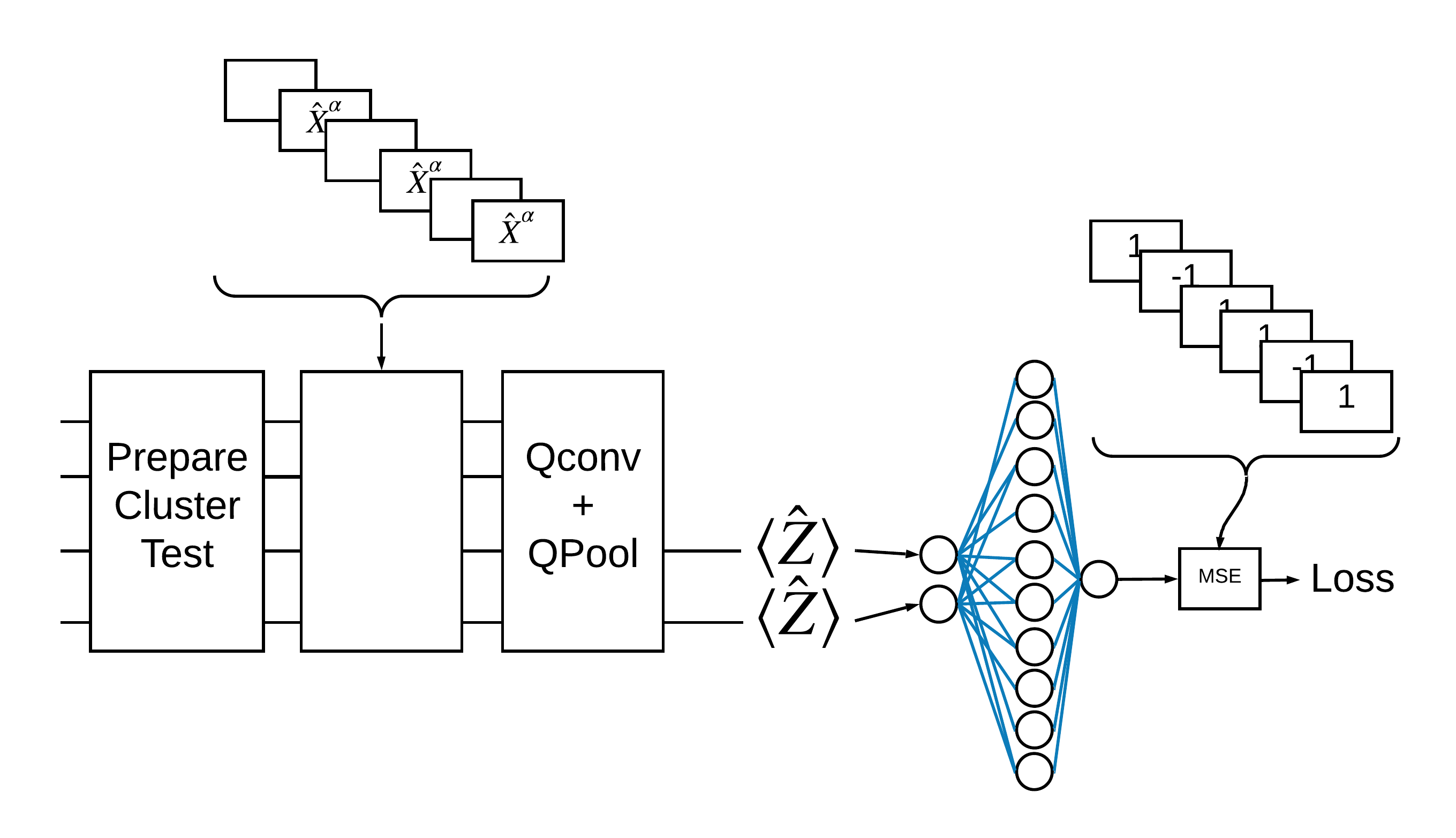}
\caption{A simple hybrid architecture in which the outputs of a truncated QCNN are fed into a classical neural network.}
    \label{fig:hqcnn1}
\end{figure}

We now consider a hybrid classifier. Instead of using quantum layers to pool all the way down to 1 qubit, we can truncate the QCNN and measure a vector of operators on the remaining qubits.  The resulting vector of expectation values is then fed into a classical neural network. This hybrid model is shown schematically in Fig.~\ref{fig:hqcnn1}.

This can be achieved in TFQ with a few simple modifications to the previous model, implemented with the code below:
\begin{lstlisting}
# Build multi-readout quantum layer
readouts = [cirq.Z(bit) for bit in cluster_state_bits[4:]]
quantum_model_dual = tfq.layers.PQC(
    multi_readout_model_circuit(cluster_state_bits),
    readouts)(cluster_state)
# Build classical neural network layers
d1_dual = tf.keras.layers.Dense(8)(quantum_model_dual)
d2_dual = tf.keras.layers.Dense(1)(d1_dual)
hybrid_model = tf.keras.Model(inputs=[excitation_input], outputs=[d2_dual])
\end{lstlisting}
In the code, \Colorbox{bkgd}{\lstinline{multi_readout_model_circuit}} applies just one round of convolution and pooling, reducing the system size from 8 to 4 qubits.  This hybrid model can be trained using the same Keras tools as the purely quantum model.  Accuracy plots can be seen in the example notebook.

The third architecture we will explore creates three independent quantum filters, and combines the outputs from all three with a single classical neural network.  This architecture is shown in Fig.~\ref{fig:hqcnn2}.  This multi-filter architecture can be implemented in TFQ as below:

\begin{figure}
    \centering
    \includegraphics[width=\columnwidth]{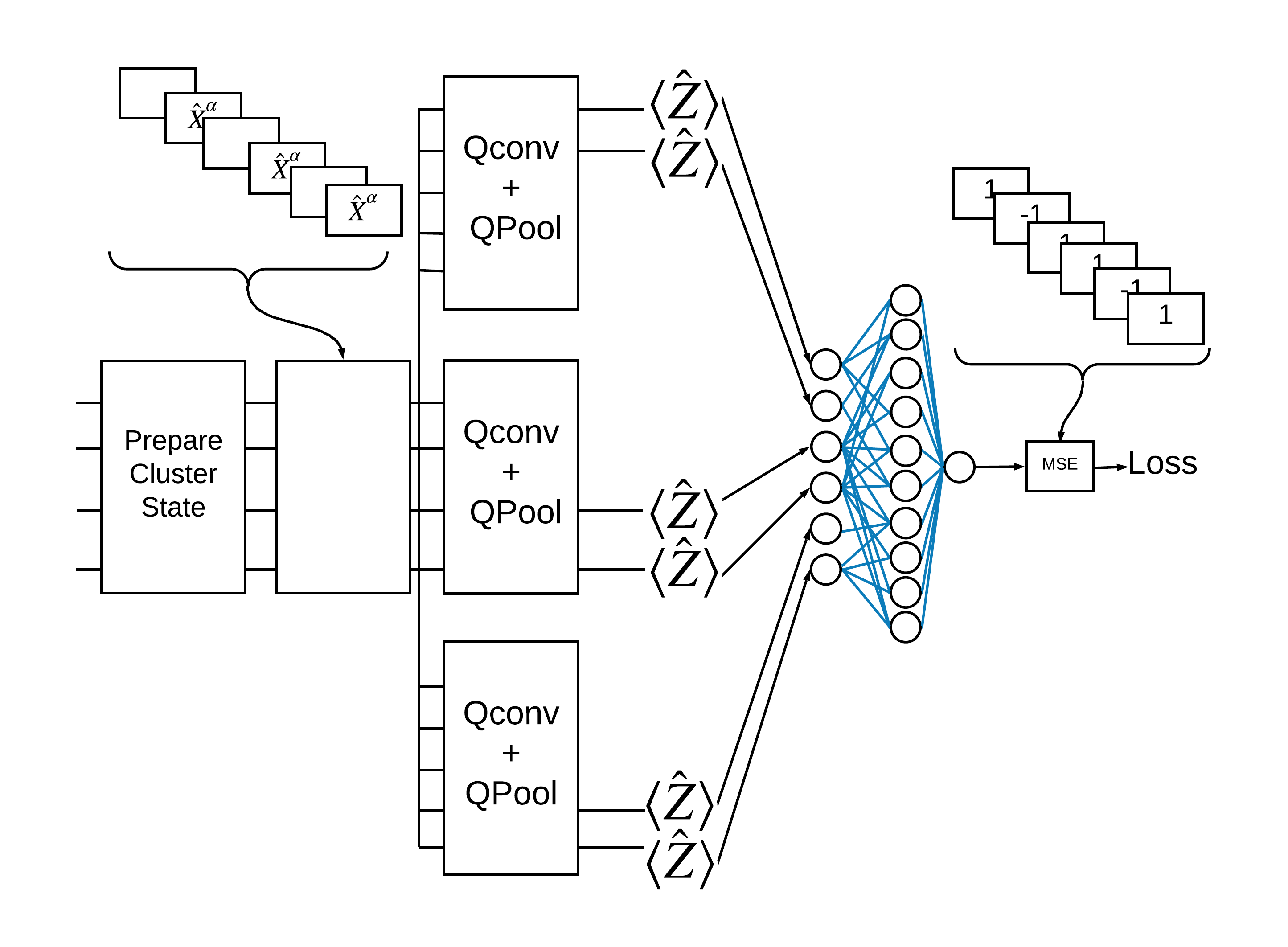}
\caption{A hybrid architecture in which the outputs of 3 separate truncated QCNNs are fed into a classical neural network.}
    \label{fig:hqcnn2}
\end{figure}

\begin{figure}[H]
    \centering
    \includegraphics[width=0.8\columnwidth]{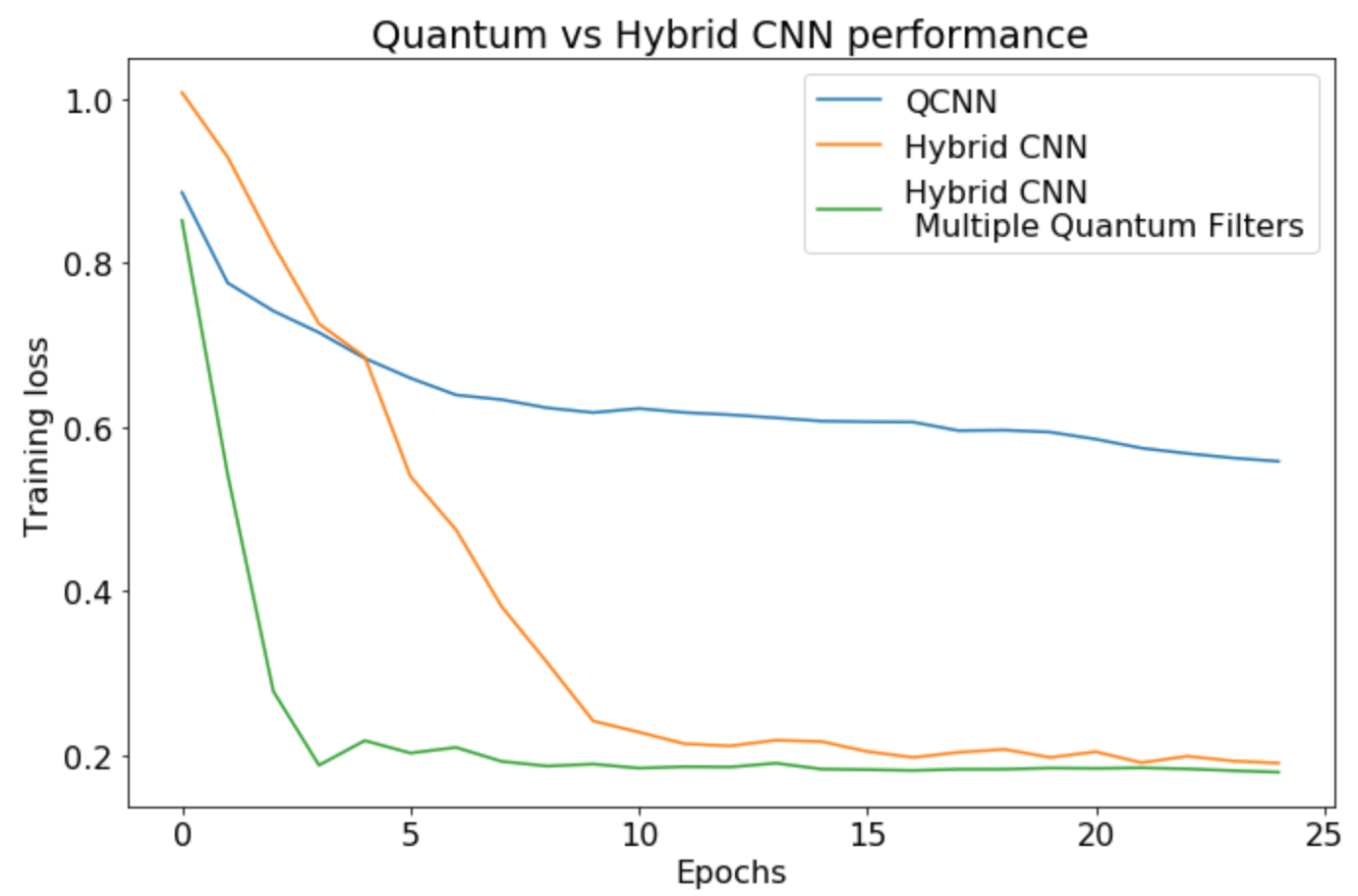}
\caption{Mean squared error loss as a function of training epoch for three different hybrid classifiers.  We find that the purely quantum classifier trains the slowest, while the hybrid architecture with multiple quantum filters trains the fastest.}
    \label{fig:hybrid_qcnn_loss}
\end{figure}

\begin{lstlisting}
# Build 3 quantum filters
QCNN_1 = tfq.layers.PQC(
    multi_readout_model_circuit(cluster_state_bits),
    readouts)(cluster_state)
QCNN_2 = tfq.layers.PQC(
    multi_readout_model_circuit(cluster_state_bits),
    readouts)(cluster_state)
QCNN_3 = tfq.layers.PQC(
    multi_readout_model_circuit(cluster_state_bits),
    readouts)(cluster_state)
# Feed all QCNNs into a classical NN
concat_out = tf.keras.layers.concatenate(
    [QCNN_1, QCNN_2, QCNN_3])
dense_1 = tf.keras.layers.Dense(8)(concat_out)
dense_2 = tf.keras.layers.Dense(1)(dense_1)
multi_qconv_model = tf.keras.Model(
    inputs=[excitation_input],
    outputs=[dense_2])
\end{lstlisting}
We find that that for the same optimization settings, the purely quantum model trains the slowest, while the three-quantum-filter hybrid model trains the fastest.  This data is shown in Fig.~\ref{fig:hybrid_qcnn_loss}.  This demonstrates the advantage of exploring hybrid quantum-classical architectures for classifying quantum data.

\subsection{Hybrid Machine Learning for Quantum Control} \label{sec:qctrl}
To run this example in the browser through Colab, follow the link:
\fancylink{https://github.com/tensorflow/quantum/blob/research/control/control.ipynb}{research/control/control.ipynb}

Recently, neural networks  have been successfully   deployed for solving quantum control   problems ranging from optimizing gate decomposition,   error correction subroutines,  to continuous  Hamiltonian controls. To fully leverage the power of neural networks without being hobbled by possible  computational overhead, it is essential to obtain a deeper understanding of the  connection between various neural network representations and different types of quantum control dynamics.  We demonstrate tailoring machine learning architectures to underlying quantum dynamics using TFQ in \cite{Niu2020}. As a summary of how the unique functionalities of TFQ ease quantum control optimization, we list the problem definition and required TFQ toolboxes as follows.

\vspace{\baselineskip}
\noindent\textbf{Target problems}:
\begin{enumerate}[noitemsep]
    \item Learning quantum dynamics. 
    
    \item Optimizing quantum control signal with regard to a cost objective
    \item Error mitigation in realistic quantum device
\end{enumerate}
\noindent\textbf{ Required TFQ functionalities}:
      \begin{enumerate}[noitemsep]
        \item Hybrid quantum-classical network model
        \item Batch quantum circuit simulator
        \item Quantum expectation-based backpropagation
        \item Fast classical optimizers, both gradient based and non-gradient based
    \end{enumerate}
    
We exemplify the importance of appropriately choosing the right neural network architecture for the corresponding quantum control problems with two simple but realistic control optimizations. The two types of controls we have considered cover the full range of quantum dynamics: constant Hamiltonian evolution vs time dependent Hamiltonian evolution. In the first problem, we design a DNN to machine-learn  (noise-free) control of a single qubit. In the second problem, we design an RNN with long-term memory to learn a stochastic non-Markovian control noise model.

\subsubsection{Time-Constant Hamiltonian Control}\label{sec:time_invar_control}

\begin{figure}
    \centering
    \includegraphics[width=1\columnwidth]{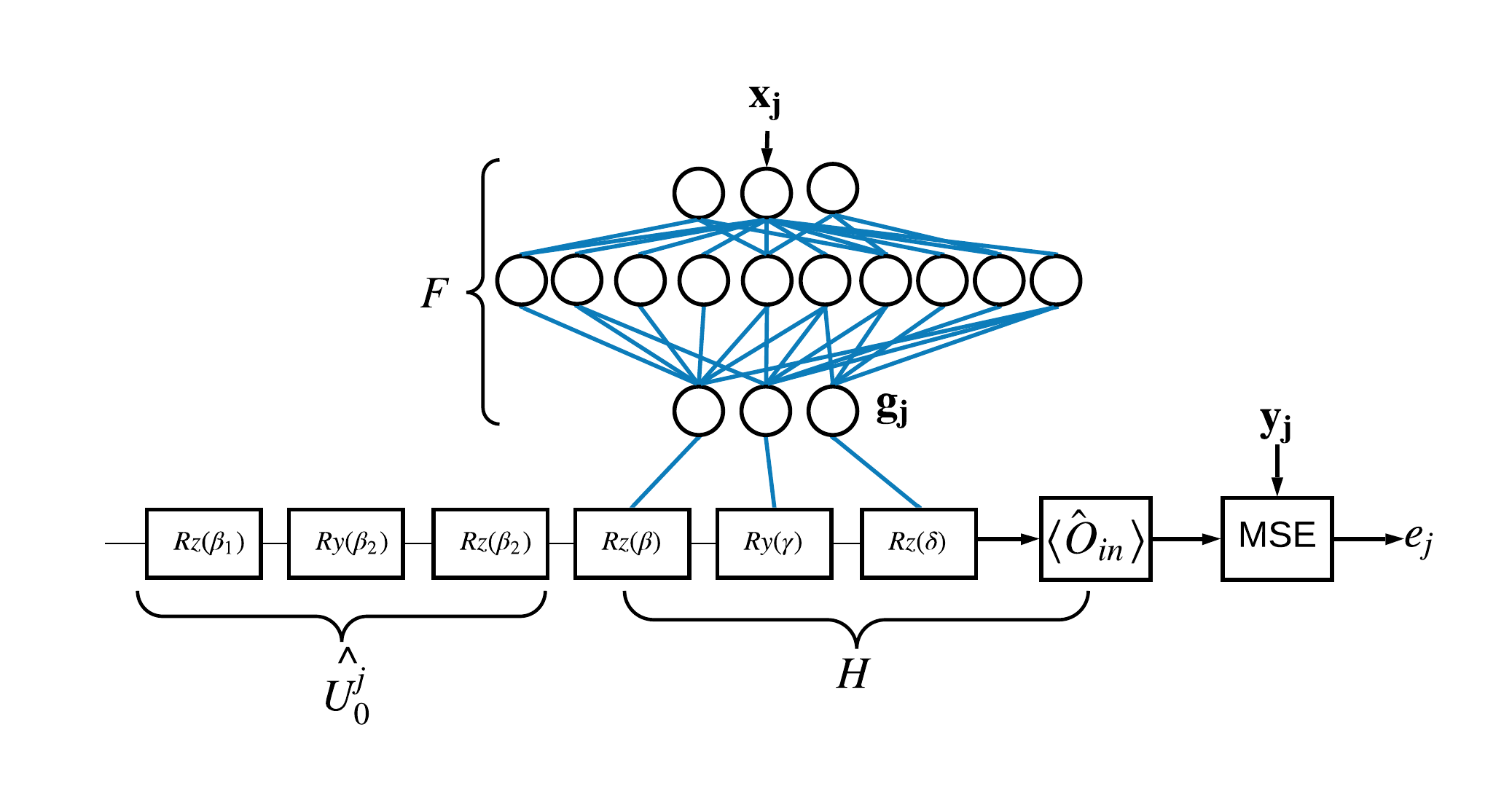}
\caption{Architecture for hybrid quantum-classical neural network model for learning a quantum control decomposition.}
    \label{fig:quantum_control_example1}
\end{figure}

If the underlying system Hamiltonian is time-invariant the task of quantum control can be simplified with open-loop optimization. Since the optimal solution is independent of instance by instance control actualization, control optimization can be done offline.  Let $\bf{x}$ be the input to a controller, which produces some control vector $\mathbf{g} = F(\mathbf{x})$. This control vector actuates a system which then produces a vector output $H(\mathbf{g})$. For a set of control inputs $\mathbf{x_j}$ and desired outputs $\mathbf{y_j}$, we can then define controller error as $e_j(F) = |\mathbf{y_j} - H(F(\mathbf{x_j}))|$. The optimal control problem can then be defined as minimizing $\sum_j e_j(F)$ for $F$. This optimal controller will produce the optimal control vector $\mathbf{g^*_j}$ given $\mathbf{x_j}$

This problem can be solved exactly if $H$ and the relationships between $\mathbf{g_j^*}$ and $\mathbf{x_j}$ are well understood and invertible. Alternatively, one can find an approximate solution using a parameterized controller $F$ in the form of a feed-forward neural network. We can calculate a  set of control pairs of size $N$: $\{\mathbf{x_i}, \mathbf{y_i}\}$ with $i\in [N]$.  We can input $\mathbf{x_i}$ into $F$ which is parameterized by its weight matrices $\{W_i\}$ and biases $\{\mathbf{b}_i\}$ of each $i$th layer. A successful training will find network parameters given by $\{W_i\}$ and   $\{\mathbf{b}_i\}$  such that for any given input  $\bf{x_i}$ the network outputs $\mathbf{g_i}$ which leads to a system output $H(\mathbf{g_i}) \approxeq \mathbf{y_i}$.  This architecture is shown schematically in Fig. \ref{fig:quantum_control_example1}
  
There are two important reasons behind the use of supervised learning for practical control optimization. Firstly, not all time-invariant Hamiltonian control problems permit analytical solutions, so inverting the control error function map can be costly in computation. Secondly, realistic deployment of even a time-constant quantum controller faces stochastic fluctuations due to noise in the classical electronics and systematic errors which cause the behavior of the system $H$ to deviate from ideal. Deploying supervised learning with experimentally measured control pairs will therefore enable the finding of  robust quantum control solutions facing systematic control offset and electronic noise. However, this necessitates seamlessly connecting the output of a classical neural network with the execution of a quantum circuit in the laboratory. This functionality is offered by TFQ through \Colorbox{bkgd}{\lstinline{ControlledPQC}}.

We showcase the use of supervised learning with hybrid quantum-classical feed-forward neural networks in TFQ for the single-qubit gate decomposition problem. A general unitary transform on one qubit can be specified by the exponential
\begin{equation}\label{eqn:gen_unitary}
    U(\phi, \theta_1, \theta_2)=e^{-i \phi (\cos\theta_1\hat{Z} + \sin\theta_1(\cos\theta_2\hat{X} + \sin\theta_2\hat{Y}))}.
\end{equation}
However, it is often preferable to enact single qubit unitaries using rotations about a single axis at a time. Therefore, given a  single qubit gate  specified by the vector of three rotations  $\{\phi,\theta_1, \theta_2 \}$, we want find the control sequence that implements this gate in the form
\begin{align}\label{eqn:zyz}
 U(\beta, \gamma, \delta)
 &=e^{i\alpha} e^{ -i \frac{\beta}{2} \hat{Z}}e^{ -i \frac{\gamma}{2} \hat{Y}}  e^{ -i \frac{\delta}{2} \hat{Z}}.
\end{align}  
This is the optimal decomposition, namely the Bloch theorem, given any unknown single-qubit unitary into hardware friendly gates which only involve the rotation along a fixed axis at a time.

The first step in the training involves preparing the training data. Since quantum control optimization only focuses on the performance in hardware deployment, the control inputs and output have to be chosen such that they are experimentally observable. We define the vector of expectation values $\mathbf{y_i}= [\langle\hat{X}\rangle_{\mathbf{x_i}}, \langle\hat{Y}\rangle_{\mathbf{x_i}}, \langle\hat{Z}\rangle_{\mathbf{x_i}}]$ of all single-qubit Pauli operators given by the quantum state prepared by the associated input $\bf{x_i}$:
\begin{align}
\ket{\psi_{0}^i} = & \hat{U}_o^i\ket{0}\\
\ket{\psi}_{\mathbf{x}} = & e^{ -i \frac{\beta}{2} \hat{Z}}e^{ -i \frac{\gamma}{2} \hat{Y}}  e^{ -i \frac{\delta}{2} \hat{Z}}\ket{\psi_{0}^i},\\
    \langle\hat{X}\rangle_{\mathbf{x}} =& \bra{\psi}_{\mathbf{x}} \hat{X}\ket{\psi}_{\mathbf{x}}, \\
        \langle\hat{Y}\rangle_{\mathbf{x}} =& \bra{\psi}_{\mathbf{x}} \hat{Y}\ket{\psi}_{\mathbf{x}},\\
            \langle\hat{Z}\rangle_{\mathbf{x}} =& \bra{\psi}_{\mathbf{x}} \hat{Z}\ket{\psi}_{\mathbf{x}}.
\end{align}
Assuming we have prepared the training dataset, each set consists of input vectors  $\mathbf{x_i}=[\phi , \theta_1, \theta_2]$ which derives from the randomly drawn $\textbf{g}$, unitaries that prepare each initial state $\hat{U}_0^i$, and the associated expectation values $\mathbf{y_i}=[ \langle\hat{X}\rangle_{\bf{x_i} }, \langle\hat{Y}\rangle_{\bf{x_i} }, \langle\hat{Z}\rangle\}_{\bf{x_i}}]$.

Now we are ready to define the hybrid quantum-classical neural network model in Keras with TensforFlow API. To start, we first define the quantum part of the hybrid neural network, which is a simple quantum circuit of three single-qubit gates as follows.
\begin{lstlisting}
control_params = sympy.symbols('theta_{1:3}')
qubit = cirq.GridQubit(0, 0)
control_circ = cirq.Circuit(
    cirq.Rz(control_params[2])(qubit),
    cirq.Ry(control_params[1])(qubit),
    cirq.Rz(control_params[0])(qubit))
\end{lstlisting}
We are now ready to finish off the hybrid network by defining the classical part, which maps the target params to the control vector $\mathbf{g}=\{\beta, \gamma, \delta\}$. Assuming we have defined the vector of observables \Colorbox{bkgd}{\lstinline{ops}}, the code to build the model is:
\begin{lstlisting}
circ_in = tf.keras.Input(
    shape=(), dtype=tf.dtypes.string)
x_in = tf.keras.Input((3,))
d1 = tf.keras.layers.Dense(128)(x_in)
d2 = tf.keras.layers.Dense(128)(d1)
d3 = tf.keras.layers.Dense(64)(d2) 
g = tf.keras.layers.Dense(3)(d3)
exp_out = tfq.layers.ControlledPQC(
    control_circ, ops)([circ_in, x_in])
\end{lstlisting}

\begin{figure}
    \centering
    \includegraphics[width=0.8\columnwidth]{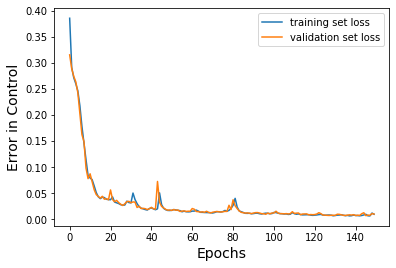}
\caption{Mean square error on training dataset, and validation dataset, each of size 5000, as a function training epoch.}
    \label{fig:quantum_control_example1result}
\end{figure}

Now, we are ready to put everything together to define and train a model in Keras.  The two axis control model is defined as follows:
\begin{lstlisting}
model = tf.keras.Model(
    inputs=[circ_in, x_in], outputs=exp_out)
\end{lstlisting}
To train this hybrid supervised model, we define an optimizer, which in our case is the Adam optimizer, with an appropriately chosen loss function:
\begin{lstlisting}
model.compile(tf.keras.Adam, loss='mse')
\end{lstlisting}
We finish off by training on the prepared supervised data in the standard way:
\begin{lstlisting}
history_two_axis = model.fit(...
\end{lstlisting}
The training converges after around 100 epochs as seen in Fig.~\ref{fig:quantum_control_example1result}, which also shows excellent generalization to validation data. 

\subsubsection{Time-dependent Hamiltonian Control}

Now we consider a second kind of quantum control problem, where the actuated system $H$ is allowed to change in time.  If the system is changing with time, the optimal control $\mathbf{g^*}$ is also generally time varying. Generalizing the discussion of section \ref{sec:time_invar_control}, we can write the time varying control error given the time-varying controller $F(t)$ as $e_j(F(t), t) = |\mathbf{y_j} - H(F(\mathbf{x_j},t),t)|$. The optimal control can then be written as $\mathbf{g^*(t)} = \bar{\mathbf{g}^*} + \mathbf{\delta(t)}$.

This task is significantly harder than the problem discussed in section \ref{sec:time_invar_control}, since we need to learn the hidden variable $ \mathbf{\delta}(t)$ which can result in potentially highly complex real-time system dynamics. We showcase how TFQ provides the perfect toolbox for such difficult control optimization with an important and realistic problem of learning and thus compensating the low frequency noise. 

One of the main contributions to time-drifting errors in realistic quantum control is $1/f^\alpha$- like errors, which encapsulate errors in the Hamiltonian amplitudes whose frequency spectrum has a large component at the low frequency regime. The origin of such low frequency noise remains largely controversial. Mathematically, we can parameterize the low frequency noise in the time domain with the amplitude of the Pauli Z Hamiltonian on each qubit as:
\begin{align}
    \hat{H}_{\text{low}}(t) = \alpha t^{e}  \hat{Z}.
 \end{align}
A simple phase control signal is given by $\omega(t) = \omega_0 t$ with the Hamiltonian $\hat{H}_0(t) = \omega_0 t \hat{Z}$. The time-dependant wavefunction is then given by
\begin{align}\label{eqn:time_dep_hammy}
    \ket{\psi(t_i)} = \mathbb{T}[e^{\int_0^{t_i}(\hat{H}_{\text{low}}(t) +\hat{H}_0(t)) dt} ]\ket{+}.
\end{align}
We can attempt to learn the noise parameters $\alpha$ and $e$ by training a recurrent neural network to perform time-series prediction on the noise. In other words, given a record of expectation values $\{ \bra{\psi(t_i)} \hat{X}\ket{\psi(t_i)}\}$ for $t_i \in \{0, \delta t, 2\delta t,\dots, T\}$ obtained on state $\ket{\psi(t)}$, we want the RNN to predict the future observables  $\{ \bra{\psi(t_i)} \hat{X}\ket{\psi(t_i)}\}$ for $t_i \in \{T, T+ \delta t, T+ 2\delta t,\dots, 2T\}$.

There are several possible ways to build such an RNN. One option is recording several timeseries on a device a-priori, later training and testing an RNN offline. Another option is an online method, which would allow for real-time controller tuning. The offline method will be briefly explained here, leaving the details of both methods to the notebook associated with this example.

\begin{figure}
    \centering
    \includegraphics[width=0.8\columnwidth]{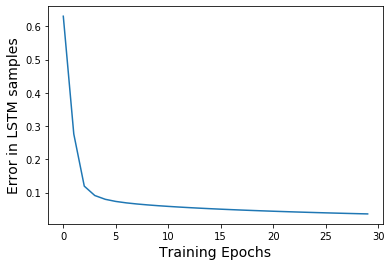}
\caption{Mean square error on LSTM predictions on 500 randomly generated inputs.}
    \label{fig:quantum_control_example2result}
\end{figure}

First, we can use TFQ or Cirq to prepare several timeseries for testing and validation. The function below performs this task using TFQ:
\begin{lstlisting}
def generate_data(end_time, timesteps, omega_0, exponent, alpha):
    t_steps = linspace(0, end_time, timesteps)
    q = cirq.GridQubit(0, 0)
    phase = sympy.symbols("phaseshift")
    c = cirq.Circuit(cirq.H(q), 
                     cirq.Rz(phase_s)(q))
    ops = [cirq.X(q)]
    phases = t_steps*omega_0 + 
        t_steps**(exponent + 1)/(exponent + 1)
    return tfq.layers.Expectation()(
      c,
      symbol_names = [phase],
      symbol_values = transpose(phases),
      operators = ops)
\end{lstlisting}
We can use this function to prepare many realizations of the noise process, which can be fed to an LSTM defined using \Colorbox{bkgd}{\lstinline{tf.keras}} as below:
\begin{lstlisting}
model = tf.keras.Sequential([
  tf.keras.layers.LSTM(
      rnn_units,
      recurrent_initializer='glorot_uniform',
      batch_input_shape=[batch_size, None, 1]),
  tf.keras.layers.Dense(1)])
\end{lstlisting}
We can then train this LSTM on the realizations and evaluate the success of our training using validation data, on which we calculate prediction accuracy. A typical example of this is shown in Fig.~\ref{fig:quantum_control_example2result}. The LSTM converges quickly to within the  accuracy from the expectation value measurements within  30 epochs.

\subsection{Simulating Noisy Circuits}\label{sec:SimulatingNoisyCircuits}
To run this example in the browser through Colab, follow the link:
\fancylink{https://github.com/tensorflow/quantum/blob/master/docs/tutorials/noise.ipynb}{docs/tutorials/qcnn.ipynb}

\subsubsection{Background}
Noise is present in modern day quantum computers. Qubits are susceptible to interference from the surrounding environment, imperfect fabrication, TLS and sometimes even cosmic rays \cite{martinis2021saving, mcewen2021resolving}. Until large scale error correction is reached, the algorithms of today must be able to remain functional in the presence of noise. This makes testing algorithms under noise an important step for validating quantum algorithms / models will function on the quantum computers of today.

\vspace{\baselineskip}
\noindent\textbf{Target problems}:
\begin{enumerate}[noitemsep]
    \item Simulating noisy circuits
    \item Compare hybrid quantum-classical model performance under different types and strengths of noise
\end{enumerate}
\noindent\textbf{ Required TFQ functionalities}:
      \begin{enumerate}[noitemsep]
      \item Serialize Cirq circuits which contain channels
      \item Quantum trajectory simulation with qsim
    \end{enumerate}

\subsubsection{Noise in Cirq and TFQ}

\begin{figure}[H]
    \centering
    \includegraphics[width=0.8\columnwidth]{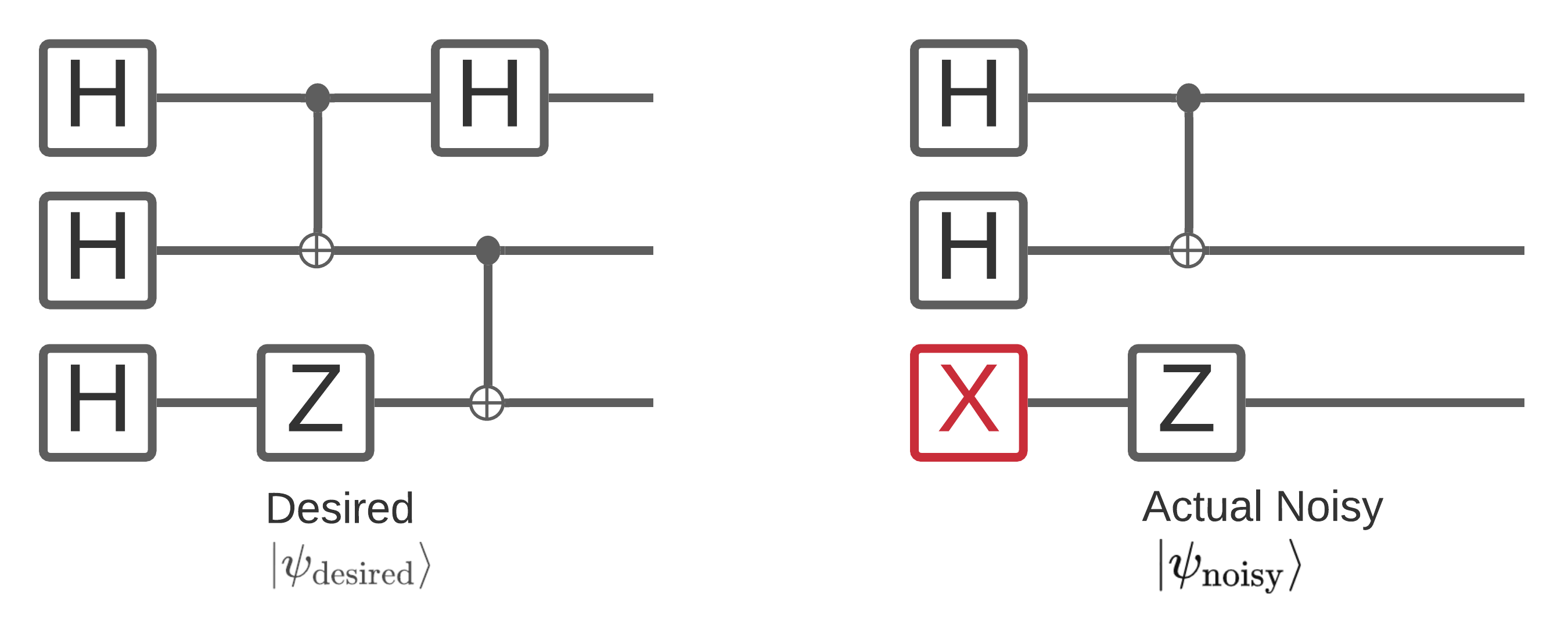}
\caption{Illustration of the effect of noise on a quantum circuit.  On the left is an example circuit on three qubits; on the right, the same circuit is shown after corruption by noise.  An initial Hadamard gate has been changed to an X gate, and the final Hadamard and CNot have been deleted.}
    \label{fig:noisy_circuit}
\end{figure}

Noise on a quantum computer impacts the bitstring samples you are able to measure from it. One intuitive way to think about noise on a quantum computer is that it will "insert", "delete" or "replace" gates in random places.  An example is shown in Fig.~\ref{fig:noisy_circuit}, where the actual circuit run has one gate corrupted and two gates deleted compared to the desired circuit.

Building off of this intuition, when dealing with noise, you are no longer using a single pure state $\ket{\psi}$ but instead dealing with an \emph{ensemble} of all possible noisy realizations of your desired circuit, $\rho = \sum_j p_j \ket{\psi_j} \bra{\psi_j}$, where $p_j$ gives the probability that the actual state of the quantum computer after your noisy circuit is $\ket{\psi_j}$ .

Revisiting Fig.~\ref{fig:noisy_circuit}, if we knew beforehand that 90\% of the time our system executed perfectly, or errored 10\% of the time with just this one mode of failure, then our ensemble would be:
\[
\rho = 0.9 \ket{\psi_\text{desired}} \bra{\psi_\text{desired}} + 0.1 \ket{\psi_\text{noisy}} \bra{\psi_\text{noisy}}.
\]
If there was more than just one way that our circuit could error, then the ensemble $\rho$ would contain more than just two terms (one for each new noisy realization that could happen). $\rho$ is referred to as the \emph{density operator} describing your noisy system.

Unfortunately, in practice, it's nearly impossible to know all the ways your circuit might error and their exact probabilities. A simplifying assumption you can make is that after each operation in your circuit a \emph{quantum channel} is applied.  Just like a quantum circuit produces pure states, a quantum channel produces density operators.  Cirq has a selection of built-in quantum channels to model the most common ways a quantum circuit can go wrong in the real world.  For example, here is how you would write down a quantum circuit with depolarizing noise:

\begin{lstlisting}
def x_circuit(qubits):
  """Produces an X wall circuit on `qubits`."""
  return cirq.Circuit(cirq.X.on_each(*qubits))

def make_noisy(circuit, p):
  """Adds a depolarization channel to all qubits in `circuit` before measurement."""
  return circuit + cirq.Circuit(
      cirq.depolarize(p).on_each(
          *circuit.all_qubits()))

my_qubits = cirq.GridQubit.rect(1, 2)
my_circuit = x_circuit(my_qubits)
my_noisy_circuit = make_noisy(my_circuit, 0.5)
\end{lstlisting}

Both \Colorbox{bkgd}{\lstinline{my_circuit}} and \Colorbox{bkgd}{\lstinline{my_noisy_circuit}} can be measured, and their outputs compared. In the noiseless case you would always expect to sample the $\ket{11}$ state. But in the noisy case there is a nonzero probability of sampling $\ket{00}$ or $\ket{01}$ or $\ket{10}$ as well.  See the notebook for a demonstration of this effect.

With this understanding of how noise can impact circuit execution, we are ready to move on to how noise works in TFQ. TensorFlow Quantum uses trajectory methods to simulate noise, as described in \ref{sec:QsimOtherFeatures}.  To enable noisy simulations, the \Colorbox{bkgd}{\lstinline{backend='noisy'}} option is available on \Colorbox{bkgd}{\lstinline{tfq.layers.Sample}}, \Colorbox{bkgd}{\lstinline{tfq.layers.SampledExpectation}} and \Colorbox{bkgd}{\lstinline{tfq.layers.Expectation}}.  For example, here is how to instantiate and call the \Colorbox{bkgd}{\lstinline{tfq.layers.Sample}} on the noisy circuit created above:
\begin{lstlisting}
"""Draw samples from `my_noisy_circuit`"""
bitstrings = tfq.layers.Sample(backend='noisy')(my_noisy_circuit, repetitions=1000)
\end{lstlisting}
See the notebook for more examples of noisy usage of TFQ layers.

\subsubsection{Training Models with Noise}

Now that you have implemented some noisy circuit simulations in TFQ, you can experiment with how noise impacts hybrid quantum-classical models.  Since quantum devices are currently in the NISQ era, it is important to compare model performance in the noisy and noiseless cases, to ensure performance does not degrade excessively under noise.

\begin{figure}[h]
    \centering
    \includegraphics[width=0.8\columnwidth]{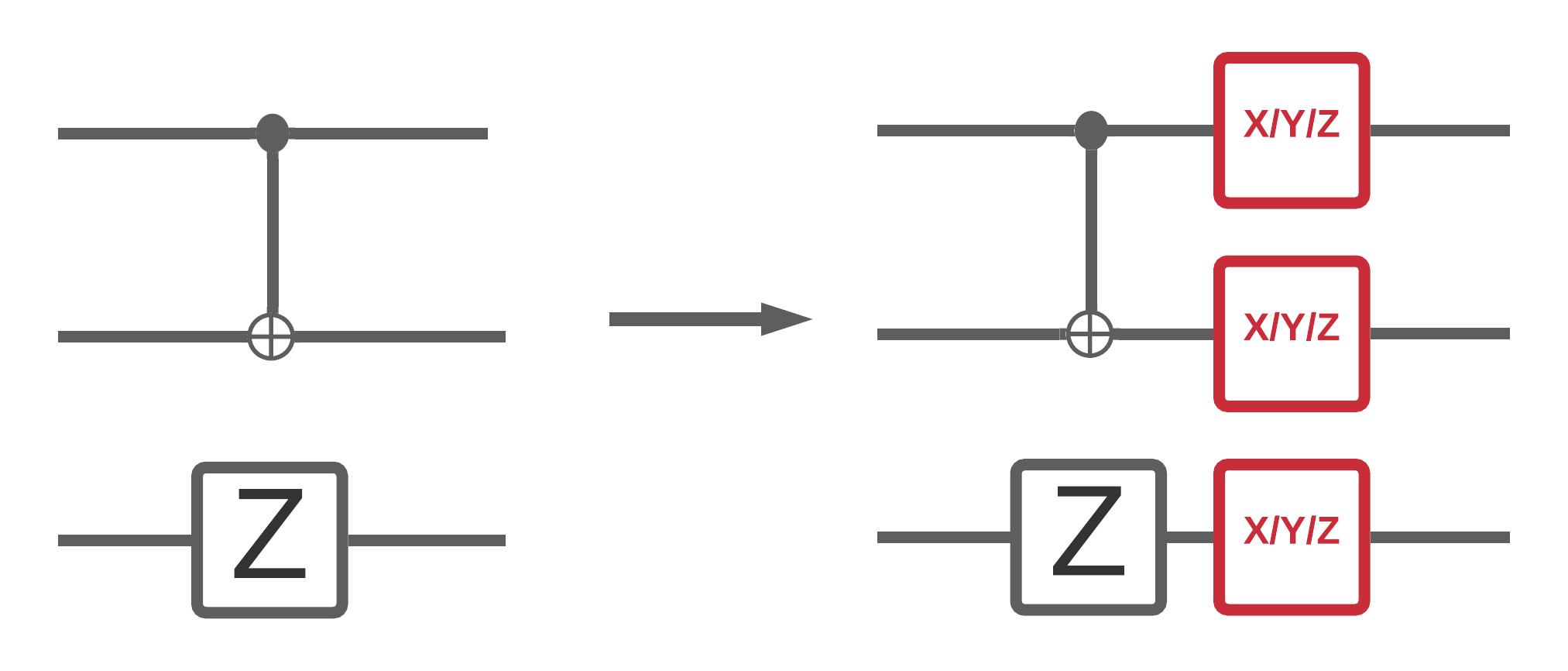}
\caption{Example circuit before and after the application of circuit-wide depolarizing noise.}
    \label{fig:noisy_quantum_data}
\end{figure}

A good first check to see if a model or algorithm is robust to noise is to test it under a circuit-wide depolarizing model.  An example circuit with depolarizing noise applied is shown in Fig.~\ref{fig:noisy_quantum_data}.  Applying a channel to a circuit using the \Colorbox{bkgd}{\lstinline{circuit.with_noise(channel)}} method creates a new circuit with the specified channel applied after each operation in the circuit.  In the case of a depolarizing channel, one of $\{X, Y, Z \}$ is applied with probability $p$ and an identity (keeping the original operation) with probability $1-p$.

\begin{figure}[h]
    \centering
    \includegraphics[width=0.8\columnwidth]{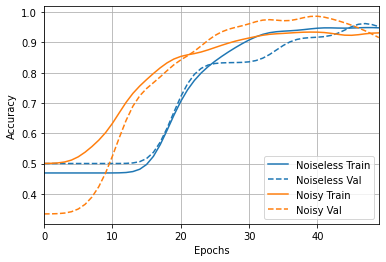}
\caption{Accuracy versus training epoch for a quantum classifier.  The blue line shows the performance of the classifier when trained on clean quantum data, while the orange line shows the performance of the classifier given noisy quantum data.  Note that the final values for accuracy are approximately equal between the noisy and noiseless cases.}
\label{fig:training_with_noise}
\end{figure}

In the notebook, we show how to create a noisy dataset based on the XXZ chain dataset described in section \ref{sec:TFQDatasets}.  This noisy dataset will be called with the function \Colorbox{bkgd}{\lstinline{modelling_circuit}}.  Then, we build a \Colorbox{bkgd}{\lstinline{tf.keras.Model}} which uses a noisy PQC layer to process the quantum data, along with Dense layers for classical post-processing.  The code to build the model is shown below:

\begin{lstlisting}
def build_keras_model(qubits, depolarize_p=0.):
  """Prepare a noisy hybrid quantum classical Keras model."""
  spin_input = tf.keras.Input(
      shape=(), dtype=tf.dtypes.string)

  circuit_and_readout = modelling_circuit(
      qubits, 4, depolarize_p)
  if depolarize_p >= 1e-5:
    quantum_model = tfq.layers.NoisyPQC(
        *circuit_and_readout,
        sample_based=False,
        repetitions=10
    )(spin_input)
  else:
    quantum_model = tfq.layers.PQC(
        *circuit_and_readout)(spin_input)

  intermediate = tf.keras.layers.Dense(
      4, activation='sigmoid')(quantum_model)
  post_process = tf.keras.layers.Dense(1)(
      intermediate)

  return tf.keras.Model(
      inputs=[spin_input],
      outputs=[post_process])
\end{lstlisting}
The goal of this model is to correctly classify the quantum datapoints according to their phase.  For the XXZ chain, the two possible phases are the critical metallic phase and the insulating phase.  Fig.~\ref{fig:training_with_noise} shows the accuracy of the model as a function of number of training epochs, for both noisy and noiseless training data.  

Notice that when the model is trained on noisy data, it achieves a similar accuracy as when it is trained on noiseless data.  This shows that QML models can still succeed at learning despite the presence of mild noise.  You can adapt this workflow to examine the performance of your own QML models in the presence of noise.

\subsection{Quantum Approximate Optimization}\label{sec:QAOA}
To run this example in the browser through Colab, follow the link:
\fancylink{https://github.com/tensorflow/quantum/blob/research/qaoa/qaoa.ipynb}{research/qaoa/qaoa.ipynb}

\subsubsection{Background}

In this section, we introduce the basics of Quantum Approximate Optimization Algorithms and show how to implement a basic case of this class of algorithms in TFQ. In the advanced applications section \ref{sec:advanced_applications}, we explore how to apply meta-learning techniques \cite{chen2016learning} to the optimization of the parameters of the algorithm.

The Quantum Approximate Optimization Algorithm was originally proposed to solve instances of the MaxCut problem \cite{farhi2014quantum}. The QAOA framework has since been extended to encompass multiple problem classes related to finding the low-energy states of Ising Hamiltonians, including Hamiltonians of higher-order and continuous-variable Hamiltonians \cite{verdon2019quantum}.

In general, the goal of the QAOA for binary variables is to find approximate minima of a pseudo Boolean function $f$ on $n$ bits, $f(\bm{z})$, $\bm{z}\in\{-1,1\}^{\times n}$. This function is often an $m^{\text{th}}$-order polynomial of  binary variables for some positive integer $m$, e.g., $f(\bm{z}) = \sum_{p\in\{0,1\}^m}\alpha_{\bm{p}}\bm{z}^{\bm{p}}$, where $\bm{z}^{\bm{p}}=\prod_{j=1}^n z_j^{p_j}$. 
QAOA has been applied to NP-hard problems such as Max-Cut \cite{farhi2014quantum} or Max-3-Lin-2 \cite{farhi2014quantumbounded}. The case where this polynomial is quadratic ($m=2$) has been extensively explored in the literature.  It should be noted that there have also been recent advances using quantum-inspired machine learning techniques, such as deep generative models, to produce approximate solutions to such problems \cite{hartnett2020_density, hartnett2020_selfsupervised}.  These 2-local problems will be the main focus in this example. In this tutorial, we first show how to utilize TFQ to solve a MaxCut instance with QAOA with $p=1$.

The QAOA approach to optimization first starts in an initial product state $\ket{\psi_0}^{\otimes n}$ and then a tunable gate sequence produces a wavefunction with a high probability of being measured in a low-energy state (with respect to a cost Hamiltonian). 

Let us define our parameterized quantum circuit ansatz. The canonical choice is to start with a uniform superposition $\ket{\psi_0}^{\otimes n} = \ket{+}^{\otimes n} = \tfrac{1}{\sqrt{2^n}}(\sum_{\bm{x}\in\{0,1\}^n}\ket{\bm{x}})$, hence a fixed state. The QAOA unitary itself then consists of applying 
\begin{equation}\label{eq:QAOA}
   \hat{U}(\bm{\eta},\bm{\gamma}) = \prod_{j=1}^{P}e^{-i\eta_{j}\hat{H}_M}e^{-i\gamma^{j} \hat{H}_C}, 
\end{equation}
onto the starter state, where $\hat{H}_M = \sum_{j\in\mathcal{V}}\hat{X}_j$ is known as the \textit{mixer Hamiltonian}, and  $\hat{H}_C \equiv f( \bm{\hat{\bm{Z}}})$ is our \textit{cost Hamiltonian}, which is a function of Pauli operators $\bm{\hat{Z}} = \{\hat{Z}_j\}_{j=1}^n$. The resulting state is given by $\ket{\Psi_{\bm{\eta}\bm{\gamma}}} = \hat{U}(\bm{\eta},\bm{\gamma})  \ket{+}^{\otimes n}$, which is our parameterized output. We define the energy to be minimized as the expectation value of the cost Hamiltonian $\hat{H}_C \equiv f( \bm{\hat{\bm{Z}}})$, where $\bm{\hat{Z}} = \{\hat{Z}_j\}_{j=1}^n$ with respect to the output parameterized state.

\vspace{\baselineskip}
\noindent\textbf{Target problems}:
\begin{enumerate}[noitemsep]
    \item Train a parameterized quantum circuit for a discrete optimization problem (MaxCut)
    \item Minimize a cost function of a parameterized quantum circuit
\end{enumerate}
\noindent\textbf{ Required TFQ functionalities}:
      \begin{enumerate}[noitemsep]
      \item Conversion of simple circuits to TFQ tensors
      \item Evaluation of gradients for quantum circuits
      \item Use of gradient-based optimizers from TF
    \end{enumerate}

\subsubsection{Implementation}
For the MaxCut QAOA, the cost Hamiltonian function $f$ is a second order polynomial of the form,
\begin{equation}\label{eq:maxcut_ham}
    \hat{H}_{C} =  f( \bm{\hat{\bm{Z}}}) = \sum_{\{j,k\}\in{\mathcal{E}}} \tfrac{1}{2}(\hat{I}- \hat{Z}_j \hat{Z}_k),
\end{equation}
where $\mathcal{G}=\{\mathcal{V},\mathcal{E}\}$ is a graph for which we would like to find the MaxCut; the largest size subset of edges (cut set) such that vertices at the end of these edges belong to a different partition of the vertices into two disjoint subsets \cite{farhi2014quantum}.

To train the QAOA, we simply optimize the expectation value of our cost Hamiltonian with respect to our parameterized output to find (approximately) optimal parameters; $\bm{\eta}^*,\bm{\gamma}^* = \text{argmin}_{\bm{\eta},\bm{\gamma}}\mathcal{L}(\bm{\eta},\bm{\gamma}) $ where $\mathcal{L}(\bm{\eta},\bm{\gamma}) = \bra{\Psi_{\bm{\eta}\bm{\gamma}}}\hat{H}_{C}\ket{\Psi_{\bm{\eta}\bm{\gamma}}} $ is our loss. Once trained, we use the QPU to sample the probability distribution of measurements of the parameterized output state at optimal angles in the standard basis,  $\bm{x} \sim p(\bm{x}) = |\braket{\bm{x}|\Psi_{\bm{\eta}^*\bm{\gamma}^*}}|^2$, and pick the lowest energy bitstring from those samples as our approximate optimum found by the QAOA.

Let us walkthrough how to implement such a basic QAOA in TFQ. The first step is to generate an instance of the MaxCut problem. For this tutorial we generate a random 3-regular graph with $10$ nodes with NetworkX \cite{hagberg2008exploring}.
\begin{lstlisting}
# generate a 3-regular graph with 10 nodes
maxcut_graph = nx.random_regular_graph(n=10,d=3)
\end{lstlisting}

The next step is to allocate $10$ qubits, to define the Hadamard layer generating the initial superposition state, the mixing Hamiltonian $H_\mathrm{M}$ and the cost Hamiltonian $H_\mathrm{P}$. 

\begin{lstlisting}
# define 10 qubits
cirq_qubits = cirq.GridQubit.rect(1, 10)

# create layer of hadamards to initialize the superposition state of all computational states
hadamard_circuit = cirq.Circuit()
for node in maxcut_graph.nodes():
    qubit = cirq_qubits[node]
    hadamard_circuit.append(cirq.H.on(qubit))

# define the two parameters for one block of QAOA
qaoa_parameters = sympy.symbols('a b')

# define the the mixing and the cost Hamiltonians
mixing_ham = 0
for node in maxcut_graph.nodes():
    qubit = cirq_qubits[node]
    mixing_ham += cirq.PauliString(cirq.X(qubit))

cost_ham = maxcut_graph.number_of_edges()/2
for edge in maxcut_graph.edges():
    qubit1 = cirq_qubits[edge[0]]
    qubit2 = cirq_qubits[edge[1]]
    cost_ham += cirq.PauliString(1/2*(cirq.Z(qubit1)*cirq.Z(qubit2)))
\end{lstlisting}
With this, we generate the unitaries representing the quantum circuit
\begin{lstlisting}
# generate the qaoa circuit
qaoa_circuit = tfq.util.exponential(operators = [cost_ham, mixing_ham], coefficients = qaoa_parameters)
\end{lstlisting}
Subsequently, we use these ingredients to build our model. We note here in this case that QAOA has no input data and labels, as we have mapped our graph to the QAOA circuit. To use the TFQ framework we specify the Hadamard circuit as input and convert it to a TFQ tensor. We may then construct a tf.keras model using our QAOA circuit and cost in a TFQ PQC layer, and use a single instance sample for training the variational parameters of the QAOA with the Hadamard gates as an input layer and a target value of $0$ for our loss function, as this is the theoretical minimum of this optimization problem.

This translates into the following code:
\begin{lstlisting}
# define the model and training data
model_circuit, model_readout = qaoa_circuit, cost_ham
input_ = [hadamard_circuit] 
input_ = tfq.convert_to_tensor(input_)
optimum = [0]

# Build the Keras model.
optimum = np.array(optimum)
model = tf.keras.Sequential()
model.add(tf.keras.layers.Input(shape=(), dtype=tf.dtypes.string))
model.add(tfq.layers.PQC(model_circuit, model_readout))
\end{lstlisting}
 
To optimize the parameters of the ansatz state, we use a classical optimization routine. In general, it would be possible to use pre-calculated parameters \cite{streif2020training} or to implement for QAOA tailored optimization routines \cite{1808.10816}. For this tutorial, we choose the Adam optimizer implemented in TensorFlow. We also choose the mean absolute error as our loss function. 

\begin{lstlisting}
model.compile(loss=tf.keras.losses.mean_absolute_error, optimizer=tf.keras.optimizers.Adam())
history = model.fit(input_,optimum,epochs=1000,verbose=1)
\end{lstlisting}

\section{Advanced Quantum Applications}\label{sec:advanced_applications}
The following applications represent how we have applied TFQ to accelerate their discovery of new quantum algorithms.
The examples presented in this section are newer research as compared to the previous section, as such they have not had as much time for feedback from the community. We include these here as they are  demonstration of the sort of advanced QML research that can be accomplished by combining several building blocks provided in TFQ.
As many of these examples involve the building and training of hybrid quantum-classical models and advanced optimizers, such research would be much more difficult to implement without TFQ. In our researchers' experience, the performance gains and the ease of use of TFQ decreased the time-to-working-prototype from \textit{weeks} to \textit{days} or even \textit{hours} when it is compared to using alternative tools.  

Finally, as we would like to provide users with advanced examples to see TFQ in action for research use-cases beyond basic implementations, along with the examples presented in this section are several notebooks accessible on Github:
\fancylink{https://github.com/tensorflow/quantum/tree/research}{github.com/tensorflow/quantum/tree/research}

 We encourage readers to read the section below for an overview of the theory and use of TFQ functions and would encourage avid readers who want to experiment with the code to visit the full notebooks.

\subsection{Meta-learning for Variational Quantum Optimization}
To run this example in the browser through Colab, follow the link:
\fancylink{https://github.com/tensorflow/quantum/blob/research/metalearning_qaoa/metalearning_qaoa.ipynb}{research/metalearning\_qaoa/metalearning\_qaoa.ipynb}

\begin{figure}
    \centering
    \includegraphics[width=0.44\textwidth]{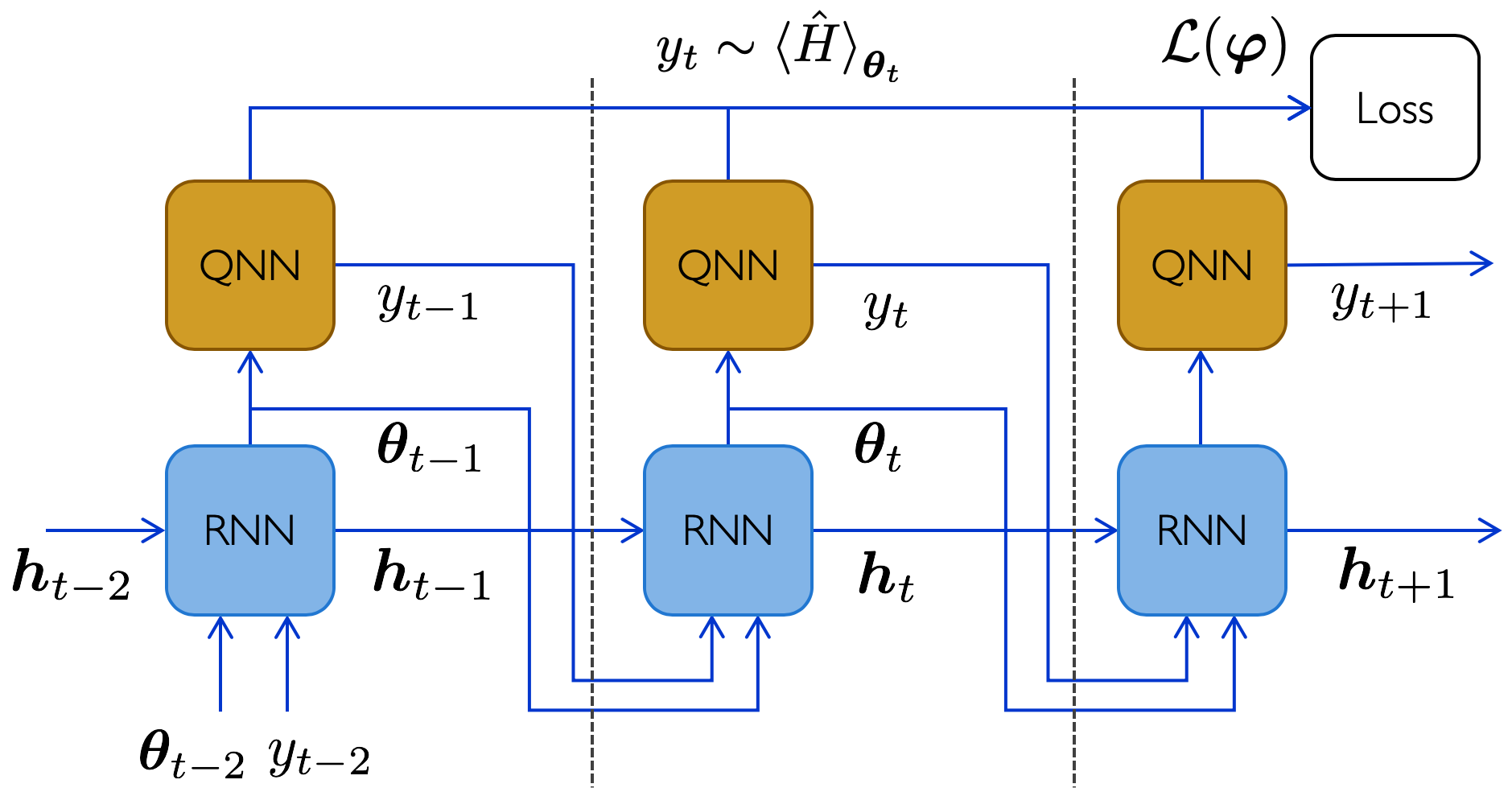}
    \caption{Quantum-classical computational graph for the meta-learning optimization of the recurrent neural network (RNN) optimizer and a quantum neural network (QNN) over several optimization iterations. The hidden state of the RNN is represented by $\bm{h}$, we represent the flow of data used to evaluate the meta-learning loss function. This meta loss function $\mathcal{L}$ is a functional of the history of expectation value estimate samples $\bm{y} = \{y_t\}_{t=1}^T$, it is not directly dependent on the RNN parameters $\bm{\varphi}$. TFQ's hybrid quantum-classical backpropagation then becomes key to train the RNN to learn to optimize the QNN, which in our particular example was the QAOA. Figure taken from \cite{Verdon2019metalearning},
    originally inspired from \cite{chen2016learning}.
    }
    \label{fig:RNN}
\end{figure}

In section \ref{sec:QAOA}, we have shown how to implement basic QAOA in TFQ and optimize it with a gradient-based optimizer, we can now explore how to leverage classical neural networks to optimize QAOA parameters.  To run this example in the browser via Colab, follow the link:

In recent works, the use of classical recurrent neural networks to learn to optimize the parameters \cite{Verdon2019metalearning} (or gradient descent hyperparameters \cite{wilson2019optimizing}) was proposed. As the choice of parameters after each iteration of quantum-classical optimization can be seen as the task of generating a sequence of parameters which converges rapidly to an approximate optimum of the landscape, we can use a type of classical neural network that is naturally suited to generate sequential data, namely, recurrent neural networks. This technique was derived from work by DeepMind \cite{chen2016learning} for optimization of classical neural networks and was extended to be applied to quantum neural networks \cite{Verdon2019metalearning}.

The application of such classical learning to learn techniques to quantum neural networks was first proposed in \cite{Verdon2019metalearning}. In this work, an RNN (long short term memory; LSTM) gets fed the parameters of the current iteration and the value of the expectation of the cost Hamiltonian of the QAOA, as depicted in Fig.~\ref{fig:RNN}.  More precisely, the RNN receives as input the previous QNN query's estimated cost function expectation $y_{t} \sim p(y|\bm{\theta}_t) $, where $y_t
$ is the estimate of $\braket{\hat{H}}_t$, as well as the parameters for which the QNN was evaluated $\bm{\theta}_t$. The RNN at this time step also receives information stored in its internal hidden state from the previous time step $\bm{h}_t$. The RNN itself has trainable parameters $\bm{\varphi}$, and hence it applies the parameterized mapping 
\begin{equation}
    \bm{h}_{t+1}, \bm{\theta}_{t+1} = \text{RNN}_{\bm{\varphi}}(\bm{h}_t,\bm{\theta}_{t},y_t)
\end{equation}
which generates a new suggestion for the QNN parameters as well as a new hidden state. Once this new set of QNN parameters is suggested, the RNN sends it to the QPU for evaluation and the loop continues.

The RNN is trained over random instances of QAOA problems selected from an ensemble of possible QAOA MaxCut problems. See the notebook for full details on the meta-training dataset of sampled problems.

The loss function we chose for our experiments is the \textit{observed improvement} at each time step, summed over the history of the optimization: 
\begin{equation}\label{eq:OI}
    \mathcal{L}(\bm{\varphi}) = \mathbb{E}_{f,\bm{y}}\left[\textstyle\sum\limits_{t=1}^T \text{min}\{f(\bm{\theta}_t) - \min_{j<t}[f(\bm{\theta}_j)], 0 \}\right],
\end{equation}
 The observed improvement at time step $t$ is given by the difference between the proposed value, $f(\bm{\theta}_t)$, and the best value obtained over the history of the optimization until that point, $\min_{j<t}[f(\bm{\theta}_j)]$. 

\begin{figure}
    \centering
    \includegraphics[width=0.2\textwidth]{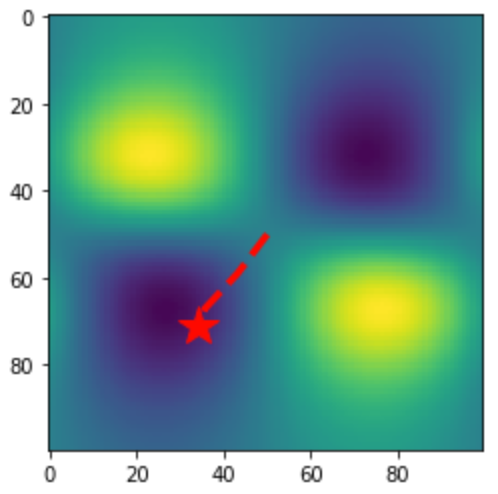}
    \caption{The path chosen by the RNN optimizer on a 12-qubit MaxCut problem after being trained on a set of random 10-qubit MaxCut problems. We see that the neural network learned to generalize its heuristic to larger system sizes, as originally pointed out in \cite{Verdon2019metalearning}.}
    \label{fig:meta_qaoa}
\end{figure}

In our particular example in this section, we will consider a time horizon of 5 time steps, hence the RNN will have to learn to very rapidly approximately optimize the parameters of the QAOA. Results are featured in Fig.~\ref{fig:meta_qaoa}. The details of the implementation are available in the Colab. Here is an overview of the problem that was tackled and the TFQ features that facilitated this implementation:

\vspace{\baselineskip}
\noindent\textbf{Target problems}:
\begin{enumerate}[noitemsep]
    \item Learning to learn with quantum neural networks via classical neural networks
    
    \item Building a neural-network-based optimizer for QAOA
    \item Lowering the number of iterations needed to optimize QAOA
\end{enumerate}
\noindent\textbf{ Required TFQ functionalities}:
      \begin{enumerate}[noitemsep]
        \item Hybrid quantum-classical networks and hybrid backpropagation
        \item Batching training over quantum data (QAOA problem instances)
        \item Integration with TF for the classical RNN
    \end{enumerate}

\vspace{1em}
\subsection{Vanishing Gradients and Adaptive Layerwise Training Strategies}
\label{sec:random_circuits}
\subsubsection{Random Quantum Circuits and Barren Plateaus}

When using parameterized quantum circuits for a learning task, inevitably one must choose an initial configuration and training strategy that is compatible with that initialization.  In contrast to problems more known structure, such as specific quantum simulation tasks~\cite{peruzzo2014variational} or optimizations~\cite{FarhiQAOA}, the structure of circuits used for learning may need to be more adaptive or general to encompass unknown data distributions.  In classical machine learning, this problem can be partially addressed by using a network with sufficient expressive power and random initialization of the network parameters.

Unfortunately, it has been proven that due to fundamental limits on quantum readout complexity in combination with the geometry of the quantum space, an exact duplication of this strategy is doomed to fail~\cite{Jarrod_QNN}.  In particular, in analog to the vanishing gradients problem that has plagued deep classical networks and historically slowed their progress~\cite{glorot2010understanding}, an exacerbated version of this problem appears in quantum circuits of sufficient depth that are randomly initialized.  This problem, also known as the problem of barren plateaus, refers to the overwhelming volume of quantum space with an exponentially small gradient, making straightforward training impossible if on enters one of these dead regions.  The rate of this vanishing increases exponentially with the number of qubits and depends on whether the cost function is global or local~\cite{cerezo2020cost}.  While strategies have been developed to deal with the challenges of vanishing gradients classically~\cite{Bengio:2012}, the combination of differences in readout complexity and other constraints of unitarity make direct implementation of these fixes challenging.  In particular, the readout of information from a quantum system has a complexity of $O(1/\epsilon^\alpha)$ where $\epsilon$ is the desired precision, and $\alpha$ is some small integer, while the complexity of the same task classically often scales as $O(\log 1/\epsilon)$ ~\cite{Knill2007Optimal}.  This means that for a vanishingly small gradient (e.g. $10^{-7}$), a classical algorithm can easily obtain at least some signal, while a quantum-classical one may diffuse essentially randomly until $\sim 10^{14}$ samples have been taken. This has fundamental consequences for the methods one uses to train, as we detail below.
The requirement on depth to reach these plateaus is only that a portion of the circuit approximates a unitary $2-$design which can occur at a depth occurring at $O(n^{1/d})$ where $n$ is the number of qubits and $d$ is the dimension of the connectivity of the quantum circuit, possibly requiring as little depth as $O(\log(n))$ in the all-to-all case~\cite{boixo2018characterizing}.  One may imagine that a solution to this problem could be to simply initialize a circuit to the identity to avoid this problem, but this incurs some subtle challenges.  First, such a fixed initialization will tend to bias results on general datasets.  This challenge has been studied in the context of more general block initialization of identity schemes~\cite{grant2019initialization}.  

Perhaps the more insidious way this problem arises, is that training with a method like stochastic gradient descent (or sophisticated variants like Adam) on the entire network, can accidentally lead one onto a barren plateau if the learning rate is too high.  This is due to the fact that the barren plateaus argument is one of volume of space and quantum-classical information exchange, and random diffusion in parameter space will tend to lead one onto a plateau.  This means that even the most clever initialization can be thwarted by the impact of this phenomenon on the training process.  In practice this severely limits learning rate and hence training efficiency of QNNs.

For this reason, one can consider training on subsets of the network which do not have the ability to completely randomize during a random walk.  This layerwise learning strategy allows one to use larger learning rates and improves training efficiency in quantum circuits~\cite{Skolik:2020}.  We advocate the use of these strategies in combination with appropriately designed local cost functions in order to circumvent the dramatically worse problems with objectives like fidelity~\cite{Jarrod_QNN,cerezo2020cost}.  TFQ has been designed to make experimenting with both of these strategies straightforward for the user, as we now document.  For an example of barren plateaus, see the notebook at the following link:
\fancylink{https://github.com/tensorflow/quantum/blob/master/docs/tutorials/barren_plateaus.ipynb}{docs/tutorials/barren\_plateaus.ipynb}

\subsubsection{Layerwise quantum circuit learning}

So far, the network training methods demonstrated in section \ref{sec:basic_app} have focused on simultaneous optimization of all network parameters. As alluded to in the section on the barren plateau effect (\ref{sec:random_circuits}), this type of strategy can lead to vanishing gradients as the number of qubits and layers in a random circuit grows. A parameter initialization strategy to avoid this effect at the initial training steps has been proposed in \cite{grant2019initialization}. This strategy initializes layers in a blockwise fashion such that pairs of layers are randomly initialized but yield an identity operation when acting on the circuit and thereby prevent initialization on a plateau. However, as described in section \ref{sec:random_circuits}, this does not avert the possibility for the circuit to drift onto a plateau during training. In this section, we will implement a layerwise training strategy that avoids initialization on a plateau as well as drifting onto a plateau during training by only training subsets of parameters in the circuit at a given update step \cite{Skolik:2020}.

The strategy consists of two training phases. In \textit{phase one}, the algorithm constructs the circuit by subsequently adding and training layers of a predefined structure. We start by picking a number of initial layers $s$, that contains parameters which are always active during training to avoid entering a regime where the number of active parameters is too small to decrease the loss \cite{campos2021abrupt}. These layers are then trained for a fixed number of iterations $e_l$, after which another set of layers is added to the circuit. How many layers this set contains is controlled by a hyperparameter $p$. Another hyperparameter $q$ determines after how many layers the parameters in previous layers are frozen. I.e., for $p=2$ and $q=4$, we add two layers at intervals of $e_l$ iterations, and only train the parameters in the last four layers, while all other parameters in the circuit are kept fixed. This procedure is repeated until a fixed, predefined circuit depth is reached.

In \textit{phase two}, we take the final circuit from phase one and now split the parameters into larger partitions. A hyperparameter $r$ specifies the percentage of parameters which are trained in the circuit simultaneously, i.e., for $r=0.5$ we split the circuit in two halves and alternate between training each of these two halves where the other half's parameters are kept fixed. By this approach, any randomization effect that occurs during training is contained to a subset of the circuit parameters and effectively prevents drifting onto a plateau, as can be seen in the original paper for the case of randomization induced by sampling noise \cite{Skolik:2020}. The size of these partitions has to be chosen with care, as overly large partitions will increase the probability of randomization, while overly small partitions may be insufficient to decrease the loss \cite{campos2021abrupt}. This alternate training of circuit partitions is then performed until the loss converges or we reach a predefined number of iterations.

In the following, we will implement phase one of the algorithm, and a complete implementation of both phases can be found in the accompanying notebook.\\

\vspace{\baselineskip}
\noindent\textbf{Target problems}:
\begin{enumerate}[noitemsep]
    \item Dynamically building circuits for arbitrary learning tasks
    \item Manipulating circuit structure and parameters during training
    \item Reducing the number of trained parameters
    \item Avoiding initialization on or drifting to a barren plateau
\end{enumerate}
\noindent\textbf{ Required TFQ functionalities}:
      \begin{enumerate}[noitemsep]
        \item Parameterized circuit layers
        \item Keras weight manipulation interface
        \item Parameter shift differentiator for exact gradient computation
    \end{enumerate}

To run this example in the browser through Colab, follow the link:
\fancylink{https://github.com/tensorflow/quantum/blob/research/layerwise_learning/layerwise_learning.ipynb}{research/layerwise\_learning/layerwise\_learning.ipynb}

As an example to show how this functionality may be explored in TFQ, we will look at randomly generated layers as shown in section \ref{sec:random_circuits}, where one layer consists of a randomly chosen rotation gate around the $X$, $Y$, or $Z$ axis on each qubit, followed by a ladder of $CZ$ gates over all qubits.

\begin{lstlisting}
def create_layer(qubits, layer_id):
    # create symbols for trainable parameters
    symbols = [
        sympy.Symbol(
            f'{layer_id}-{str(i)}') 
        for i in range(len(qubits))]
        
    # build layer from random gates
    gates = [
        random.choice([
            cirq.Rx, cirq.Ry, cirq.Rz])(
            symbols[i])(q) 
        for i, q in enumerate(qubits)]
        
    # add connections between qubits
    for control, target in zip(
        qubits, qubits[1:]):
        gates.append(cirq.CZ(control, target))
    return gates, symbols
\end{lstlisting}

We assume that we don't know the ideal circuit structure to solve our learning problem, so we start with the shallowest circuit possible and let our model grow from there. In this case we start with one initial layer $s=1$, and add a new layer after it has trained for $e_l=10$ epochs. First, we need to specify some variables:

\begin{lstlisting}
# number of qubits and layers in our circuit
n_qubits = 6
n_layers = 8

# define data and readout qubits
data_qubits = cirq.GridQubit.rect(1, n_qubits)
readout = cirq.GridQubit(0, n_qubits-1)
readout_op = cirq.Z(readout)

# symbols to parametrize circuit
symbols = []
layers = []
weights = []
\end{lstlisting}

We use the same training data as specified in the TFQ MNIST classifier example notebook available in the TFQ Github repository, which encodes a downsampled version of the digits into binary vectors. Ones in these vectors are encoded as local $X$ gates on the corresponding qubit in the register, as shown in \cite{farhi2018classification}. For this reason, we also use the readout procedure specified in that work where a sequence of $XHX$ gates is added to the readout qubit at the end of the circuit.  Now we train the circuit, layer by layer:

\begin{lstlisting}
for layer_id in range(n_layers):
    circuit = cirq.Circuit()
    layer, layer_symbols = create_layer(
        data_qubits, f'layer_{layer_id}')
    layers.append(layer)
    
    circuit += layers
    symbols += layer_symbols
    
    # set up the readout qubit
    circuit.append(cirq.X(readout))
    circuit.append(cirq.H(readout))
    circuit.append(cirq.X(readout))
    readout_op = cirq.Z(readout)
    
    # create the Keras model
    model = tf.keras.Sequential()
    model.add(tf.keras.layers.Input(
        shape=(), dtype=tf.dtypes.string))
    model.add(tfq.layers.PQC(
        model_circuit=circuit,
        operators=readout_op,
        differentiator=ParameterShift(),
        initializer=tf.keras.initializers.Zeros))
    model.compile(
        loss=tf.keras.losses.squared_hinge,
        optimizer=tf.keras.optimizers.Adam(
            learning_rate=0.01))
        
    # Update model parameters and add
    # new 0 parameters for new layers.
    model.set_weights(
        [np.pad(weights, (n_qubits, 0))])
    model.fit(x_train,
              y_train,
              batch_size=128,
              epochs=10,
              verbose=1,
              validation_data=(x_test, y_test))

    qnn_results = model.evaluate(x_test, y_test)
    
    # store weights after training a layer
    weights = model.get_weights()[0]
\end{lstlisting}

In general, one can choose many different configurations of how many layers should be trained in each step. One can also control which layers are trained by manipulating the symbols we feed into the circuit and keeping track of the weights of previous layers. The number of layers, layers trained at a time, epochs spent on a layer, and learning rate are all hyperparameters whose optimal values depend on both the data and structure of the circuits being used for learning.  This example is meant to exemplify how TFQ can be used to easily explore these choices to maximize the efficiency and efficacy of training. See our notebook linked above for the complete implementation of these features. Using TFQ to explore this type of learning strategy relieves us of manually implementing training procedures and optimizers, and autodifferentiation with the parameter shift rule. It also lets us readily use the rich functionality provided by TensorFlow and Keras. Implementing and testing all of the functionality needed in this project by hand could take up to a week, whereas all this effort reduces to a couple of lines of code with TFQ as shown in the notebook. Additionally, it lets us speed up training by using the integrated qsim simulator as shown in section \ref{sec:benchmarks}. Last but not least, TFQ provides a thoroughly tested and maintained QML framework which greatly enhances the reproducibility of our research.

\subsection{Hamiltonian Learning with Quantum Graph Recurrent Neural Networks}

\subsubsection{Motivation: Learning Quantum Dynamics with a Quantum Representation}

Quantum simulation of time evolution was one of the original envisioned applications of quantum computers when they were first proposed by Feynman \cite{Feynman1982}. Since then, quantum time evolution simulation methods have seen several waves of great progress, from the early days of Trotter-Suzuki methods, to methods of qubitization and randomized compiling \cite{Campbell_2019}, and finally recently with some methods for quantum variational methods for approximate time evolution \cite{cirstoiu2019variational}.

The reason that quantum simulation has been such a focus of the quantum computing community is because we have some indications to believe that quantum computers can demonstrate a quantum advantage when evolving quantum states through unitary time evolution; the classical simulation overhead scales exponentially with the depth of the time evolution.

As such, it is natural to consider if such a potential quantum simulation advantage can be extended to the realm of quantum machine learning as an inverse problem, that is, given access to some black-box dynamics, can we learn a Hamiltonian such that time evolution under this Hamiltonian replicates the unknown dynamics. This is known as the problem of Hamiltonian learning, or quantum dynamics learning, which has been studied in the literature \cite{Wiebe_2014,Carolan2020}. Here, we use a Quantum Neural Network-based approach to learn the Hamiltonian of a quantum dynamical process, given access to quantum states at various time steps.

As was pointed out in the barren plateaus section \ref{sec:random_circuits}, attempting to do QML with no prior on the physics of the system or no imposed structure of the ansatz hits the quantum version of the no free lunch theorem; the network has too high a capacity for the problem at hand and is thus hard to train, as evidenced by its vanishing gradients. Here, instead, we use a highly structured ansatz, from work featured in \cite{verdon2019quantumgraph}. First of all, given that we know we are trying to replicate quantum dynamics, we can structure our ansatz to be based on Trotter-Suzuki evolution \cite{suzuki1990fractal} of a learnable parameterized Hamiltonian. This effectively performs a form of parameter-tying in our ansatz between several layers representing time evolution. In a previous example on quantum convolutional networks \ref{sec:HQCNN_bg}, we performed parameter tying for spatial translation invariance, whereas here, we will assume the dynamics remain constant through time, and perform parameter tying across time, hence it is akin to a quantum form of recurrent neural networks (RNN). More precisely, as it is a parameterization of a Hamiltonian evolution, it is akin to a quantum form of recently proposed models in classical machine learning called Hamiltonian neural networks \cite{greydanus2019hamiltonian}.

Beyond the quantum RNN form, we can impose further structure. We can consider a scenario where we know we have a one-dimensional quantum many-body system. As Hamiltonians of physical have local couplings, we can use our prior assumptions of locality in the Hamiltonian and encode this as a graph-based parameterization of the Hamiltonian. As we will see below, by using a Quantum Graph Recurrent Neural network \cite{verdon2019quantumgraph} implemented in TFQ, we will be able to learn the effective Hamiltonian topology and coupling strengths quite accurately, simply from having access to quantum states at different times and employing mini-batched gradient-based training. 

Before we proceed, it is worth mentioning that the approach featured in this section is quite different from the learning of quantum dynamics using a classical RNN feature in previous example section \ref{sec:qctrl}. As sampling the output of a quantum simulation at different times can become exponentially hard, we can imagine that for large systems, the Quantum RNN dynamics learning approach could have primacy over the classical RNN approach, thus potentially demonstrating a quantum advantage of QML over classical ML for this problem.

\vspace{\baselineskip}
\noindent\textbf{Target problems}:
\begin{enumerate}[noitemsep]
    \item Preparing low-energy states of a quantum system
    \item Learning Quantum Dynamics using a Quantum Neural Network Model
\end{enumerate}
\noindent\textbf{ Required TFQ functionalities}:
      \begin{enumerate}[noitemsep]
        \item Quantum compilation of exponentials of Hamiltonians
        \item Training multi-layered quantum neural networks with shared parameters
        \item Batching QNN training data (input-output pairs and time steps) for supervised learning of quantum unitary map
\end{enumerate}

\subsubsection{Implementation}

Please see the tutorial notebook for full code details:
\fancylink{https://github.com/tensorflow/quantum/blob/research/qgrnn_ising/qgrnn_ising.ipynb}{research/qgrnn\_ising/qgrnn\_ising.ipynb}

Here we provide an overview of our implementation.  We can define a general Quantum Graph Neural Network as a repeating sequence of exponentials of a Hamiltonian defined on a graph, \(\hat{U}_{\textsc{qgnn}}(\mbox{\boldmath$\eta$}, \mbox{\boldmath$\theta$}) = \prod_{p=1}^{P}\big[\prod_{q=1}^{Q}e^{-i\eta_{pq}\hat{H}_q(\mbox{\boldmath$\theta$})}\big]\) where the $\hat{H}_{q}(\mbox{\boldmath$\theta$})$ are generally 2-local Hamiltonians whose coupling topology is that of an assumed graph structure.

In our Hamiltonian learning problem, we aim to learn a target 
$\hat{H}_{\text{target}}$ which will be an Ising model Hamiltonian with $J_{jk}$ as couplings and $B_{v}$ for site bias term of each spin, i.e.,  $\hat{H}_{\text{target}} = \sum_{j,k} J_{jk} \hat{Z}_j \hat{Z}_k + \sum_{v} B_v \hat{Z}_v + \sum_{v} \hat{X}_v$, given access to pairs of states at different times that were subjected to the target time evolution operator $\hat{U}(T) = e^{-i\hat{H}_{\text{target}}T}$.

We will use a recurrent form of QGNN, using Hamiltonian generators $\hat{H}_1(\bm{\theta}) = \sum_{v\in\mathcal{V}} \alpha_v \hat{X}_v$ and $\hat{H}_2(\bm{\theta}) = \sum_{\{j,k\}\in\mathcal{E}} \theta_{jk} \hat{Z}_j \hat{Z}_k + \sum_{v\in\mathcal{V}} \phi_v \hat{Z}_v$, with trainable parameters\footnote{ For simplicity we set $\alpha_{v}$ to constant 1's in this example.} \{$\theta_{jk},\phi_{v},\alpha_{v}\}$, for our choice of graph structure prior $\mathcal{G} = \{\mathcal{V},\mathcal{E}\}$. The QGRNN is then resembles applying a Trotterized time evolution of a parameterized Ising Hamiltonian $\hat{H}(\bm{\theta})  = \hat{H}_1(\bm{\theta}) + \hat{H}_2(\bm{\theta}) $ where $P$ is the number of Trotter steps. This is a good parameterization to learn the effective Hamiltonian of the black-box dynamics as we know from quantum simulation theory that Trotterized time evolution operators can closely approximate true dynamics in the limit of $|\eta_{jk}|\rightarrow0$ while $P\rightarrow\infty$.

For our TFQ software implementation, we can initialize Ising model \& QGRNN model parameters as random values on a graph. It is very easy to construct this kind of graph structure Hamiltonian by using Python NetworkX library.
\begin{lstlisting}
N = 6
dt = 0.01
# Target Ising model parameters
G_ising = nx.cycle_graph(N)
ising_w = [dt * np.random.random() for _ in G.edges]
ising_b = [dt * np.random.random() for _ in G.nodes]
\end{lstlisting}
Because the target Hamiltonian and its nearest-neighbor graph structure is unknown to the QGRNN, we need to initialize a new random graph prior for our QGRNN. In this example we will use a random 4-regular graph with a cycle as our prior. Here, \Colorbox{bkgd}{\lstinline{params}} is a list of trainable parameters of the QGRNN. 

\begin{lstlisting}
# QGRNN model parameters
G_qgrnn = nx.random_regular_graph(n=N, d=4)
qgrnn_w = [dt] * len(G_qgrnn.edges)
qgrnn_b = [dt] * len(G_qgrnn.nodes)
theta = ['theta{}'.format(e) for e in G.edges]
phi = ['phi{}'.format(v) for v in G.nodes]
params = theta + phi
\end{lstlisting}

Now that we have the graph structure, weights of edges \& nodes, we can construct Cirq-based Hamiltonian operator which can be directly calculated in Cirq and TFQ. To create a Hamiltonian by using \Colorbox{bkgd}{\lstinline{cirq.PauliSum}}'s or \Colorbox{bkgd}{\lstinline{cirq.PauliString}}'s, we need to assign appropriate qubits on them. Let's assume \Colorbox{bkgd}{\lstinline{Hamiltonian()}} is the Hamiltonian preparation function to generate cost Hamiltonian from interaction weights and mixer Hamiltonian from bias terms. We can bring qubits of Ising \& QGRNN models by using \Colorbox{bkgd}{\lstinline{cirq.GridQubit}}.

\begin{lstlisting}
qubits_ising = cirq.GridQubit.rect(1, N)
qubits_qgrnn = cirq.GridQubit.rect(1, N, 0, N)
ising_cost, ising_mixer = Hamiltonian(
    G_ising, ising_w, ising_b, qubits_ising)
qgrnn_cost, qgrnn_mixer = Hamiltonian(
    G_qgrnn, qgrnn_w, qgrnn_b, qubits_qgrnn)
\end{lstlisting}

To train the QGRNN, we need to create an ensemble of states which are to be subjected to the unknown dynamics. We chose to prepare a low-energy states by first performing a Variational Quantum Eigensolver (VQE) \cite{mcclean2016theory} optimization to obtain an approximate ground state. Following this, we can apply different amounts of simulated time evolutions onto to this state to obtain a varied dataset. This emulates having a physical system in a low-energy state and randomly picking the state at different times.
First things first, let us build a VQE model

\begin{lstlisting}
def VQE(H_target, q)
  # Parameters
  x = ['x{}'.format(i) for i, _ in enumerate(q)]
  z = ['z{}'.format(i) for i, _ in enumerate(q)]
  symbols = x + z
  circuit = cirq.Circuit() 
  circuit.append(cirq.X(q_)**sympy.Symbol(x_) for q_, x_ in zip(q, x))
  circuit.append(cirq.Z(q_)**sympy.Symbol(z_) for q_, z_ in zip(q, z))
\end{lstlisting}  

Now that we have a parameterized quantum circuit, we can minimize the expectation value of given Hamiltonian. Again, we can construct a Keras model with \Colorbox{bkgd}{\lstinline{Expectation}}. Because the output expectation values are calculated respectively, we need to sum them up at the last.

\begin{lstlisting}
  circuit_input = tf.keras.Input(
      shape=(), dtype=tf.string)
  output = tfq.layers.Expectation()(
      circuit_input,
      symbol_names=symbols,
      operators=tfq.convert_to_tensor(
          [H_target]))
  output = tf.math.reduce_sum(
      output, axis=-1, keepdims=True)
\end{lstlisting}  

Finally, we can get approximated lowest energy states of the VQE model by compiling and training the above Keras model.\footnote{ Here is some tip for training. Setting the output true value to theoretical lower bound, we can minimize our expectation value in the Keras model fit framework. That is, we can use the inequality $\langle \hat{H}_{\text{target}} \rangle = \sum_{jk}J_{jk}\langle Z_jZ_k\rangle + \sum_{v}B_{v}\langle Z_v\rangle + \sum_{v}\langle X_v\rangle \ge \sum_{jk}(-)|J_{jk}| -\sum_{v}|B_{v}| - N $.}

\begin{lstlisting}
  model = tf.keras.Model(
      inputs=circuit_input, outputs=output)
  adam = tf.keras.optimizers.Adam(
      learning_rate=0.05)
  
  low_bound = -np.sum(np.abs(ising_w + ising_b)) - N
  inputs = tfq.convert_to_tensor([circuit])
  outputs = tf.convert_to_tensor([[low_bound]])
  model.compile(optimizer=adam, loss='mse')
  model.fit(x=inputs, y=outputs,
            batch_size=1, epochs=100)
  params = model.get_weights()[0]
  res = {k: v for k, v in zip(symbols, params)}
  return cirq.resolve_parameters(circuit, res)
\end{lstlisting}

Now that the VQE function is built, we can generate the initial quantum data input with the low energy states near to the ground state of the target Hamiltonian for both our data and our input state to our QGRNN.

\begin{lstlisting}
H_target = ising_cost + ising_mixer
low_energy_ising = VQE(H_target, qubits_ising)
low_energy_qgrnn = VQE(H_target, qubits_qgrnn)
\end{lstlisting}

The QGRNN is fed the same input data as the true process. We will use gradient-based training over minibatches of randomized timesteps chosen for our QGRNN and the target quantum evolution. We will thus need to aggregate the results among the different timestep evolutions to train the QGRNN model. To create these time evolution exponentials, we can use the \Colorbox{bkgd}{\lstinline{tfq.util.exponential}} function to exponentiate our target and QGRNN Hamiltonians\footnote{Here, we use the terminology \textit{cost} and \textit{mixer} Hamiltonians as the Trotterization of an Ising model time evolution is very similar to a QAOA, and thus we borrow nomenclature from this analogous QNN.}

\begin{lstlisting}
exp_ising_cost = tfq.util.exponential(
    operators=ising_cost)
exp_ising_mix = tfq.util.exponential(
    operators=ising_mixer)
exp_qgrnn_cost = tfq.util.exponential(
    operators=qgrnn_cost, coefficients=params)
exp_qgrnn_mix = tfq.util.exponential(
    operators=qgrnn_mixer)
\end{lstlisting}

Here we randomly pick the 15 timesteps and apply the Trotterized time evolution operators using our above constructed exponentials. We can have a quantum dataset $\{(|\psi_{T_j}\rangle, |\phi_{T_j}\rangle) | j = 1..M\}$ where $M$ is the number of data, or batch size (in our case we chose $M=15$), $|\psi_{T_j}\rangle=\hat{U}^j_{\text{target}}|\psi_0\rangle$ and $|\phi_{T_j}\rangle=\hat{U}^j_{\textsc{qgrnn}}|\psi_0\rangle$.

\begin{lstlisting}
def trotterize(inp, depth, cost, mix):
    add = tfq.layers.AddCircuit()
    outp = add(cirq.Circuit(), append=inp)
    for _ in range(depth):
        outp = add(outp, append=cost)
        outp = add(outp, append=mix)
    return outp

batch_size = 15
T = np.random.uniform(0, T_max, batch_size)
depth = [int(t/dt)+1 for t in T]
true_states = []
pred_states = []
for P in depth:
    true_states.append(
        trotterize(low_energy_ising, P,
            exp_ising_cost, exp_ising_mix))
    pred_states.append(
        trotterize(low_energy_qgrnn, P,
            exp_qgrnn_cost, exp_qgrnn_mix))
\end{lstlisting}

Now we have both quantum data from (1) the true time evolution of the target Ising model and (2) the predicted data state from the QGRNN.
In order to maximize overlap between these two wavefunctions, we can aim to maximize the fidelity between the true state and the state output by the QGRNN, averaged over random choices of time evolution.
To evaluate the fidelity between two quantum states (say $\ket{A}$ and $\ket{B}$) on a quantum computer, a well-known approach is to perform the \textit{swap test} \cite{Cincio_2018}. In the swap test, an additional observer qubit is used, by putting this qubit in a superposition and using it as control for a Fredkin gate (controlled-SWAP), followed by a Hadamard on the observer qubit, the observer qubit's expectation value in the encodes the fidelity of the two states,  $|\braket{A|B}|^2$. Thus, right after Fidelity Swap Test, we can measure the swap test qubit with Pauli $\hat{Z}$ operator with \Colorbox{bkgd}{\lstinline{Expectation}}, $\langle\hat{Z}_{test}\rangle$, and then we can calculate the average of fidelity (inner product) between a batch of two sets of quantum data states, which can be used as our classical loss function in TensorFlow.

\begin{lstlisting}
# Check class SwapTestFidelity in the notebook.
fidelity = SwapTestFidelity(
    qubits_ising, qubits_qgrnn, batch_size)

state_true = tf.keras.Input(shape=(),
                            dtype=tf.string)
state_pred = tf.keras.Input(shape=(),
                            dtype=tf.string)
fid_output = fidelity(state_true, state_pred)
fid_output = tfq.layers.Expectation()(
    fid_output,
    symbol_names=symbols,
    operators=fidelity.op)

model = tf.keras.Model(
    inputs=[state_true, state_pred],
    outputs=fid_output)
\end{lstlisting}

Here, we introduce the average fidelity and implement this with custom Keras loss function.

$\begin{matrix}L(\theta, \phi) & = & 1 - \frac{1}{B} \sum^{B}_{j=1} |\langle \psi_{T_j} | \phi_{T_j}\rangle|^2 \\ & = &1 - \frac{1}{B} \sum^{B}_{j=1} \langle \hat{Z}_{test} \rangle_j\end{matrix}$

\begin{lstlisting}
def average_fidelity(y_true, y_pred):
    return 1 - K.mean(y_pred)
\end{lstlisting}

Again, we can use Keras model fit. To feed a batch of quantum data, we can use \Colorbox{bkgd}{\lstinline{tf.concat}} because the quantum circuits are already in \Colorbox{bkgd}{\lstinline{tf.Tensor}}. In this case, we know that the lower bound of fidelity is 0, but the \Colorbox{bkgd}{\lstinline{y_true}} is not used in our custom loss function \Colorbox{bkgd}{\lstinline{average_fidelity}}. We set learning rate of Adam optimizer to $0.05$.

\begin{lstlisting}
y_true = tf.concat(true_states, axis=0)
y_pred = tf.concat(pred_states, axis=0)

model.compile(
    loss=average_fidelity,
    optimizer=tf.keras.optimizers.Adam(
        learning_rate=0.05))
model.fit(x=[y_true, y_pred],
          y=tf.zeros([batch_size, 1]),
          batch_size=batch_size,
          epochs=500)
\end{lstlisting}

The full results are displayed in the notebook, we see for this example that our time-randomized gradient-based optimization of our parameterized class of quantum Hamiltonian evolution ends up learning the target Hamiltonian and its couplings to a high degree of accuracy.

\begin{figure}[H]
    \centering
    \includegraphics[width=0.32\columnwidth]{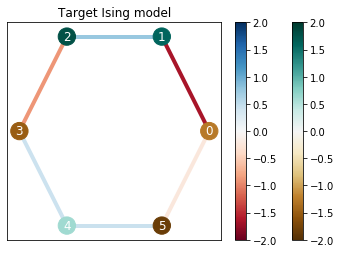}    \includegraphics[width=0.32\columnwidth]{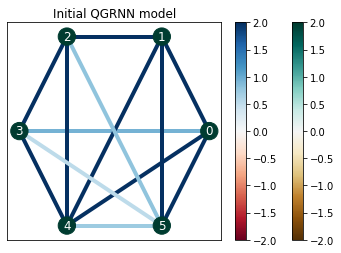}    \includegraphics[width=0.32\columnwidth]{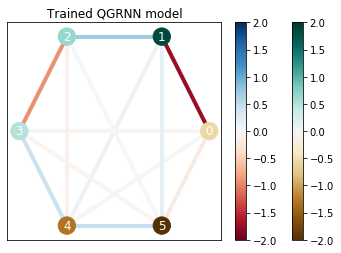}
\caption{Left: True (target) Ising Hamiltonian with edges representing couplings and nodes representing biases. Middle: randomly chosen initial graph structure and parameter values for the QGRNN. Right: learned Hamiltonian from the trained QGRNN.}
    \label{fig:qgrnn_result}
\end{figure}

\subsection{Generative Modelling of Quantum Mixed States with Hybrid Quantum-Probabilistic Models}
\subsubsection{Background}
Often in quantum mechanical systems, one encounters so-called \textit{mixed states}, which can be understood as probabilistic mixtures over pure quantum states \cite{MichaelIsaacQC}. Typical cases where such mixed states arise are when looking at finite-temperature quantum systems, open quantum systems, and subsystems of pure quantum mechanical systems. As the ability to model mixed states are thus key to understanding quantum mechanical systems, in this section, we focus on models to learn to represent and mimic the statistics of quantum mixed states.  

As mixed states are a combination of a classical probability distribution and quantum wavefunctions, their statistics can exhibit both classical and quantum forms of correlations (e.g., entanglement). As such, if we wish to learn a representation of such mixed state which can generatively model its statistics, one can expect that a hybrid representation combining classical probabilistic models and quantum neural networks can be an ideal. Such a decomposition is ideal for near-term noisy devices, as it reduces the overhead of representation on the quantum device, leading to lower-depth quantum neural networks. Furthermore, the quantum layers provide a valuable addition in representation power to the classical probabilistic model, as they allow the addition of quantum correlations to the model. 

Thus, in this section, we cover some examples where one learns to generatively model mixed states using a hybrid quantum-probabilistic model \cite{verdon2019quantum}.  
Such models use a parameterized ansatz of the form
\begin{align}\label{eq:qhbm_model}
\hat{\rho}_{\bm{\theta\phi}} = \hat{U}(\bm{\phi}) \hat{\rho}_{\bm{\theta}}\hat{U}^\dagger(\bm{\phi}), \quad  \hat{\rho}_{\bm{\theta}}=\sum_{\bm{x}} p_{\bm{\theta}}(\bm{x})\ket{\bm{x}}\!\bra{\bm{x}}
\end{align}
where $\hat{U}(\bm{\phi})$ is a unitary quantum neural network with parameters $\bm{\phi}$ and $p_{\bm{\theta}}(\bm{x})$ is a classical probabilistic model with parameters $\bm{\theta}$. We call  $\hat{\rho}_{\bm{\theta\phi}}$ the \textit{visible} state and $\hat{\rho}_{\bm{\theta}}$ the \textit{latent} state. Note the latent state is effectively a classical distribution over the standard basis states, and its only parameters are those of the classical probabilistic model.

As we shall see below, there are methods to train both networks simultaneously. In terms of software implementation, as we have to combine probabilistic models and quantum neural networks, we will use a combination of TensorFlow Probability \cite{dillon2017tensorflow} along with TFQ. A first class of application we will consider is the task of generating a thermal state of a quantum system given its Hamiltonian. A second set of applications is given several copies of a mixed state, learn a generative model which replicates the statistics of the state.

\vspace{\baselineskip}
\noindent\textbf{Target problems}:
\begin{enumerate}[noitemsep]

    \item Incorporating probabilistic and quantum models
    \item Variational Quantum Simulation of Quantum Thermal States
    \item Learning to generatively model mixed states from data
\end{enumerate}
\noindent\textbf{ Required TFQ functionalities}:
      \begin{enumerate}[noitemsep]
        \item Integration with TF Probability \cite{dillon2017tensorflow}
        \item Sample-based simulation of quantum circuits
        \item Parameter shift differentiator for gradient computation
    \end{enumerate}

\subsubsection{Variational Quantum Thermalizer}

Full notebook of the implementations below are available at: 
\fancylink{https://github.com/tensorflow/quantum/blob/research/vqt_qmhl/vqt_qmhl.ipynb}{research/vqt\_qmhl/vqt\_qmhl.ipynb}

Consider the task of preparing a thermal state: given a Hamiltonian $\hat{H}$ and a target inverse temperature $\beta = 1/T$, we want to variationally approximate the state
\begin{align}
\hat{\sigma}_{\beta} = \tfrac{1}{\mathcal{Z}_{\beta}}e^{-\beta \hat{H}},\quad  \mathcal{Z}_\beta = \text{tr}(e^{-\beta \hat{H}}),
\end{align}
using a state of the form presented in equation \eqref{eq:qhbm_model}. That is, we aim to find a value of the hybrid model parameters $\{\bm{\theta}^*,\bm{\phi}^*\}$ such that $\hat{\rho}_{\bm{\theta}^*\bm{\phi}^*} \approx \hat{\sigma}_{\beta}$. In order to converge to this approximation via optimization of the parameters, we need a loss function to optimize which quantifies statistical distance between these quantum mixed states. If we aim to minimize the discrepancy between states in terms of quantum relative entropy $D(\hat{\rho}_{\bm{\theta\phi}}\Vert \hat{\sigma}_\beta) = -S(\hat{\rho}_{\bm{\theta\phi}}) - \text{tr}(\hat{\rho}_{\bm{\theta\phi}}\log \hat{\sigma}_\beta)$, (where $ S(\hat{\rho}_{\bm{\theta\phi}}) = -\tr(\hat{\rho}_{\bm{\theta\phi}}\log \hat{\rho}_{\bm{\theta\phi}})$ is the entropy), then, as described in the full paper \cite{verdon2019quantumVQT} we can equivalently minimize the free energy\footnote{More precisely, the loss function here is in fact the inverse temperature multiplied by the free energy, but this detail is of little import to our optimization.}, and hence use it as our loss function:
\begin{equation}
    \mathcal{L}_{\textsc{fe}}(\bm{\theta},\bm{\phi}) =\beta \text{tr}(\hat{\rho}_{\bm{\theta\phi}}\hat{H}) -S(\hat{\rho}_{\bm{\theta\phi}}).  
\end{equation}
The first term is simply the expectation value of the energy of our model, while the second term is the entropy. Due to the structure of our quantum-probabilistic model, the entropy of the visible state is equal to the entropy of the latent state, which is simply the classical entropy of the distribution, $S(\hat{\rho}_{\bm{\theta\phi}})= S(\hat{\rho}_{\bm{\theta}}) = - \sum_{\bm{x}} p_{\bm{\theta}}(\bm{x})\log p_{\bm{\theta}}(\bm{x})$. This comes in quite useful during the optimization of our model.

Let us implement a simple example of the VQT model which minimizes free energy to achieve an approximation of the thermal state of a physical system. Let us consider a two-dimensional Heisenberg spin chain 
\begin{align}
    \hat{H}_{\textsc{heis}} = &\sum_{\langle ij\rangle_h} J_{h} \hat{\bm{S}}_i \cdot \hat{\bm{S}}_j +  \sum_{\langle ij\rangle_v} J_{v} \hat{\bm{S}}_i \cdot \hat{\bm{S}}_j
\end{align}
where $h$ ($v$) denote horizontal (vertical) bonds, while $\langle \cdot \rangle $ represent nearest-neighbor pairings. First, we define this Hamiltonian on a grid of qubits:

\begin{lstlisting}
def get_bond(q0, q1):
  return cirq.PauliSum.from_pauli_strings([
    cirq.PauliString(cirq.X(q0), cirq.X(q1)),
    cirq.PauliString(cirq.Y(q0), cirq.Y(q1)),
    cirq.PauliString(cirq.Z(q0), cirq.Z(q1))])

def get_heisenberg_hamiltonian(qubits, jh, jv):
  heisenberg = cirq.PauliSum()
  # Apply horizontal bonds
  for r in qubits:
    for q0, q1 in zip(r, r[1::]):
      heisenberg += jh * get_bond(q0, q1)
  # Apply vertical bonds
  for r0, r1 in zip(qubits, qubits[1::]):
    for q0, q1 in zip(r0, r1):
      heisenberg += jv * get_bond(q0, q1)
  return heisenberg
\end{lstlisting}
For our QNN, we consider a unitary consisting of general single qubit rotations and powers of controlled-not gates.  Our code returns the associated symbols so that these can be fed into the \Colorbox{bkgd}{\lstinline{Expectation}} op: \\

\begin{lstlisting}
def get_rotation_1q(q, a, b, c):
  return cirq.Circuit(
    cirq.X(q)**a, cirq.Y(q)**b, cirq.Z(q)**c)

def get_rotation_2q(q0, q1, a):
  return cirq.Circuit(
    cirq.CNotPowGate(exponent=a)(q0, q1))

def get_layer_1q(qubits, layer_num, L_name):
  layer_symbols = []
  circuit = cirq.Circuit()
  for n, q in enumerate(qubits):
    a, b, c = sympy.symbols(
        "a{2}_{0}_{1} b{2}_{0}_{1} c{2}_{0}_{1}".format(layer_num, n, L_name))
    layer_symbols += [a, b, c]
    circuit += get_rotation_1q(q, a, b, c)
  return circuit, layer_symbols

def get_layer_2q(qubits, layer_num, L_name):
  layer_symbols = []
  circuit = cirq.Circuit()
  for n, (q0, q1) in enumerate(zip(qubits[::2], qubits[1::2])):
    a = sympy.symbols("a{2}_{0}_{1}".format(layer_num, n, L_name))
    layer_symbols += [a]
    circuit += get_rotation_2q(q0, q1, a)
  return circuit, layer_symbols
\end{lstlisting}

It will be convenient to consider a particular class of probabilistic models where the estimation of the gradient of the model parameters is straightforward to perform. This class of models are called \textit{exponential families} or \textit{energy-based models} (EBMs). If our parameterized probabilistic model is an EBM, then it is of the form: 
\begin{equation}\label{eq:EBM}
p_{\bm{\theta}}(x) = \tfrac{1}{\mathcal{Z}_{\bm{\theta}}} e^{-E_{\bm{\theta}}(\bm{x})}, \quad \mathcal{Z}_{\bm{\theta}} \equiv\textstyle \sum_{\bm{x}\in\Omega}e^{-E_{\bm{\theta}}(\bm{x})}.\end{equation} 

For gradients of the VQT free energy loss function with respect to the QNN parameters,
\(\partial_{\bm{\phi}} \mathcal{L}_{\textsc{fe}}(\bm{\theta},\bm{\phi}) =\beta\partial_{\bm{\phi}} \text{tr}(\hat{\rho}_{\bm{\theta\phi}}\hat{H})\), this is simply the gradient of an expectation value, hence we can use TFQ parameter shift gradients or any other method for estimating the gradients of QNN's outlined in previous sections.  

As for gradients of the classical probabilistic model, one can readily derive that they are given by the following covariance:
\spliteq{\partial_{\bm{\theta}} &\mathcal{L}_{\textsc{fe}} \!=\!\mathbb{E}_{\bm{x}\sim p_{\bm{\theta}}(\bm{x})}\Big[(E_{\bm{\theta}}(\bm{x})-\beta H_{\bm{\phi}}(\bm{x}) ) \nabla_{\bm{\theta}}E_{\bm{\theta}}(\bm{x}) \Big]\\&\!\!\!-\!
( \mathbb{E}_{\bm{x}\sim p_{\bm{\theta}}(\bm{x})}\big[E_{\bm{\theta}}(\bm{x})\!-\!\beta H_{\bm{\phi}}(\bm{x}) \big]) ( \mathbb{E}_{\bm{y}\sim p_{\bm{\theta}}(\bm{y})}\big[\nabla_{\bm{\theta}}E_{\bm{\theta}}(\bm{y})\big] )   ,}\label{eq:fe-classical-gradient}
where $H_{\bm{\phi}}(\bm{x}) \equiv \bra{\bm{x}}\hat{U}^\dagger({\bm{\phi}})\hat{H}\hat{U}({\bm{\phi}})\ket{\bm{x}}$ is the expectation value of the Hamiltonian at the output of the QNN with the standard basis element $\ket{\bm{x}}$ as input. Since the energy function and its gradients can easily be evaluated as it is a neural network, the above gradient is straightforward to estimate via sampling of the classical probabilistic model and the output of the QPU. 

\begin{figure}[H]
    \centering
    \includegraphics[width=0.65\columnwidth]{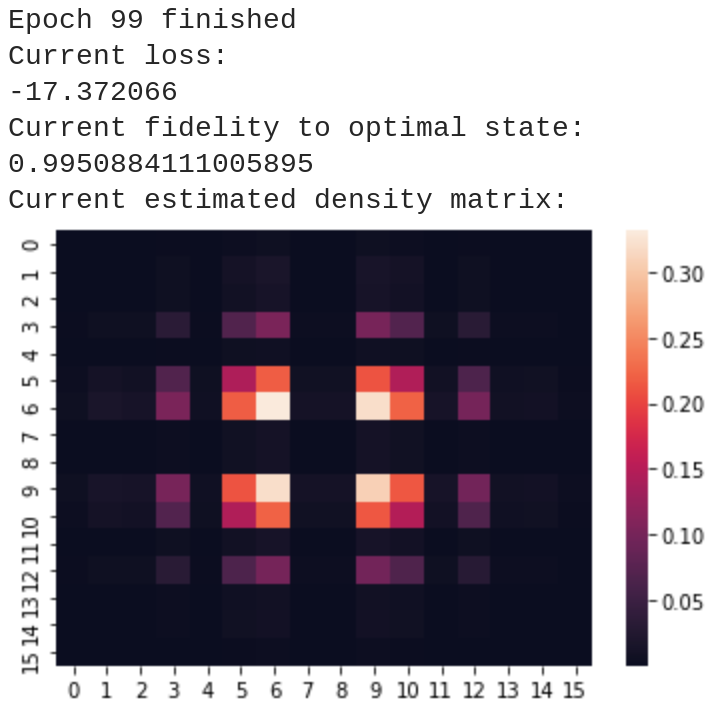}
\caption{Final density matrix output by the VQT algorithm run with a factorized Bernoulli distribution as classical latent distribution, trained via a gradient-based optimizer. See notebook for details. }
    \label{fig:vqt_result}
\end{figure}

For our classical latent probability distribution $p_{\bm{\theta}}(\bm{x})$, as a first simple case, we can use the product of independent Bernoulli distributions $p_{\bm{\theta}}(\bm{x}) = \prod_j p_{{\theta}_j}(x_j) =\prod_j \theta_j^{x_j}(1-\theta_j)^{1-x_j} $, where $x_j\in\{0,1\}$ are binary values.  We can re-phrase this distribution as an energy based model to take advantage of equation \eqref{eq:fe-classical-gradient}.  We move the parameters into an exponential, so that the probability of a bitstring becomes 
\(p_{\bm{\theta}}(\bm{x}) = \prod_je^{\theta_jx_j}/(e^{\theta_j} + e^{-\theta_j})\).
Since this distribution is a product of independent variables, it is easy to sample from.  We can use the TensorFlow Probability library \cite{dillon2017tensorflow} to produce samples from this distribution, using the \Colorbox{bkgd}{\lstinline{tfp.distributions.Bernoulli}} object:
\begin{lstlisting}
def bernoulli_bit_probability(b):
  return np.exp(b)/(np.exp(b) + np.exp(-b))

def sample_bernoulli(num_samples, biases):
  prob_list = []
  for bias in biases.numpy():
    prob_list.append(
        bernoulli_bit_probability(bias))
  latent_dist = tfp.distributions.Bernoulli(
    probs=prob_list, dtype=tf.float32)
  return latent_dist.sample(num_samples)
\end{lstlisting}
After getting samples from our classical probabilistic model, we take gradients of our QNN parameters.  Because TFQ implements gradients for its expectation ops, we can use \Colorbox{bkgd}{\lstinline{tf.GradientTape}} to obtain these derivatives.  Note that below we used $\Colorbox{bkgd}{\lstinline{tf.tile}}$ to give our Hamiltonian operator and visible state circuit the correct dimensions: 
\begin{lstlisting}
bitstring_tensor = sample_bernoulli(
    num_samples, vqt_biases)
with tf.GradientTape() as tape:
  tiled_vqt_model_params = tf.tile(
    [vqt_model_params], [num_samples, 1])
  sampled_expectations = expectation(
    tiled_visible_state,
    vqt_symbol_names,
    tf.concat([bitstring_tensor,
      tiled_vqt_model_params], 1),
    tiled_H)
  energy_losses = beta*sampled_expectations
  energy_losses_avg = tf.reduce_mean(energy_losses)
vqt_model_gradients = tape.gradient(
    energy_losses_avg, [vqt_model_params])
\end{lstlisting}
Putting these pieces together, we train our model to output thermal states of the 2D Heisenberg model on a 2x2 grid.  The result after 100 epochs is shown in Fig.~\ref{fig:vqt_result}.

A great advantage of this approach to optimization of the probabilistic model is that the partition function $\mathcal{Z}_{\bm{\theta}}$ does not need to be estimated. As such, more general more expressive models beyond factorized distributions can be used for the probabilistic modelling of the latent classical distribution. In the advanced section of the notebook, we show how to use a Boltzmann machine as our energy based model. Boltzmann machines are EBM's where for bitstring $x \in \{0, 1\}^n$, the energy is defined as $E(x) = -\sum_{i, j}w_{ij} x_i x_j - \sum_i b_i x_i$.

It is worthy to note that our factorized Bernoulli distribution is in fact a special case of the Boltzmann machine, one where only the so-called \textit{bias terms} in the energy function are present:  $E(x) = - \sum_i b_i x_i$. In the notebook, we start with this simpler Bernoulli example of the VQT, the resulting density matrix converges to the known exact result for this system, as shown in Fig.~\ref{fig:vqt_result}. We also provide a more advanced example with a general Boltzmann machine. In the latter example, we picked a fully visible, fully-connected classical Ising model energy function, and used MCMC with Metropolis-Hastings \cite{robert2015metropolishastings} to sample from the energy function.

\subsubsection{Quantum Generative Learning from Quantum Data}

Now that we have seen how to prepare thermal states from a given Hamiltonian, we can consider how we can learn to generatively model mixed quantum states using quantum-probabilistic models in the case where we are given several copies of a mixed state rather than a Hamiltonian. That is, we are given access to a data mixed state $ \hat{\sigma}_{\mathcal{D}} $, and we would like to find optimal parameters $\{\bm{\theta}^*,\bm{\phi}^*\}$ such that $\hat{\rho}_{\bm{\theta}^*\bm{\phi}^*} \approx \hat{\sigma}_{\mathcal{D}}$, where the model state is of the form described in \eqref{eq:qhbm_model}. Furthermore, for reasons of convenience which will be apparent below, it is useful to posit that our classical probabilistic model is of the form of an \textit{energy-based model} as in equation \eqref{eq:EBM}.

If we aim to minimize the quantum relative entropy between the data and our model (in reverse compared to the VQT) i.e., $D(\hat{\sigma}_{\mathcal{D}} \Vert \hat{\rho}_{\bm{\theta\phi}}) $  then it suffices to minimize the quantum cross entropy as our loss function
\[\mathcal{L}_{\textsc{xe}}(\bm{\theta,\phi}) \equiv - \text{tr}(\hat{\sigma}_{\mathcal{D}} \log \hat{\rho}_{\bm{\theta}\bm{\phi}} ).\]

By using the energy-based form of our latent classical probability distribution, as can be readily derived (see \cite{verdon2019quantumVQT}), the cross entropy is given by
\[\mathcal{L}_{\textsc{xe}}(\bm{\theta,\phi}) = \mathbb{E}_{\bm{x}\sim \sigma_{\bm{\phi}}(\bm{x})}[E_{\bm{\theta}}(\bm{x})] + \log \mathcal{Z}_{\bm{\theta}},\]
where $\sigma_{\bm{\phi}}(\bm{x}) \equiv \bra{\bm{x}}\hat{U}^\dagger(\bm{\phi})\hat{\sigma}_{\mathcal{D}}\hat{U}(\bm{\phi})\ket{\bm{x}}$ is the distribution obtained by feeding the data state $\hat{\sigma}_{\mathcal{D}}$ through the inverse QNN circuit $\hat{U}^\dagger(\bm{\phi})$ and measuring in the standard basis. 

As this is simply an expectation value of a state propagated through a QNN, for gradients of the loss with respect to QNN parameters we can use standard TFQ differentiators, such as the parameter shift rule presented in section \ref{sec:theory}. As for the gradient with respect to the EBM parameters, it is given by

\[\partial_{\bm{\theta}}\mathcal{L}_{\textsc{xe}}(\bm{\theta,\phi})=\mathbb{E}_{\bm{x}\sim \sigma_{\bm{\phi}}(\bm{x})}[\nabla_{\bm{\theta}} E_{\bm{\theta}}(\bm{x})] - \mathbb{E}_{\bm{y}\sim p_{\theta}(\bm{y})} 
[\nabla_{\bm{\theta}}E_{\bm{\theta}}(\bm{y})].\]

Let us implement a scenario where we were given the output density matrix from our last VQT example as data, let us see if we can learn to replicate its statistics from data rather than from the Hamiltonian.
For simplicity we focus on the Bernoulli EBM defined above.  We can efficiently sample bitstrings from our learned classical distribution and feed them through the learned VQT unitary to produce our data state.  These VQT parameters are assumed fixed; they represent a quantum datasource for QMHL.

We use the same ansatz for our QMHL unitary as we did for VQT, layers of single qubit rotations and exponentiated CNOTs.  We apply our QMHL model unitary to the output of our VQT to produce the pulled-back data distribution.  Then, we take expectation values of our current best estimate of the modular Hamiltonian:
\begin{lstlisting}
def get_qmhl_weights_grad_and_biases_grad(
    ebm_deriv_expectations, bitstring_list, biases):
  bare_qmhl_biases_grad = tf.reduce_mean(
    ebm_deriv_expectations, 0)
  c_qmhl_biases_grad = ebm_biases_derivative_avg(bitstring_list)
  return tf.subtract(bare_qmhl_biases_grad, c_qmhl_biases_grad)
\end{lstlisting}
Note that we use the \Colorbox{bkgd}{\lstinline{tf.GradientTape}} functionality to obtain the gradients of the QMHL model unitary.  This functionality is enabled by our TFQ differentiators module.

The classical model parameters can be updated according to the gradient formula above. See the VQT notebook for the results of this training.\\

\subsection{Subspace-Search Variational Quantum Eigensolver for Excited States: Integration with OpenFermion}
\subsubsection{Background}

The Variational Quantum Eigensolver (VQE) \cite{peruzzo2014variational} is a heuristic algorithm for preparing the ground state of a many-body quantum system. It is often used for modelling strongly correlated systems that appear challenging to simulate classically but contain enough structure that ground state preparation is believed to be tractable quantum mechanically (at least for some interesting instances); e.g., it is often studied in the context of the molecular electronic structure Hamiltonian of interest in quantum chemistry. The basic idea of VQE is that one can use a quantum computer with limited resources to prepare a highly entangled parameterized state that can be optimized via classical feedback to provide a classically inaccessible description of the ground states of these systems. Specifically, given the time independent Schr{\"o}dinger equation $\hat{H}|\Psi\rangle = E|\Psi \rangle$, with Hamiltonian $\hat{H} = \sum_{i = 0}^N \lambda_i |\psi_i \rangle \langle \psi_i |$, where $\lambda_i$ are the eigenvalues of increasing size and $\psi_i$ the eigenvectors in the spectral decomposition, VQE provides an approximation of $\lambda_0$. Measurements of the Hamiltonian yield real eigenvalues (due to the Hermitian nature of quantum operators), therefore $\lambda_0 \leq \langle \Psi | \hat{H} | \Psi \rangle$. Given optimization parameters $\theta$ and $\Psi = U(\theta)|0^{\otimes N} \rangle$, the classical optimization problem associated with VQE is defined as $min_\theta \; \langle 0 | U^\dagger (\theta) \hat{H} U(\theta) | 0 \rangle$. This formulation is limited to finding the ground state energy, but higher energy states are often chemically and physically important \cite{mcardle2020quantum}. In order to expand the VQE algorithm to higher energy states, exploitation of the orthogonality of different quantum energy levels has been proposed \cite{mcclean2017hybrid, higgott2019variational, nakanishi2019subspace}. 

Here, we focus on describing how TFQ can be used to investigate an example of a state-of-the-art variational algorithm known as the Subspace-Search Variational Quantum Eigensolver (SSVQE) for excited states \cite{nakanishi2019subspace}. The SSVQE modifies the VQE optimization problem to be: 
\begin{equation*}
\begin{aligned}
\min_\theta \quad & \sum_{i = 0}^k w_i \langle \Psi_i | U^\dagger (\theta) \hat{H} U(\theta) | \Psi_i \rangle \\
\textrm{s.t.} \quad & w_{i+1} \leq w_i, \; \langle \Psi_i | \Psi_j \rangle = \delta_{ij}
\end{aligned}
\end{equation*}

The goal is to minimize the expectation value of the Hamiltonian, with the initial state of each energy level's evaluation being orthogonal. In SSVQE, as in VQE, there are a number of choices for anstaz, i.e. the circuit structure of $U(\theta)$, and optimization method \cite{kandala_hardware_efficient_2017, grimsley2019adaptive, ostaszewski2021structure, barkoutsos2020improving, yamamoto2019natural}. 

Tensorflow-Quantum has several features that make it appealing for work with VQE algorithms. The adjoint differentiator enables rapid code execution, which is especially important given the number of Hamiltonians encountered in quantum chemistry experiments. The ability to work with any optimization target via custom layers and custom objectives enables straightforward implementation of and experimentation with any VQE based algorithm. Additionally, OpenFermion \cite{mcclean2020openfermion} has native methods for working with Cirq, enabling the quantum chemistry methods OpenFermion provides to be easily combined with the quantum machine learning methods Tensorflow-Quantum provides. 

\noindent\textbf{Target problems}:
\begin{enumerate}[noitemsep]
    \item Implement SSVQE \cite{nakanishi2019subspace} in Tensorflow-Quantum
    \item Find the ground state and first excited state energies for $H_2$ at increasing bond lengths
\end{enumerate}
\noindent\textbf{Required TFQ functionalities}:
  \begin{enumerate}[noitemsep]
    \item Integration with OpenFermion \cite{mcclean2020openfermion}        
    \item Custom Tensorflow-Quantum Layers
\end{enumerate}

\subsubsection{Implementation}

The full notebook and implementation is available at: 

\fancylink{https://github.com/tensorflow/quantum/tree/research/ssvqe}{research/ssvqe}

First, we need to create the VQE anstaz. Here, we construct one similar to the Hardware Efficient Ansatz \cite{kandala_hardware_efficient_2017}. We create layers of parameterized $Ry$ and $Rz$ gates, entangled with CNOTs. 

\begin{lstlisting}
def layer(circuit, qubits, parameters):
    for i in range(len(qubits)):
        circuit += cirq.ry(parameters[3*i]).on(qubits[i])
        circuit += cirq.rz(parameters[3*i+1]).on(qubits[i])
        circuit += cirq.ry(parameters[3*i+2]).on(qubits[i])
    for i in range(len(qubits)-1):
        circuit += cirq.CNOT(qubits[i], qubits[i+1])
    circuit += cirq.CNOT(qubits[-1], qubits[0])
    return circuit

def ansatz(circuit, qubits, layers, parameters):
    for i in range(layers):
        params = parameters[3*i*len(qubits):3*(i+1)*len(qubits)]
        circuit = layer(circuit, qubits, params)
    return circuit
\end{lstlisting}

Now we create the readout operators for each qubit, which are equivalent to the associated terms in the Hamiltonian. This will yield the expectation value of the Hamiltonian. 
 
\begin{lstlisting}
def exp_val(qubits, hamiltonian):
    return prod([op(qubits[i]) for i, op in enumerate(hamiltonian) if hamiltonian[i] != 0])
\end{lstlisting}

Since the Hamiltonian can be expressed as a sum of simple operators, i.e.,
$$ \hat{H} = \sum_{\ell=1}^L a_\ell P_\ell $$
for real scalars $a_\ell$ and Pauli strings $P_\ell$. The Hamiltonian is sparse in this representation; typically $L={\cal O}(N^4)$ in the number of spin-orbitals (or equivalently, qubits) $N$. We create custom VQE layers with different readout operators but shared parameters. The SSVQE is then implemented as a collection of these VQE layers, with each input being orthogonal. This is implemented by applying a Pauli X gate prior to creating $U(\theta)$. 

\begin{lstlisting}
class VQE(tf.keras.layers.Layer):
    def __init__(self, circuits, ops):
        super(VQE, self).__init__()
        self.layers = [tfq.layers.ControlledPQC(circuits[i], ops[i], differentiator=tfq.differentiators.Adjoint()) for i in range(len(circuits))]

    def call(self, inputs):
        return sum([self.layers[i]([inputs[0], inputs[1]]) for i in range(len(self.layers))])

class SSVQE(tf.keras.layers.Layer):
    def __init__(self, num_weights, circuits, ops, k, const):
        super(SSVQE, self).__init__()
        self.theta = tf.Variable(np.random.uniform(0, 2 * np.pi, (1, num_weights)), dtype=tf.float32)
        self.hamiltonians = []
        self.k = k
        self.const = const
        for i in range(k):
            self.hamiltonians.append(VQE(circuits[i], ops[i]))

    def call(self, inputs):
        total = 0
        energies = []
        for i in range(self.k):
            c = self.hamiltonians[i]([inputs, self.theta]) + self.const
            energies.append(c)
            if i == 0:
               total += c
            else:
                total += ((0.9 - i * 0.1) * c)
        return total, energies
\end{lstlisting}

Next we need to set up the training function for the SSVQE. The optimization loop directly minimizes the output, as we have already encoded the constraints.

\begin{lstlisting}
def train_ssvqe(ssvqe, opt, tol=5e-6, patience=10):
    ssvqe_model = ssvqe[0]
    energies = ssvqe[1]
    prev_loss = 100
    counter = 0
    inputs = tfq.convert_to_tensor([cirq.Circuit()])
    while True:
        with tf.GradientTape() as tape:
            loss = ssvqe_model(inputs)
        grads = tape.gradient(loss, ssvqe_model.trainable_variables)
        opt.apply_gradients(zip(grads, ssvqe_model.trainable_variables))
        loss = loss.numpy()[0][0]
        if abs(loss - prev_loss) < tol:
            counter += 1
        if counter > patience:
            break
        prev_loss = loss

    energies.theta = ssvqe_model.trainable_variables[0]
    energies = [i.numpy()[0][0] for i in energies(inputs)[1]]
    return energies[0], energies[1]
\end{lstlisting}

With the SSVQE set up, we now need to generate the inputs. In the code below, we load the data that has already been generated. These files were generated using OpenFermion \cite{mcclean2020openfermion} and PySCF \cite{sun2018pyscf}. Using the OpenFermionPySCF function \Colorbox{bkgd}{\lstinline{generate_molecular_hamiltonian}} to generate the Hamiltonians for each bond length, we then converted this to a form compatible with quantum circuits using the Jordan-Wigner Transform, which maps fermionic annihilation operators to qubits via: $a_p \mapsto \frac{1}{2} \left ( X_p + iY_p \right ) Z_1 \cdots Z_{p-1}$ and $a_p^\dagger \mapsto \frac{1}{2} \left ( X_p - iY_p \right ) Z_1 \cdots Z_{p-1}$. This yields a Hamiltonian of the form: $\hat{H} = g_0 \mathbb{1} + g_1 X_0 X_1 Y_2 Y_3 + g_2 X_0 Y_1 Y_2 X_3 + g_3 Y_0 X_1 X_2 Y_3 + g_4 Y_0 Y_1 X_2 X_3 + g_5 Z_0 + g_6 Z_0 Z_1 + g_7 Z_0 Z_2 + g_8 Z_0 Z_3 + g_9 Z_1 + g_{10} Z_1 Z_2 + g_{11} Z_1 Z_3 + g_{12} Z_2 + g_{13} Z_2 Z_3 + g_{13} Z_3$, where $g_n$ is determined by the bond length. It is this PauliSum object which is then saved and loaded. With the Hamiltonians created, we iterate over the bond lengths and compute the predicted energies of the states. 

\begin{lstlisting}
diatomic_bond_length = 0.2
interval = 0.1
max_bond_length = 2.0   
k = 2

# VQE Hyperparameters
layers = 4
n_qubits = 4
optimizer = tf.keras.optimizers.Adam(lr=0.1)

step = 0
while diatomic_bond_length <= max_bond_length:
    eigs = real[step]
    # Load the Data
    ham_name = "mol_hamiltonians_" + str(step)
    coef_name = "coef_hamiltonians_" + str(step)
    with open(ham_name, "rb") as ham_file:
        hamiltonians = load(ham_file)
    with open(coef_name, "rb") as coeff_file:
        coefficients = load(coeff_file)
    # Create the SSVQE and Approximate the Energies
    ssvqe = make_ssvqe(n_qubits, layers, coefficients, hamiltonians, k)
    ground, excited = train_ssvqe(ssvqe, optimizer)
    diatomic_bond_length += interval
    step += 1

\end{lstlisting}

\begin{figure}[H]
    \centering
    \includegraphics[scale=0.5]{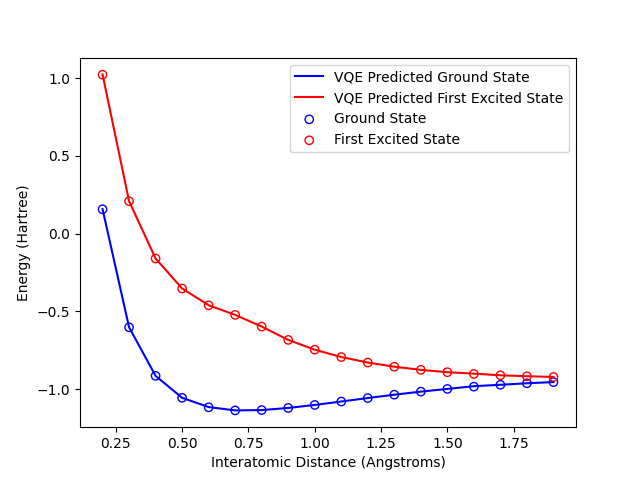}
    \caption{Comparison of SSVQE predicted energy for the ground state and first excited state compared with the true values}
    \label{fig:ssvqe}
\end{figure}

This implementation takes several minutes to run, even with the Adjoint differentiator. This is because there are 4 layers * 3 parameters per qubit * 4 qubits = 48 parameters per layer and because we must optimize these parameters for 20 different Hamiltonians. We can then plot and compare the actual energies with the VQE predicted energies, as done in Figure \ref{fig:ssvqe}. 

\vspace{0.5 em}

\subsection{Classification of Quantum Many-Body Phase Transitions}
To run this example see the Colab notebook at:
\fancylink{https://github.com/tensorflow/quantum/blob/research/phase_classifier/phase_classification.ipynb}{research/phase\_classifier/phase\_classification.ipynb}
\subsubsection{Background}

In quantum many-body systems we are interested in studying different phases of matter and the transitions between these different phases. This requires a detailed characterization of the observables that change as a function of the order values of the system. To estimate these observables, we have to rely on quantum Monte Carlo techniques or tensor network approaches, that each come with their respective advantages and drawbacks \cite{Sandvik2010qmc, Verstraete2008mps}. 

Recently, work on classifying phases with supervised machine learning has been proposed as a method for understanding phase transitions \cite{carrasquilla2017machine, vanNieuwenburg2017confusion}. For such a supervised approach, we label the different phases of the system and use classical data obtained from the system of interest and train a machine learning model to predict the labels. A natural extension of these ideas is a quantum machine learning model for phase classification \cite{Uvarov2020qc_phase_trans}, where the information that characterizes the state is encoded in a variational quantum circuit. This would require preparing high fidelity quantum many-body states on NISQ devices, an area of active research in the field of quantum computing \cite{Smith2019qmbssim, Ho2019qmbssim, Wierichs2020avoiding, Wiersema2020exploring}. In this section, we show how a quantum classifier can be trained end-to-end in Tensorflow quantum by using a data set of quantum many-body states.

\begin{figure}[htb!]
    \centering
    \includegraphics[width=\columnwidth]{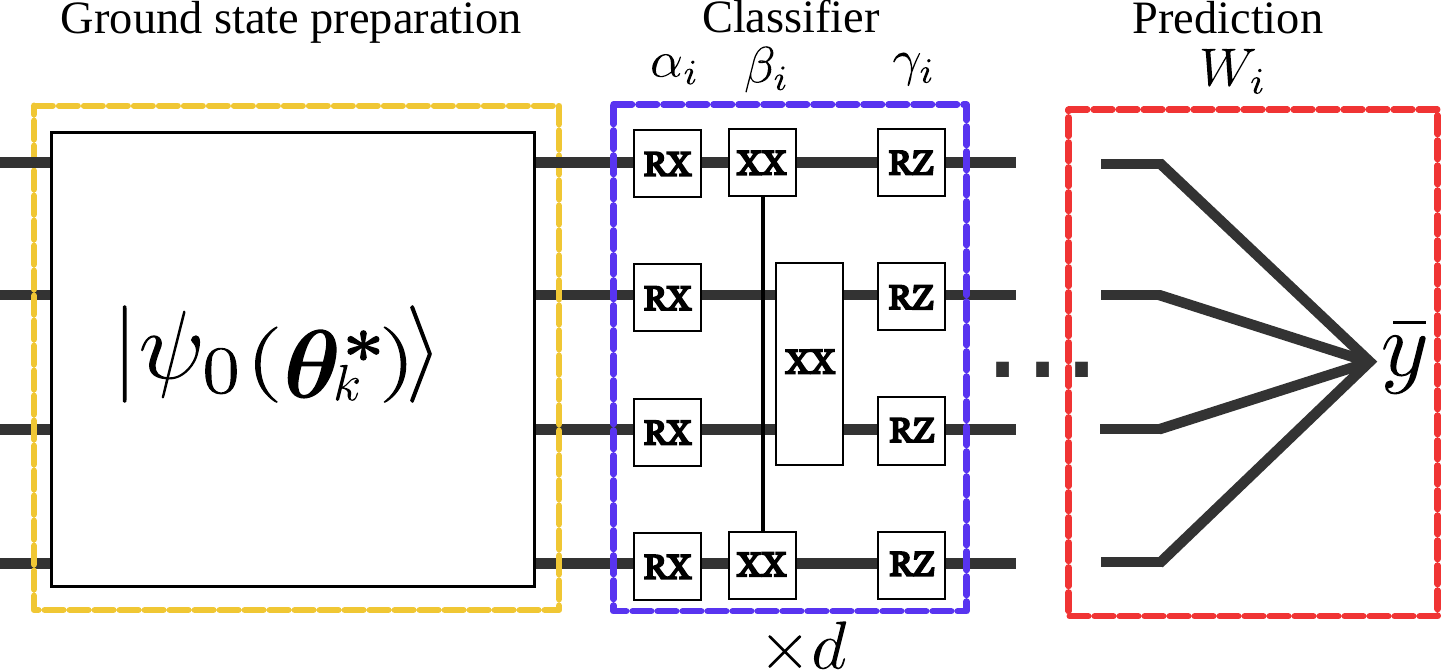}
    \caption{The first part of the circuit prepares the ground state of a Hamiltonian  $H(\lambda)$ with a fixed ansatz $\ket{\psi_0(\bm{\theta})}$. For each order value $\lambda_k$ there is a corresponding set of fixed parameters $\bm{\theta}_k^*$ such that $\ket{\psi_0(\bm{\theta}^*_k)}$ is the ground state of $H(\lambda_k)$. The second part of the circuit is the quantum classifier, a hardware efficient ansatz with trainable parameters for each gate $(\bm{\alpha}_i, \bm{\beta}_i, \bm{\gamma}_i)$, where $i = 1, \ldots, d$ indicates the layer. Finally, we obtain the expectation value $\langle Z \rangle_i$ on all qubits and combine these into a the output label $\bar{y}$ after a rescaled sigmoid activation function.}
    \label{fig:qclassif}
\end{figure}

The quantum classifier we consider here is a made up of three parts: a circuit that prepares a set of states of interest, a variational quantum circuit consisting of multiple trainable layers and a collection of measurements that can be combined to create a predictor for the classification. In \cref{fig:qclassif} we give a schematic depiction of the full model.

Tensorflow Quantum has a data set module that contains quantum circuits for different quantum many-body ground states. This makes it easy to set up a workflow where we want to load a set of quantum states and perform a quantum machine learning task on these data. 

As an example, we study the two-dimensional transverse field Ising-model (TFIM) on a torus,
\begin{equation*}
    H_{\text{TFIM}} = -\sum_{\langle i,j\rangle} \sigma_i^z \sigma_{j}^z - g \sum_i^N \sigma_i^x = H_{zz} + g H_x,
\end{equation*}
where $\langle i,j\rangle$ indicates the set of nearest neighbor indices for each point in the lattice. On the torus, this system has a phase transition at $g\approx3.04$ from an ordered to a disordered phase \cite{deJongh1998dmrgtfi2d, Blote2002tfi2d}. 

After the state preparation circuit, we use the hardware efficient ansatz to train the quantum classifier. This ansatz consists of two layers of parameterized single qubit gates and a single layer of two-qubit parameterized gates \cite{kandala_hardware_efficient_2017}. Note that the parameters in the state preparation circuit stay fixed during training. Finally, we measure the observables $\langle Z_i \rangle$ on each qubit and apply a rescaled sigmoid to the linear predictor of the outcomes,
\begin{align*}
    \bar{y} = \tanh{\sum_i \langle Z_i \rangle W_i},
\end{align*}
where $W_i \in\mathbb{R}^n$ is a weight vector. Hence, given an input ground state $\ket{\psi_0(\bm{\theta}^*_k)}$, our classifier will output a single scalar $\bar{y}\in(-1,1)$.

We use the Hinge loss function to train the model to output the correct labels. This loss function is given by
\begin{align}
    \ell = \max(0, t - y \cdot \bar{y})
\end{align}
where, $y\in\{-1, 1\}$ are the ground truth labels and $\bar{y} \in (-1,1)$ the model predictions. The entire end-to-end model contains several non-trivial components, such as loading quantum states as data or backpropagating classical gradients through quantum gates.

\vspace{\baselineskip}

\noindent\textbf{Target problems}:
\begin{enumerate}[noitemsep]
    \item Training a quantum circuit classifier to detect a phase transition in a condensed matter physics system
\end{enumerate}
\noindent\textbf{Required TFQ functionalities}:
      \begin{enumerate}[noitemsep]
        \item Hybrid quantum-classical optimization of a variational quantum circuit and logistic function.
        \item Batching training over quantum data (multiple ground states and labels)
    \end{enumerate}

\subsubsection{Implementation}
For the two-dimensional TFIM on a rectangular lattice, the available data sets are a $3\times3$, $4\times 3$, and $4\times4$ lattice for $g\in [2.5,3.5]$. To load the data, we use the \Colorbox{bkgd}{\lstinline{tfi_rectangular}} function to obtain both the circuits and labels indicating the ordered ($y=1$) and disordered ($y=-1$) phase. Here, we consider the $4\times4$ lattice.
\begin{lstlisting}
nspins = 16
qubits = cirq.GridQubit.rect(nspins, 1)

circuits, labels, _, _ = tfq.datasets.tfi_rectangular(qubits)

labels = np.array(labels)
labels[labels >= 1] = 1.0
labels = labels * 2 - 1
x_train_tfcirc = tfq.convert_to_tensor(circuits)
\end{lstlisting}
\Colorbox{bkgd}{\lstinline{circuits}} now contains 51 quantum circuits that prepare ${>0.999}$ fidelity ground states of the two-dimensional TFIM at the respective order values $g$. Next, we add the layers of our hardware efficient ansatz. We parameterize each gate in the classifier individually.

\begin{figure}[htb!]
    \centering
    \includegraphics[width=0.8\columnwidth]{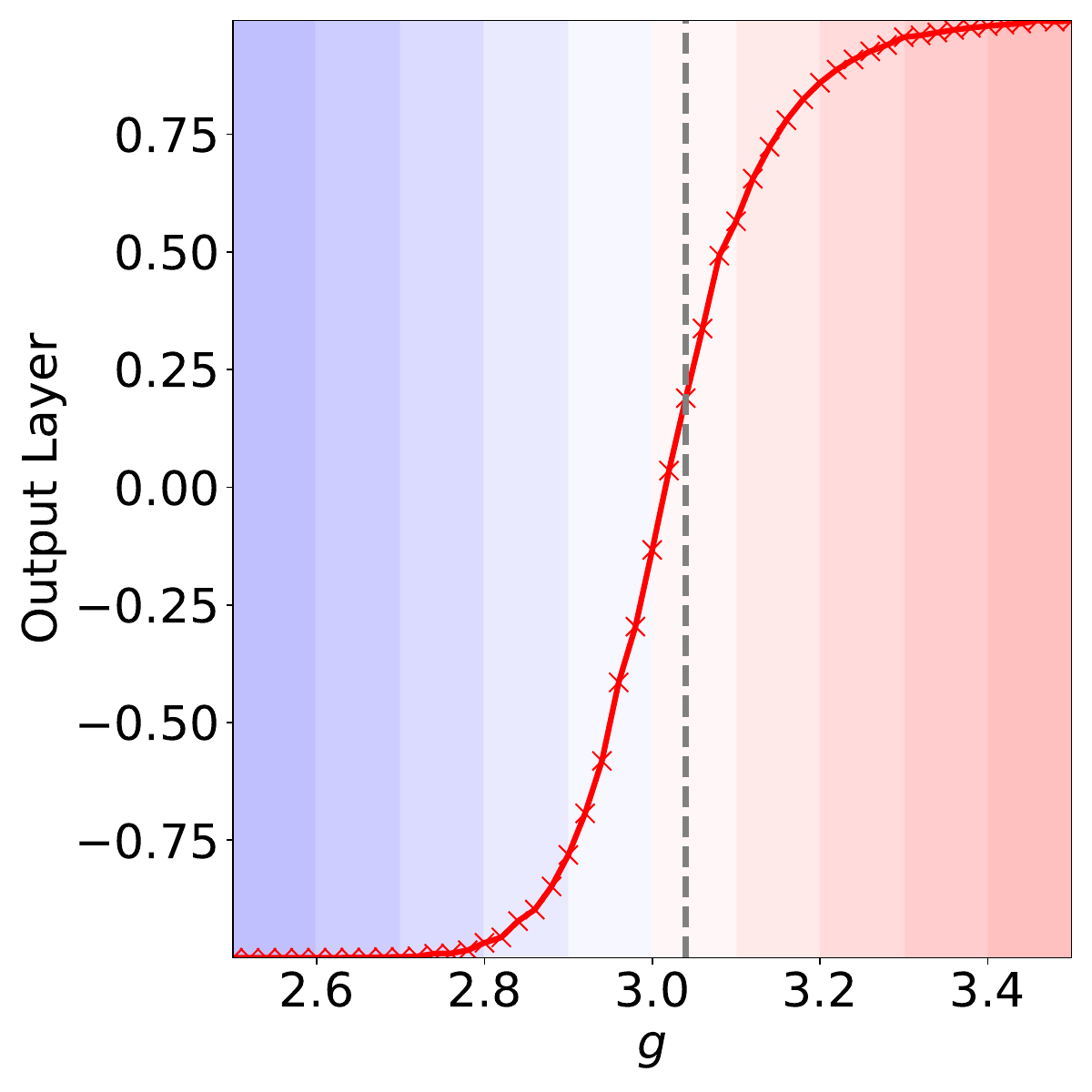}
    \caption{Quantum classifier output versus order value $g$ of the two-dimensional TFIM. The dashed line indicates the critical point $g\approx3.04$.}
    \label{fig:qclass}
\end{figure}

\begin{lstlisting}
def add_layer_nearest_neighbours(circuit, qubits, gate, prefix):
    for i, q in enumerate(zip(qubits, qubits[1:])):
        symbol = sympy.Symbol(prefix + '-' + str(i))
        circuit.append(gate(*q) ** symbol)


def add_layer_single(circuit, qubits, gate, prefix):
    for i, q in enumerate(qubits):
        symbol = sympy.Symbol(prefix + '-' + str(i))
        circuit.append(gate(q) ** symbol)


def create_quantum_model(N, num_layers):
    qubits = cirq.GridQubit.rect(N, 1)
    circuit = cirq.Circuit()
    for l in range(num_layers):
        add_layer_single(circuit, qubits, cirq.X, f"x_{l}")
        add_layer_nearest_neighbours(circuit, qubits, cirq.XX, f"xx_{l}")
        add_layer_single(circuit, qubits, cirq.Z, f"z_{l}")

    readout = [cirq.Z(q) for q in qubits]
    return circuit, readout
\end{lstlisting}
We can seamlessly integrate our quantum classifier into a Keras model.
\begin{lstlisting}
model = tf.keras.Sequential([
    tf.keras.layers.Input(shape=(), dtype=tf.string),
    tfq.layers.PQC(circuit, output),
    tf.keras.layers.Dense(1, activation=tf.keras.activations.tanh)
])
\end{lstlisting}
Additionally, we can compile the model with other metrics to track during training. In our case, we use the hinge accuracy to count the number of missclasified data points.
\begin{lstlisting}
def hinge_accuracy(y_true, y_pred):
    y_true = tf.squeeze(y_true) > 0.0
    y_pred = tf.squeeze(y_pred) > 0.0
    result = tf.cast(y_true == y_pred, tf.float32)

    return tf.reduce_mean(result)

model.compile(
    loss=tf.keras.losses.Hinge(),
    optimizer=tf.keras.optimizers.Adam(learning_rate=0.01),
    metrics=[hinge_accuracy])
\end{lstlisting}
If we set the maximum number of epochs and batch size, we are ready to fit the model. We add an early stopping callback to make sure that the training can terminate when the loss no longer decreases consistently.
\begin{lstlisting}
epochs = 25
batch_size = 32

qnn_history = model.fit(
    x_train_tfcirc, labels,
    batch_size=BATCH_SIZE,
    epochs=EPOCHS,
    verbose=1,
    callbacks=[tf.keras.callbacks.EarlyStopping('loss', patience=5)])
predictions = model.predict(x_train_tfcirc)

\end{lstlisting}
After the model is trained, we can predict the labels of the phases from the model output $\bar{y}$ and visualize how well the model has captured the phase transition. In \cref{fig:qclass}, we see that the inflection point at $\bar{y}\approx0$ coincides with the phase transition at $g\approx3.04$, as expected. Although we have not explored this here, we could train the classifier on a subset of the data, for instance around the critical point, and then see how well the classifier generalizes to states outside of the seen data.
\vspace{1em}

\vspace{0.5 em}

\subsection{Quantum Generative Adversarial Networks}
\subsubsection{Background}
Generative adversarial networks (GANs)~\cite{goodfellow2014generative} have met widespread success in classical machine learning. While classical data can be seen as a special case of data corresponding to a probabilistic mixture, we consider the task of adversarially learning most general form of data: quantum states. A \emph{generator} seeks to create a circuit that reproduces a given quantum state, while a \emph{discriminator} circuit is presented either with the true data or with fake data from the generator. In this section, we present viable applications of a new quantum GAN architecture for both purely quantum datasets and quantum states built from classical datasets.

Recent work on a quantum GAN (QuGAN)~\cite{PhysRevA.98.012324,lloyd2018quantum} has proposed a direct analogy of the classical GAN architecture in designing the generator and discriminator circuits. However, the QuGAN does not always converge but rather in certain cases oscillates between a finite set of states due to mode collapse, and in general suffers from a non-unique Nash equilibrium~\cite{eqgan}. This motivates a new \emph{entangling} quantum GAN (EQ-GAN) with a uniquely quantum twist: rather than providing the discriminator with \emph{either} true \emph{or} fake data, we allow the discriminator to entangle \emph{both} true and fake data. The convergence of the EQ-GAN to the global optimal Nash equilibrium is theoretically verified and numerical experiments confirm that the EQ-GAN converges on problem instances that the QuGAN failed on~\cite{eqgan}.

A pervasive issue in near-term quantum computing is noise: when performing an entangling operation, the required two-qubit gate introduces phase errors that are difficult to fully calibrate against due to a time-dependent noise model. However, such operations are required to measure the overlap between two quantum states, including the assessment of how close the fake data is to the true data. This issue provides further motivation for the use of \emph{adversarial} generative learning. Without adversarial learning, one may freeze the discriminator to perform an exact fidelity measurement between the true and fake data. While this would replicate the original state in the absence of noise, gate errors in the implementation of the discriminator will cause convergence to the incorrect optimum. As seen in the first example below, the adversarial approach of the EQ-GAN is more robust to such errors than the simpler supervised learning approach. Since training quantum machine learning models can require extensive time to compute gradients on current quantum hardware, resilience to gate errors drifting during the training process is especially valuable in the noisy intermediate-scale quantum era of quantum computing.

Most proposals for quantum machine learning algorithms on \emph{classical} data require a quantum random access memory (QRAM)~\cite{2017Natur.549..195B}. A QRAM typically stores the classical dataset in a superposition of quantum states, allowing a quantum device to efficiently access the dataset. However, a QRAM is difficult to achieve experimentally, posing a further roadblock for near-term quantum machine learning applications. We provide an example application of the EQ-GAN to create an approximate QRAM by learning a shallow quantum circuit that generates a superposition of classical data. In particular, the QRAM is applied to quantum neural networks~\cite{farhi2018classification}, improving the performance of a quantum neural network for a classification task.

\vspace{\baselineskip}
\noindent\textbf{Target problems}:
\begin{enumerate}[noitemsep]
    \item Suppress noise in a generative learning task for quantum states
    \item Prepare an approximate quantum memory for faster training of a quantum neural network
\end{enumerate}
\noindent\textbf{Required TFQ functionalities}:
      \begin{enumerate}[noitemsep]
        \item Use of a quantum hardware backend
        \item Shared variables between quantum neural network layers
        \item Training on a noisy quantum circuit
    \end{enumerate}

\subsubsection{Noise Suppression with Adversarial Learning}
To run this example, see the Colab notebook at:
\fancylink{https://github.com/tensorflow/quantum/blob/research/eq_gan/noise_suppression.ipynb}{research/eq\_gan/noise\_suppression.ipynb}

The entangling quantum generative adversarial network (EQ-GAN) uses a minimax cost function
\begin{align}
  \min_{\mathbf{\theta}_g}\max_{\mathbf{\theta}_d} V(\theta_g, \theta_d)  &=\min_{\mathbf{\theta}_g}\max_{\mathbf{\theta}_d} [1-  D_\sigma(\mathbf{\theta}_d, \rho(\theta_g))]
\end{align}
to learn the true data density matrix $\sigma$. The generator produces a fake quantum state $\rho(\theta_g)$, while the discriminator performs a parameterized swap test (fidelity measurement) $D_\sigma(\theta_d, \rho(\theta_g))$ between the true data $\sigma$ and the fake data $\rho$. The swap test is parameterized such that there exist parameters $\theta_d^\mathrm{opt}$ that realize a perfect swap test, i.e. $D_\sigma(\theta_d^\mathrm{opt}, \rho(\theta_g)) = \frac{1}{2} + \frac{1}{2}D_\sigma^\mathrm{fid}(\rho(\theta_g))$ where
\begin{align}
D_\sigma^\mathrm{fid}(\rho(\theta_g)) = \left(\text{Tr}\sqrt{ \sigma^{1/2} \, \rho(\theta_g) \, \sigma^{1/2}} \, \right)^2.
\end{align}

In the presence of noise, it is generally difficult to determine a circuit ansatz for $D_\sigma(\theta_d, \rho)$ such that parameters $\theta_d^\mathrm{opt}$ exist. When implementing a $CZ$ gate, gate parameters such as the conditional $Z$ phase, single qubit $Z$ phase and swap angles in two-qubit entangling gate can drift and oscillate over the time scale of $O(10)$ minutes~\cite{arute2020supp,arute2020observation}. Such unknown systematic and time-dependent coherent errors provides significant challenges for applications in quantum machine learning where gradient computation and update requires many measurements, especially because of the use of entangling gates in a swap test circuit. However, the large deviations in single-qubit and two-qubit $Z$ rotation angles can largely be mitigated by including additional single-qubit $Z$ phase compensations. In learning the discriminator circuit that is closest to a true swap test, the adversarial learning of EQ-GAN provides a useful paradigm that may be broadly applicable to improving the fidelity of other near-term quantum algorithms.

To operate under this noise model, we can define a \Colorbox{bkgd}{\lstinline{cirq.NoiseModel}} that is compatible with TFQ, implementing the \Colorbox{bkgd}{\lstinline{noisy_operation}} method to add $CZ$ and $Z$ phase errors on all $CZ$ gates.

\begin{lstlisting}
def noisy_operation(self, op):
  if isinstance(op.gate, cirq.ops.CZPowGate):
    error_2q = cirq.ops.CZPowGate(exponent=np.random.normal(self.mean[0], self.stdev[0]))(*op.qubits)
    error_1q_0 = cirq.ops.ZPowGate(exponent=self.single_errors[op.qubits[0]])(op.qubits[0])
    error_1q_1 = cirq.ops.ZPowGate(exponent=self.single_errors[op.qubits[1]])(op.qubits[1])
    return [op, error_2q, error_1q_0, error_1q_1]
  return op
\end{lstlisting}

\begin{figure}[H]
\begin{center}
\includegraphics[width=0.6\linewidth]{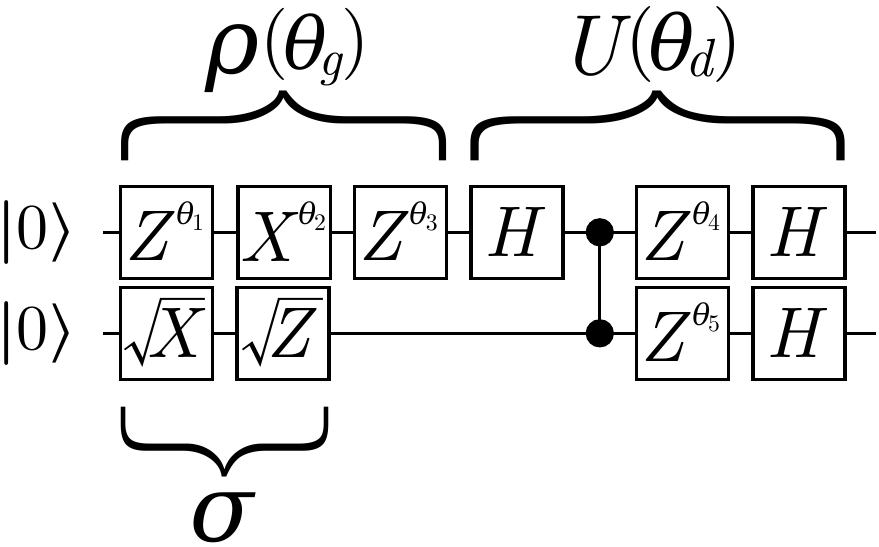}
\caption{EQ-GAN experiment for learning a single-qubit state. The discriminator $(U(\theta_d)$ is constructed with free $Z$ rotation angles to suppress CZ gate errors, allowing the generator $\rho(\theta_g)$ to converge closer to the true data state $\sigma$ by varying $X$ and $Z$ rotation angles.}
\label{fig:experiment}
\end{center}
\end{figure}

To add the noise to TFQ, we can simply convert the circuit with noise into a tensor. For instance, to generate swap tests over \Colorbox{bkgd}{\lstinline{n_data}} datapoints:

\begin{lstlisting}
swap_test_circuit = swap_test_circuit.with_noise(noise_model)
swap_test_circuit = tf.tile(tfq.convert_to_tensor([swap_test_circuit]), tf.constant([n_data]))
\end{lstlisting}

As an example, we consider the task of learning the superposition $\frac{1}{\sqrt{2}}(\ket{0} + \ket{1})$ on a quantum device with noise (Fig.~\ref{fig:experiment}). The discriminator is defined by a swap test with CZ gate providing the necessary two-qubit operation. To learn to correct gate errors, however, the discriminator adversarially learns the angles of single-qubit $Z$ rotations insert directly after the $CZ$ gate. Hence, the EQ-GAN obtains a state overlap significantly better than that of the perfect swap test.

While the noise suppression experiment can be evaluated with a simulated noise model as described above, the full model can also be run by straightforwardly changing the backend via the Google Quantum Engine API.

\begin{lstlisting}
engine = cirq.google.Engine(project_id=project_id)
backend = engine.sampler(processor_id=['rainbow'], gate_set=cirq.google.XMON)
\end{lstlisting}

The hardware backend can then be directly applied in a model.

\begin{lstlisting}
expectation_output = tfq.layers.Expectation(backend=backend)(
  full_swaptest,
  symbol_names=generator_symbols,
  operators=circuits.swap_readout_op(generator_qubits, data_qubits),
  initializer=tf.constant_initializer(generator_initialiation))
\end{lstlisting}

When evaluating the adversarial swap test on a calibrated quantum device, we found that that the error was reduced by a factor of around four compared to a direct implementation of the standard swap test~\cite{eqgan}.

\subsubsection{Approximate Quantum Random Access Memory}
To run this example, see the Colab notebook at:
\fancylink{https://github.com/tensorflow/quantum/blob/research/eq_gan/variational_qram.ipynb}{research/eq\_gan/variational\_qram.ipynb}

When applying quantum machine learning to classical data, most algorithms require access to a quantum random access memory (QRAM) that stores the classical dataset in superposition. More particularly, a set of classical data can be described by the empirical distribution $\{P_i\}$ over all possible input data $i$. Most quantum machine learning algorithms require the conversion from  $\{P_i\}$ into a quantum state $\sum_i \sqrt{P_i}\ket{\psi_i}$, i.e. a superposition of orthogonal basis states $\ket{\psi_i}$ representing each single classical data entry with an amplitude proportional to the square root of the classical probability $P_i$. Preparing such a superposition of an arbitrary set of $n$ states takes $O(n)$ operations at best, which ruins the exponential speedup. Given a suitable ansatz, we may use an EQ-GAN to learn a state approximately equivalent to the superposition of data.

\begin{figure}[H]
\begin{center}
\includegraphics[width=\linewidth]{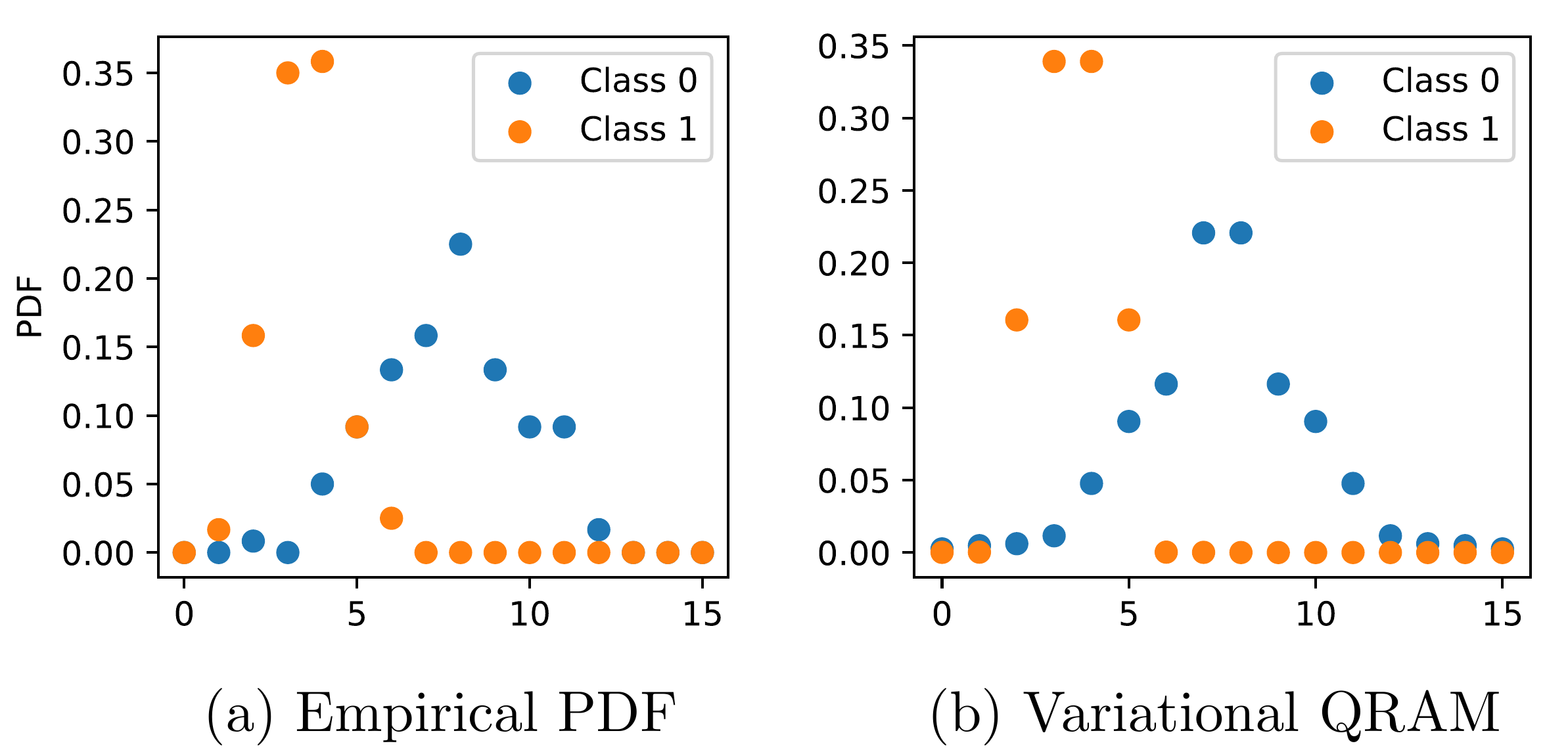}
\caption{Two-peak total dataset (sampled from normal distributions, $N=120$) and variational QRAM of the training dataset ($N=60$). The variational QRAM is obtained by training an EQ-GAN to generate a state $\rho$ with the shallow peak ansatz to approximate an exact superposition of states $\sigma$. The training and test datasets (each $N=60$) are both balanced between the two classes.}
\label{fig:pencode}
\end{center}
\end{figure}

To demonstrate a variational QRAM, we consider a dataset of two peaks sampled from different Gaussian distributions. Exactly encoding the empirical probability density function requires a very deep circuit and multiple-control rotations; similarly, preparing a Gaussian distribution on a device with planar connectivity requires deep circuits. Hence, we select shallow circuit ansatzes that generate concatenated exponential functions to approximate a symmetric peak~\cite{PhysRevA.102.012612}. Once trained to approximate the empirical data distribution, the variational QRAM closely reproduces the original dataset (Fig.~\ref{fig:pencode}).

As a proof of principle for using such QRAM in a quantum machine learning context, we demonstrate in the example code an application of the QRAM to train a quantum neural network~\cite{farhi2018classification}. The loss function is computed either by considering each data entry individually (encoded as a quantum circuit) or by considering each class individually (encoded as a superposition in variational QRAM). Given the same number of circuit evaluations to compute gradients, the superposition converges to a better accuracy at the end of training despite using an approximate distribution.

\vspace{0.5em}

\subsection{Reinforcement Learning with Parameterized Quantum Circuits}

\subsubsection{Background}

Reinforcement learning (RL) \cite{sutton1998introduction} is one of the three main paradigms of machine learning. It pertains to a learning setting where an agent interacts with an environment as to solve a sequential decision task set by this environment, e.g., playing Atari games \cite{mnih2015human} or navigating cities without a map \cite{mirowski2018learning}. This interaction is commonly divided into episodes $(s_0, a_0, r_0, s_1, a_1, r_1, \ldots)$ of states, actions and rewards exchanged between agent and environment. For each state $s$ the agent perceives, it performs an action $a$ sampled from its policy $\pi(a|s)$, i.e., a probability distribution over actions given states. The environment then rewards the agent's action with a real-valued reward $r$ and updates the state of the agent. This interaction repeats cyclically until episode termination. The goal of the agent here is to optimize its policy $\pi(a|s)$ as to maximize its expected rewards in an episode, and hence solve the task at hand. More precisely, the agent's figure of merit is defined by a so-called value function
\begin{equation}\label{eq:value-function}
    V_{\pi}(s_0) = \mathbb{E}_{\pi} \left[\sum_{t=0}^{H} \gamma^t r_t\right]
\end{equation}
where $H$ is the horizon (or length) of an episode, and $\gamma$ is a discount factor in $[0,1]$ that adjusts the relative value of immediate versus long-term rewards.\\

\begin{figure}[H]
\begin{center}
\includegraphics[width=\linewidth]{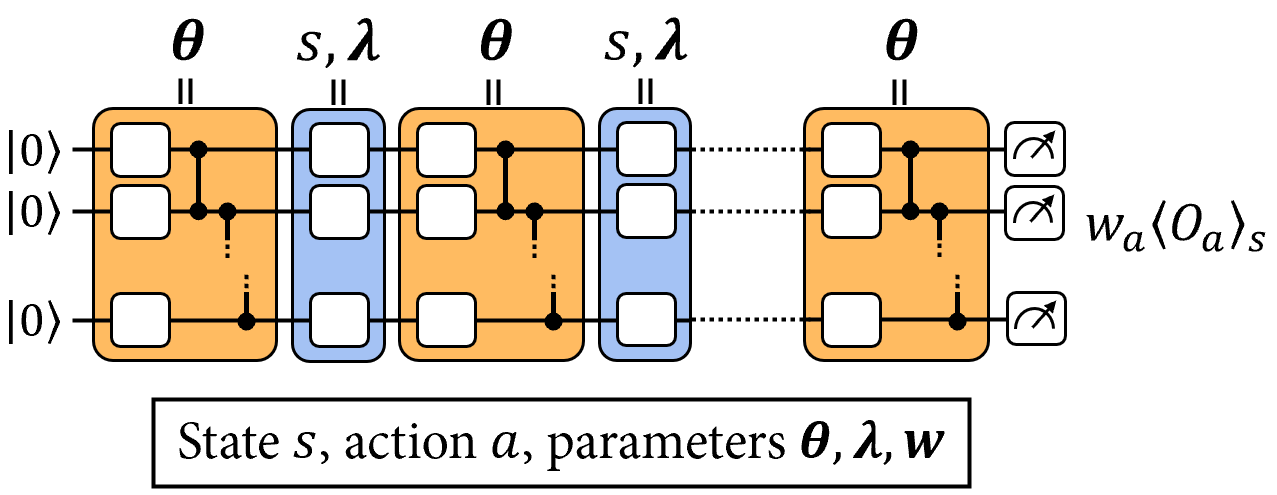}
\caption{A parametrized quantum circuit with data re-uploading as used in quantum reinforcement learning. Each white gate represents a single-qubit rotation, parametrized either by a variational angle $\theta$ in the variational layers (orange), or by a data component $s$ (re-scaled by a parameter $\lambda$) in encoding layers (blue).}
\label{fig:data-reuploading}
\end{center}
\end{figure}

Traditionally, RL algorithms belong to two families:
\begin{itemize}[leftmargin=3mm]
    \item Policy-based algorithms, where an agent's policy is defined by a parametrized model $\pi_{\bm{\theta}}(a|s)$ and is updated via steepest ascent on its resulting value function.
    \item Value-based algorithms, where a parametrized model $V_{\bm{\theta}}(s)$ is used to approximate the optimal value function, i.e., associated to an optimal policy. The agent interacts with the environment with a policy that generally depends on $V_{\bm{\theta}}(s)$ and its experienced rewards are used to improve the quality of the approximation. 
\end{itemize}
When facing environments with large (or continuous) state/action spaces, DNNs recently became the standard models used in both policy-based and value-based approaches. Deep RL has achieved a number of unprecedented achievements, such as superhuman performance in Go \cite{silver2017mastering}, StarCraft II \cite{vinyals2019grandmaster} and Dota 2 \cite{berner2019dota}. More recently, several proposals have been made to enhance deep RL agents with quantum analogues of DNNs (QNNs, or PQCs), both in policy-based \cite{jerbi2021variational} and value-based RL \cite{chen2020variational,lockwood2020reinforcement,skolik2021quantum,lockwood2021playing}. These works essentially described how to adapt deep RL algorithms to work with PQCs as either policies or value-function approximators, and tested their performance on benchmarking environments \cite{brockman2016openai}. As pointed out in \cite{jerbi2021variational,skolik2021quantum}, certain design choices of PQCs can lead to significant gaps in learning performance. The most crucial of these choices is arguably the data encoding strategy. As supported by theoretical investigations \cite{schuld2021effect,perez2020data}, data re-uploading (see Fig.~\ref{fig:data-reuploading}) where encoding layers and variational layers of parametrized gates are sequentially alternated, stands out as a way to get highly expressive models.

\subsubsection{Policy-Gradient Reinforcement Learning with PQCs}

In the following, we focus on the implementation of a policy-based approach to RL using PQCs \cite{jerbi2021variational}. For that, let us define the parametrized policy of our agent as:
\begin{equation}\label{eq:PQC-policy}
    \pi_{\bm{\theta}} (a|s) = \frac{e^{w_a\braket{O_a}_{s,\bm{\theta},\bm{\lambda}}}} {e^{w_{a'}\braket{O_{a'}}_{s,\bm{\theta},\bm{\lambda}}}}
\end{equation}
where $\braket{O_a}_{s,\bm{\theta},\bm{\lambda}}$ is the expectation value of an observable $O_a$ (e.g., Pauli $Z$ on first qubit) associated to action $a$, as measured at the output of the PQC. This expectation value is also augmented by a trainable weight $w_a$, which is also action specific (see Fig.~\ref{fig:data-reuploading}).

As stated above, the parameters of this policy are updated as to perform gradient ascent on their corresponding value function (see \eqref{eq:value-function}). Thanks to the so-called policy gradient theorem \cite{sutton1999policy}, we know that another formulation of this objective function is given by the following loss function:
\begin{equation*}
\mathcal{L}(\bm{\theta}) = -\frac{1}{|\mathcal{B}|}\sum_{s_0,a_0,r_0, \ldots \in \mathcal{B}} \left[\sum_{t=0}^{H-1} \log(\pi_{\bm{\theta}}(a_t|s_t)) \sum_{t'=1}^{H-t} \gamma^{t'} r_{t+t'} \right]
\end{equation*}
for a batch $\mathcal{B}$ of episodes $(s_0,a_0,r_0, ...)$ sampled by following $\pi_{\bm{\theta}}$ in the environment. This expression has the advantage that it avoids numerical differentiation to compute the gradient of the loss with respect to $\bm{\theta}$.\\

\noindent\textbf{Target problems}:
\begin{enumerate}[noitemsep]
    \item Implement a parametrized quantum circuit with data re-uploading
    \item Train a reinforcement learning agent based on this quantum circuit
\end{enumerate}
\noindent\textbf{Required TFQ functionalities}:
      \begin{enumerate}[noitemsep]
        \item Parametrized circuit layers
        \item Custom Keras layers
        \item Automatic differentiation w.r.t.\ a custom loss using \Colorbox{bkgd}{\lstinline{tf.GradientTape}}
    \end{enumerate}

\subsubsection{Implementation}
To run this example in the browser through Colab, follow the link:
\fancylink{https://github.com/tensorflow/quantum/blob/master/docs/tutorials/quantum_reinforcement_learning.ipynb}{docs/tutorials/quantum\_reinforcement\_learning.ipynb}

We demonstrate the implementation of the quantum policy-gradient algorithm on CartPole-v1, a benchmarking task from OpenAI Gym \cite{brockman2016openai}. This environment has a 4-dimensional continuous state space and a discrete action space of size 2. We hence use a PQC acting on 4 qubits, on which we evaluate 2 expectation values. 

We start by implementing the quantum circuit used as the agent's PQC. It is composed of alternating layers of variational single-qubit rotations, entangling CZ operators, and data-encoding single-qubit rotations.

\begin{lstlisting}
def one_qubit_rotation(q, symbols):
  return [cirq.X(q)**symbols[0], cirq.Y(q)**symbols[1], cirq.Z(q)**symbols[2]]

def entangling_layer(qubits):
  cz_ops = [cirq.CZ(q0, q1) for q0, q1 in zip(qubits, qubits[1:])]
  cz_ops += ([cirq.CZ(qubits[0], qubits[-1])] if len(qubits) != 2 else [])
  return cz_ops
    
def generate_circuit(qubits, n_layers):
  n_qubits = len(qubits)
  
  params = sympy.symbols(f'theta(0:{3*(n_layers+1)*n_qubits})')
  params = np.asarray(params).reshape((n_layers+1, n_qubits, 3))
  
  inputs = sympy.symbols(f'x(0:{n_qubits})'+f'(0:{n_layers})')
  inputs = np.asarray(inputs).reshape((n_layers, n_qubits))
  
  circuit = cirq.Circuit()
  for l in range(n_layers):
    circuit += cirq.Circuit(one_qubit_rotation(q, params[l,i]) for i,q in enumerate(qubits))
    circuit += entangling_layer(qubits)
    circuit += cirq.Circuit(cirq.rx(inputs[l,i])(q) for i,q in enumerate(qubits))
  circuit += cirq.Circuit(one_qubit_rotation(q, params[n_layers,i]) for i,q in enumerate(qubits))

  return circuit, list(params.flat), list(inputs.flat)
\end{lstlisting}

We use this quantum circuit to define a \Colorbox{bkgd}{\lstinline{ControlledPQC}} layer. To sort between variational and encoding angles in the data re-uploading scheme, we include the \Colorbox{bkgd}{\lstinline{ControlledPQC}} in a custom Keras layer. This \Colorbox{bkgd}{\lstinline{ReUploadingPQC}} layer will  manage the trainable parameters (variational angles $\bm{\theta}$ and input-scaling parameters $\bm{\lambda}$) and resolve the input values (input state $s$) into the appropriate symbols in the circuit.

\begin{lstlisting}
class ReUploadingPQC(tf.keras.layers.Layer):
  def __init__(self, qubits, n_layers, observables, activation='linear', name="re-uploading_PQC"):
    super(ReUploading, self).__init__(name=name)
    self.n_layers = n_layers

    circuit, theta_symbols, input_symbols = generate_circuit(qubits, n_layers)
    self.computation_layer = tfq.layers.ControlledPQC(circuit, observables)

    theta_init = tf.random_uniform_initializer(minval=0., maxval=np.pi)
    self.theta = tf.Variable(initial_value=theta_init(shape=(1, len(theta_symbols)), dtype="float32"), trainable=True, name="thetas")
    lmbd_init = tf.ones(shape=(len(qubits)*n_layers,))
    self.lmbd = tf.Variable(initial_value=lmbd_init, dtype="float32", trainable=True, name="lambdas")
    symbols = [str(x) for x in theta_symbols+input_symbols]
    self.indices = tf.constant([sorted(symbols).index(a) for a in symbols])
    self.activation = activation
    self.empty_circuit = tfq.convert_to_tensor([cirq.Circuit()])

  def call(self, inputs):
    batch_dim = tf.gather(tf.shape(inputs[0]), 0)
    tiled_up_circuits = tf.repeat(self.empty_circuit, repeats=batch_dim)
    tiled_up_thetas = tf.tile(self.theta, multiples=[batch_dim, 1])
    tiled_up_inputs = tf.tile(inputs[0], multiples=[1, self.n_layers])
    scaled_inputs = tf.einsum("i,ji->ji", self.lmbd, tiled_up_inputs)
    squashed_inputs = tf.keras.layers.Activation(self.activation)(scaled_inputs)
    joined_vars = tf.concat([tiled_up_thetas, squashed_inputs], axis=1)
    joined_vars = tf.gather(joined_vars, self.indices, axis=1)
	
    return self.computation_layer([tiled_up_circuits, joined_vars])
\end{lstlisting}

We also implement a custom Keras layer to post-process the expectation values $\braket{O_a}_{s,\bm{\theta},\bm{\lambda}}$ at the output of the PQC. Here, the \Colorbox{bkgd}{\lstinline{Alternating}} layer multiplies a single $\braket{Z_0Z_1Z_2Z_3}_{s,\bm{\theta},\bm{\lambda}}$ expectation value by weights $(w_0,w_1)$ initialized to $(1,-1)$.

\begin{lstlisting}
class Alternating(tf.keras.layers.Layer):
  def __init__(self, output_dim):
    super(Alternating, self).__init__()
    self.w = tf.Variable(initial_value=tf.constant([[(-1.)**i for i in range(output_dim)]]), dtype="float32", trainable=True, name="obs-weights")

  def call(self, inputs):
    return tf.matmul(inputs, self.w)

ops = [cirq.Z(q) for q in qubits]
observables = [reduce((lambda x, y: x*y), ops)]
\end{lstlisting}

We now put together all these layers to define a Keras model of the policy in equation \eqref{eq:PQC-policy}.

\begin{lstlisting}
def generate_model_policy(qubits, n_layers, n_actions, beta, observables):
  input_tensor = tf.keras.Input(shape=(len(qubits), ), dtype=tf.dtypes.float32, name='input')
  re_uploading = ReUploading(qubits, n_layers, observables)([input_tensor])
  process = tf.keras.Sequential([
    Alternating(n_actions),
    tf.keras.layers.Lambda(lambda x: x * beta),
    tf.keras.layers.Softmax()
  ], name="observables-policy")
  policy = process(re_uploading)
  model = tf.keras.Model(inputs=[input_tensor], outputs=policy)
  return model
\end{lstlisting}

We now move on to the implementation of the learning algorithm. We start by defining two helper functions: a \Colorbox{bkgd}{\lstinline{gather_episodes}} function that gathers a batch of episodes of interaction with the environment, and a \Colorbox{bkgd}{\lstinline{compute_returns}} function that computes the discounted sums of rewards appearing in the agent's loss function.

\begin{lstlisting}
def gather_episodes(state_bounds, n_actions, model, n_episodes, env_name):
  trajectories = [defaultdict(list) for _ in range(n_episodes)]
  envs = [gym.make(env_name) for _ in range(n_episodes)]
  done = [False for _ in range(n_episodes)]
  states = [e.reset() for e in envs]

  while not all(done):
    unfinished_ids = [i for i in range(n_episodes) if not done[i]]
    normalized_states = [s/state_bounds for i, s in enumerate(states) if not done[i]]
    for i, state in zip(unfinished_ids, normalized_states):
      trajectories[i]['states'].append(state)

    # Compute policy for all unfinished envs in parallel
    states = tf.convert_to_tensor(normalized_states)
    action_probs = model([states])

    # Store action and transition all environments to the next state
    states = [None for i in range(n_episodes)]
    for i, policy in zip(unfinished_ids, action_probs.numpy()):
      action = np.random.choice(n_actions, p=policy)
      states[i], reward, done[i], _ = envs[i].step(action)
      trajectories[i]['actions'].append(action)
      trajectories[i]['rewards'].append(reward)
  return trajectories

def compute_returns(rewards_history, gamma):
  returns = []
  discounted_sum = 0
  for r in rewards_history[::-1]:
    discounted_sum = r + gamma * discounted_sum
    returns.insert(0, discounted_sum)
  return returns
\end{lstlisting}

To train the policy, we need a function that updates the model parameters. This is done via gradient descent on the loss of the agent. Since the loss function in policy-gradient approaches is more involved than, for instance, a supervised learning loss, we use a \Colorbox{bkgd}{\lstinline{tf.GradientTape}} to store the contributions of our model evaluation to the loss. When all contributions have been added, this tape can then be used to backpropagate the loss on all the model evaluations and hence compute the required gradients.

\begin{lstlisting}
@tf.function
def reinforce_update(states, actions, returns, model):
  states = tf.convert_to_tensor(states)
  actions = tf.convert_to_tensor(actions)
  returns = tf.convert_to_tensor(returns)

  with tf.GradientTape() as tape:
    tape.watch(model.trainable_variables)
    logits = model(states)
    p_actions = tf.gather_nd(logits, actions)
    log_probs = tf.math.log(p_actions)
    loss = tf.math.reduce_sum(-log_probs * returns) / batch_size
  grads = tape.gradient(loss, model.trainable_variables)
  for optimizer, w in zip([optimizer_in, optimizer_var, optimizer_out], [w_in, w_var, w_out]):
    optimizer.apply_gradients([(grads[w], model.trainable_variables[w])])
\end{lstlisting}

With this, we can implement the main training loop of the agent.

\begin{lstlisting}
env_name = "CartPole-v1"

for batch in range(n_episodes // batch_size):
  # Gather episodes
  episodes = gather_episodes(state_bounds, n_actions, model, batch_size, env_name)
  
  # Group states, actions and returns in arrays
  states = np.concatenate([ep['states'] for ep in episodes])
  actions = np.concatenate([ep['actions'] for ep in episodes])
  rewards = [ep['rewards'] for ep in episodes]
  returns = np.concatenate([compute_returns(ep_rwds, gamma) for ep_rwds in rewards])
  returns = np.array(returns, dtype=np.float32)
  id_action_pairs = np.array([[i, a] for i, a in enumerate(actions)])
  
  # Update model parameters
  reinforce_update(states, id_action_pairs, returns, model)
\end{lstlisting}

Thanks to the parallelization of model evaluations and fast automatic differentiation using the \Colorbox{bkgd}{\lstinline{tfq.differentiators.Adjoint()}} differentiator, the execution of this training loop for $500$ episodes takes about $15$ minutes on a regular laptop (for a PQC acting on $4$ qubits and with $5$ re-uploading layers).

\subsubsection{Value-Based Reinforcement Learning with PQCs}

The example above provides the code for a policy-based approach to RL with PQCs. The tutorial under this link:
\fancylink{https://github.com/tensorflow/quantum/blob/master/docs/tutorials/quantum_reinforcement_learning.ipynb}{docs/tutorials/quantum\_reinforcement\_learning.ipynb}
also implements the value-based algorithm introduced in \cite{skolik2021quantum}. Aside from the learning mechanisms specific to deep value-based RL (e.g., a replay memory to re-use past experience and a target model to stabilize learning), these two methods essentially differ in the role played by the expectation values $\braket{O_a}_{s,\bm{\theta},\bm{\lambda}}$ of the PQC. In our quantum approach to value-based RL, these correspond to approximations of the values $Q(s,a)$ (i.e., the value function $V(s)$ taken as a function of a state and an action), trained using a loss function of the form:
\begin{equation*}
    \mathcal{L}(\bm{\theta}) = \frac{1}{|\mathcal{B}|}\sum_{s,a,r,s' \in \mathcal{B}} \left(Q_{\bm{\theta}}(s,a) - [r +\max_{a'} Q_{\bm{\theta'}}(s',a')]\right)^2
\end{equation*}
derived from Q-learning \cite{sutton1998introduction}. We refer to \cite{skolik2021quantum,lockwood2020reinforcement} and the tutorial for more details.

\subsubsection{Quantum Environments}

In the initial works, it was proven that there exist learning environments that can only be efficiently tackled by quantum agents (barring well-established computational assumptions) \cite{jerbi2021variational}. However, these problems are artificial and contrived, and it is a key question in the field of Quantum RL whether there exist natural environments for which quantum agents can have a large learning advantage over their classical counterparts. Intuitive candidates are RL environments that are themselves quantum in nature. In this setting, early results have already demonstrated that variational quantum methods can be advantageous when the data perceived by a learning agent stems from measurements on a quantum system \cite{jerbi2021variational}. Going a step further, one could also consider a setting where the states perceived by the agent are genuinely quantum and can therefore be processed directly by a PQC, as was recently explored in a quantum control environment \cite{wu2020quantum}. 

\vspace{0.5em}

\section{Closing Remarks}

The rapid development of quantum hardware represents an impetus for the equally rapid development of quantum applications. In October 2017, the Google AI Quantum team and collaborators released its first software library, OpenFermion, to accelerate the development of quantum simulation algorithms for chemistry and materials sciences. Likewise, TensorFlow Quantum is intended to accelerate the development of quantum machine learning algorithms for a wide array of applications. Quantum machine learning is a very new and exciting field, so we expect the framework to change with the needs of the research community, and the availability of new quantum hardware. We have open-sourced the framework under the commercially friendly Apache2 license, allowing future commercial products to embed TFQ royalty-free. If you would like to participate in our community, visit us at:

\fancylink{https://github.com/tensorflow/quantum/}{github.com/tensorflow/quantum/}
\vspace{0.5 em}

\section{Acknowledgements}
The authors would like to thank Google Research for supporting this project. In particular, M.B., G.V., T.M., and A.J.M. would like to thank the Google Quantum AI team for their support during their respective internships, and several useful discussions with Matt Harrigan, John Platt, and Nicholas Rubin. The authors would like to also thank Achim Kempf from the University of Waterloo for sponsoring this project. M.B. and J.Y. would like to thank the Google Brain and Core team for supporting this project, in particular Christian Howard, Billy Lamberta, Tom O'Malley, Francois Chollet, Yifei Feng, David Rim, Justin Hong, and Megan Kacholia. G.V., A.J.M. and J.Y. would like to thank Stefan Leichenauer, Jack Hidary and the rest of the Sandbox@Alphabet team for their support during their respective Quantum Residencies. G.V. acknowledges support from NSERC. D.B. is an Associate Fellow in the CIFAR program on Quantum Information Science. A.S. and M.S. were supported by the USRA Feynman Quantum Academy funded by the NAMS R\&D Student Program at NASA Ames Research Center and by the Air Force Research Laboratory (AFRL), NYSTEC-USRA Contract (FA8750-19-3-6101). A.Z. acknowledges support from Caltech's Intelligent Quantum Networks and Technologies (INQNET) research program and by the DOE/HEP QuantISED program grant, Quantum Machine Learning and Quantum Computation Frameworks (QMLQCF) for HEP, award number DE-SC0019227. S.J. acknowledges support from the Austrian Science Fund (FWF) through the projects DK-ALM:W1259-N27 and SFB BeyondC F7102, and from the Austrian Academy of Sciences as a recipient of the DOC Fellowship. V.D. acknowledges support from the Dutch Research Council (NWO/OCW), as part of the Quantum Software Consortium programme (project number 024.003.037).

\bibliography{lib.bib}

\end{document}